\pdfoutput=1

\documentclass[12pt,reqno]{article}
\usepackage{jheppub}
\usepackage{amsmath,amssymb,amsfonts}

\usepackage{url}

\usepackage{natbib}
\usepackage{graphicx}

\usepackage[usenames,dvipsnames]{xcolor}
\usepackage{dcolumn}
\usepackage{tikz}
\usetikzlibrary{shapes.geometric, arrows}
\usepackage{upgreek}
\usepackage{setspace}
\usepackage{enumitem}
\usepackage{array,multirow,bigdelim,arydshln}
\usepackage{mathrsfs}

\def\be{\begin{equation}}
\def\ee{\end{equation}}
\def\ba{\begin{eqnarray}}
\def\ea{\end{eqnarray}}

\def\a{\alpha}
\def\b{\beta}

\def\b#1{\overline{#1}}

\def\CP1{\mathbb{CP}^1}
\def\SL2C{\mathrm{SL}(2,\mathbb{C})}

\def\Z2{\mathbb{Z}_2}

\def\su2{{SU(2)}}

\def\a{{\alpha}}

\def\[{\left[}
\def\]{\right]}

\def\s{\sigma}
\def\a{\alpha}
\def\b{\beta}

\def\({\left(}
\def\){\right)}
\def\[{\left[}
\def\]{\right]}

\def\<{\langle}
\def\>{\rangle}

\def\i2{\frac{i}{2}}

\title{Computation of Contour Integrals on ${\cal M}_{0,n}$}

\author[\spadesuit]{Freddy Cachazo,}
\author[\spadesuit,\clubsuit]{Humberto Gomez}
\affiliation[\spadesuit]{Perimeter Institute for Theoretical Physics, Waterloo, ON N2L 2Y5, Canada}
\affiliation[\clubsuit]{Instituto de Fisica Teorica UNESP -- Universidade Estadual Paulista,\\
Caixa Postal 70532-2 01156-970 Sao Paulo, SP, Brazil}
\emailAdd{fcachazo@pitp.ca}
\emailAdd{humgomzu@gmail.com}

\abstract{Contour integrals of rational functions over ${\cal M}_{0,n}$, the moduli space of $n$-punctured spheres, have recently appeared at the core of the tree-level S-matrix of massless particles in arbitrary dimensions. The contour is determined by the critical points of a certain Morse function on ${\cal M}_{0,n}$. The integrand is a general rational function of the puncture locations with poles of arbitrary order as two punctures coincide. In this note we provide an algorithm for the analytic computation of any such integral. The algorithm uses three ingredients: an operation we call general KLT, Petersen's theorem applied to the existence of a 2-factor in any 4-regular graph and Hamiltonian decompositions of certain 4-regular graphs. The procedure is iterative and reduces the computation of a general integral to that of simple building blocks. These are integrals which compute double-color-ordered partial amplitudes in a bi-adjoint cubic scalar theory.}

\begin{document}
{\setstretch{1}
\maketitle
}

%\setstretch{1.2}
\onehalfspacing

%%%%%%%%%%%%%%%%%%%%%%%%
\section{Introduction}
%%%%%%%%%%%%%%%%%%%%%%%

The complete tree-level S-matrix of a large variety of field theories of massless particles are now known (or conjectured) to have a description in terms of contour integrals over ${\cal M}_{0,n}$, the moduli space of $n$-punctured Riemann sphere
\cite{Cachazo:2013hca,Cachazo:2013iea,Mason:2013sva,Dolan:2013isa,Berkovits:2013xba,Adamo:2013tsa,Gomez:2013wza,Kalousios:2013eca,Dolan:2014ega,
Geyer:2014fka,Cachazo:2014nsa,Cachazo:2014xea,Ohmori:2015}. Some of these theories are Yang-Mills, Einstein gravity, Dirac-Born-Infeld, and the $U(N)$ non-linear sigma model \cite{Cachazo:2014xea,Ohmori:2015}. The new formulas for the scattering of $n$ particles are given as a sum over multidimensional residues \cite{harris} on ${\cal M}_{0,n}$.

The position of $n$ punctures on a sphere can be given using inhomogenous coordinates as $\{ \sigma_1,\sigma_2,\ldots ,\sigma_n\}$. Three of them can be fixed using $PSL(2,\mathbb{C})$ transformations, say $\sigma_1,\sigma_2,\sigma_3$. Therefore the space is $n-3$ dimensional and we are working locally on a patch isomorphic to $\mathbb{C}^{n-3}$. The next step in the construction is a rational map from ${\mathbb C}^{n-3}\to {\mathbb C}^{n-3}$ which is a function of the entries of a symmetric $n\times n$ matrix, $s_{ab}$, with vanishing diagonal, i.e., $s_{aa}=0$, and all rows adding up to zero. These are the coordinates of the space of kinematic invariant for the scattering of $n$ massless particles. The explicit form of the map is $\{\sigma_4,\sigma_5,\ldots, \sigma_n\}\to \{ E_4,E_5,\ldots ,E_n\}$ with
\be
E_a(\sigma) =\sum_{b=1,b\neq a}^n\frac{s_{ab}}{\sigma_a-\sigma_b} \quad {\rm for} \quad a\in \{1,2,\ldots ,n\}.
\ee

Using this map, scattering amplitudes, denoted as $M_n$, are defined as the sum over the residues of
\be\label{form1}
 \int \prod_{a=4}^{n}d\sigma_a |123|^2 \frac{H(\sigma,k,\epsilon )}{E_4(\sigma)E_5(\sigma)\cdots E_n(\sigma)}
\ee
over all the zeroes of the map $\{E_4,E_5,\ldots ,E_n\}$. Here $|123|\equiv (\sigma_1-\sigma_2)(\sigma_2-\sigma_3)(\sigma_3-\sigma_1)$ and $H(\sigma, k,\epsilon)$ is a rational function that depends on the theory under consideration and contains all information regarding wave functions of the particles such as polarization vectors $\epsilon_a^\mu$ and momenta $k_a^\mu$. The equations defining the zeroes, $E_4=E_5=\cdots E_n= 0$, are known as the scattering equations \cite{Fairlie:1972,Roberts:1972,Fairlie:2008dg,Gross:1987ar,Witten:2004cp,Caputa:2011zk,Caputa:2012pi,Makeenko:2011dm,Cachazo:2012da} More explicitly,
\be
M_n = \sum_{\sigma^*\in Z(E)}\frac{|123|^2H(\sigma^*,k,\epsilon )}{{\rm det}\left.\left(\frac{\partial (E_4,\ldots E_n)}{\partial (\sigma_4\ldots \sigma_n)}\right)\right|_{\sigma^*}}
\ee
where $Z(E)$ is the set of all zeroes of the map. This representation of scattering amplitudes is known as the Cachazo-He-Yuan (CHY) approach \cite{Cachazo:2013hca,Cachazo:2013iea,Cachazo:2014nsa,Cachazo:2014xea}

The zeroes are generically isolated and are the values of $\sigma'$s for which the Morse function on ${\cal M}_{0,n}$
\be
\phi(\s,\bar{\s}) =\frac{1}{2}\,
\sum_{a\,< \,b}\,s_{ab}\,\ln|\s_a-\s_b|^2
\ee
has local extremes\footnote{A Morse function is a real function with non-degenerate critical points \cite{morse}.} \cite{Gross:1987ar,Ohmori:2015}. %Different choices of $\phi(z)$ labeled by $s_{ab}$ can be thought of as a family of complex structure deformations \cite{Dhoker} and the solutions to $E_4=E_5=\cdots E_n= 0$ are the fixed points of the deformation.

In this paper we are not concerned with particular theories. Instead, our aim is to provide an algorithm for the analytic computation of any integral of the form
\be\label{intt}
\int_{\Gamma}\prod_{a=4}^n d\sigma_a \frac{|123|^2}{E_4(\sigma)E_5(\sigma)\cdots E_n(\sigma)}F(\sigma ),
\ee
where $\Gamma$ is the same contour as above, i.e., a sum over all residues at $Z(E)$. Here $F(\sigma)$ is any rational function of {\it only} the puncture coordinates $\sigma$'s which transforms as
\be
F(\sigma ) \to \prod_{a=1}^n(\textsf{c}\,\sigma_a + \textsf{d})^4 F(\sigma ), \quad {\rm under} \quad \sigma_a \to \frac{\textsf{a}\,\sigma_a + \textsf{b}}{\textsf{c}\,\sigma_a + \textsf{d}},
\ee
with $\textsf{a}\textsf{d}-\textsf{b}\textsf{c}=1 $, i.e., under an $PSL(2,\mathbb{C})$ transformation.

The transformation of $F(\sigma )$ ensures that the integral \eqref{intt} is independent of both the choice of which puncture coordinates to fix and their values. The transformation also implies that $F(\sigma)$ is only a function of differences $\sigma_a-\sigma_b$ which we denote as $\sigma_{ab}$. Clearly $\sigma_{ab}=-\sigma_{ba}$. The only other condition we impose on $F(\sigma_{ab})$ is that all its poles are of the form $\sigma_{ab}^m$ for some integer $m\geq 0$.

The simplest kind of integrals are defined in terms of the so-called Parke-Taylor factors \cite{Parke:1986gb} defined for a particular ordering of $n$ labels $(\alpha(1)\alpha(2)\cdots \alpha(n))$ with $\alpha\in S_n$ as
\be\label{ptf}
\frac{1}{(\alpha(1)\alpha(2)\cdots \alpha(n))} \equiv \frac{1}{\sigma_{\alpha(1)\alpha(2)}\,\sigma_{\alpha(2)\alpha(3)}\cdots \sigma_{\alpha(n-1)\alpha(n)}\,\sigma_{\alpha(n)\alpha(1)}}.
\ee
Clearly, any Parke-Taylor factor has half the $PSL(2,\mathbb{C})$ weight needed to construct a valid $F(\sigma_{ab})$ \cite{Cachazo:2013hca,Cachazo:2014xea}. One can define integrals labeled by a pair a permutations $\alpha,\beta\in S_n$ using
\be
F^{\alpha,\beta}(\sigma_{ab}) = \frac{1}{(\alpha(1)\alpha(2)\cdots \alpha(n))}\times \frac{1}{(\beta(1)\beta(2)\cdots \beta(n))},
\ee
or more explicitly \cite{Cachazo:2013iea}
\be
m(\alpha|\beta)\equiv \int_\Gamma d\mu_n \frac{1}{(\alpha(1)\alpha(2)\cdots \alpha(n))}\frac{1}{(\beta(1)\beta(2)\cdots \beta(n))},
\ee
where we have introduced a shorthand notation for the measure
\be\label{meas}
d\mu_n \equiv \prod_{a=4}^n d\sigma_a \frac{|123|^2}{E_4(\sigma)E_5(\sigma)\cdots E_n(\sigma)}.
\ee

Integrals of the form $m(\alpha|\beta)$ have been studied in the literature and are known to evaluate to a sum over connected tree Feynman graphs with only cubic (trivalent) interactions which are compatible with the two planar orderings defined by $\alpha$ and $\beta$ \cite{Cachazo:2013iea}. We review this result in detail in section \ref{bb} and explain how to explicitly evaluate them as a rational function of the variables $s_{ab}$. Here it suffices to say that these known integrals form the basic building blocks of our construction and the main result of this work is an algorithm for writing 
\be\label{result}
\int_\Gamma d\mu_n F(\sigma_{ab}) = R(m(\alpha|\beta)),
\ee
where $R$ is a rational function of its variables with only numerical coefficients.

The reason general integrals are of interest can be seen, for example, in the evaluation of an $n$ graviton amplitude which contains a term of the form \cite{Cachazo:2013hca,Cachazo:2013iea,Cachazo:2014nsa}
\be\label{mons}
\int_\Gamma d\mu_n \,\frac{(\epsilon_1\cdot\epsilon_2)^2}{\sigma_{12}^4}\prod_{a=3}^n\left(\sum_{b=2,b\neq a}^n\epsilon_a\cdot k_b\frac{\sigma_{1b}}{\sigma_{ab}\sigma_{1a}}\right)^2.
\ee
In this formula $\epsilon_c,k_c$ are fixed data and after fully expanding \eqref{mons} they can be factored out leaving arbitrarily complicated integrals of the form \eqref{result} to be evaluated. Also motivated by the same physical problem, Kalousios developed a technique, different from the one presented here, for the computation of general five-point integrals in \cite{Kalousios:2015fya}.

The algorithm we develop is based on three key constructions. The first is a generalization of the Kawai-Lewellel-Tye (KLT) relation \cite{Kawai:1985xq,Berends:1988,Bern:1998sv}. The KLT relation was originally discovered as a relation among closed and open string theory amplitudes but since then it has inspired similar relations in field theory and more recently it found a natural set up which allows vast generalizations in the CHY representation of amplitudes. We present the general KLT construction in section \ref{kltgen}.

The second result is a classic one from graph theory \cite{graph1}. Consider an integrand $F(\sigma)$ such that it does not have any zeroes. This means that it is only the product of $2n$ factors $\sigma_{ab}$ in the denominator with a trivial numerator that can be set to unity. Representing each puncture by a vertex and each $\sigma_{ab}$ by an undirected edge connecting vertices $a$ and $b$ one finds that each $F(\sigma)$ leads to a unique 4-regular graph $G_F$ (not necessarily simple). A classic result of Petersen guarantees that any 4-regular graph with $n$ vertices is 2-factorable. This means that $G_F$ it is always the union of two 2-regular graphs with $n$ vertices. Petersen's result is reviewed in section \ref{peter}.

The third and final ingredient is an observation regarding the existence of a Hamiltonian decomposition of graphs \cite{graph1,graph2}. In order to state the observation let us choose any 2-regular multigraph\footnote{In this work we use the terminology graph and multigraph interchangeably. In fact, the restriction to simple graphs is never necessary.} $G$ with $n$-vertices and no loops. We say that a connected 2-regular graph with $n$-vertices, $H^{\rm conn}$, is compatible with $G$ if the 4-regular graph obtained from the union of $G$ and $H^{\rm conn}$ contains two edge-disjoint Hamilton cycles. The observation is that out of the $(n-1)!$ possible connected graphs the number of compatible graphs with $G$ is always larger than $(n-3)!$. This is explained in section \ref{HD}.

In section \ref{mainalgo} all ingredients are combined to produce the final algorithm for computing the rational function $R$ in \eqref{result}. The algorithm is general but in particular cases it can be modified to make it much more efficient.

Section \ref{allsix} is devote to examples that not only illustrate the use of the algorithm but also give the explicit Hamiltonian decompositions needed for the computation of the most general six-point integral.

In section \ref{disc} we end with discussions including future directions and physical applications in the form of novel relations among amplitudes. The appendix has a detailed explanation of how to implement Petersen's theorem. The implementation is not far from being the actual proof so it a good way to gain intuition on why the theorem holds.

%%%%%%%%%%%%%%%%%%%%%%%%%%%%%%%%%%%%%%%%%%%%%%%%%%%%
\section{Definition of Building Blocks}\label{bb}
%%%%%%%%%%%%%%%%%%%%%%%%%%%%%%%%%%%%%%%%%%%%%%%%%%%%

The aim of this work is to provide an algorithm for the reduction of contour integrals on the moduli space of an n-punctured sphere of the form
\be
\int d\mu_n F(\sigma )
\ee
in terms of a basis of known integrals. Ensuring that the integrand is $PSL(2,\mathbb{C})$ invariant implies that $F(\sigma )$ has the form
\be
F(\sigma ) = \frac{1}{(12\cdots n)(\gamma(1)\gamma(2)\cdots \gamma(n))}f(r_{ijkl}),
\ee
where $(12\cdots n)$ is the canonical Parke-Taylor and $(\gamma(1)\gamma(2)\cdots \gamma(n))$ is a Parke-Taylor factor with a $\gamma\in S_n$ ordering (see \eqref{ptf} for the Parke-Taylor factor definition). $f$ is a rational function of $r_{ijkl}$ which are general cross ratios, i.e.,
\be
r_{ijkl} \equiv \frac{\sigma_{ij}\sigma_{kl}}{\sigma_{il}\sigma_{jk}}.
\ee
Of course, the choice of Parte-Taylor factor is completely arbitrary and can be conveniently made depending on the case. The measure $d\mu_n$ was defined in \eqref{meas} and is reviewed below.

In this section we discuss the basic building blocks which are special contour integrals with $f(r_{ijkl})=1$ and whose values are explicitly known \cite{Cachazo:2013iea,Dolan:2013isa,Dolan:2014ega}. The building blocks are labeled by a pair of permutations $\alpha,\beta\in S_n/\mathbb{Z}_n$. The reason one has to mod out by cyclic permutations $\mathbb{Z}_n$ is obvious from the definition
\be\label{mdef}
m(\alpha|\beta)\equiv \int_\Gamma d\mu_n \frac{1}{(\alpha(1)\alpha(2)\cdots \alpha(n))}\,\frac{1}{(\beta(1)\beta(2)\cdots \beta(n))}.
\ee
 Recall that $(\alpha(1)\alpha(2)\cdots \alpha(n)) \equiv \sigma_{\alpha(1)\alpha(2)}\sigma_{\alpha(2)\alpha(3)}\cdots \sigma_{\alpha(n-1)\alpha(n)}\sigma_{\alpha(n)\alpha(1)}$ and the measure is
\be
d\mu_n \equiv \prod_{a=4}^nd\sigma_a \frac{|123|^2}{E_4E_5\cdots E_n}.
\ee
An explicit evaluation of the integral $m(\alpha|\beta)$ would involve solving the equations \cite{Cachazo:2013gna,Dolan:2013isa,Kalousios:2013eca,
Dolan:2014ega,Weinzierl:2014vwa,Kalousios:2015fya}
\be
E_a(\sigma) =\sum_{b=1,b\neq a}^n\frac{s_{ab}}{\sigma_a-\sigma_b} = 0 \quad {\rm for} \quad a\in \{4,5,\ldots ,n\}.
\ee
These equations have $(n-3)!$ solutions as proven in \cite{Cachazo:2013hca,Cachazo:2013gna} and the data $s_{ab}$ can be taken to be the components of a symmetric $n\times n$ matrix of complex entries such that $s_{11}=s_{22}=\cdots s_{nn}=0$ and
\be\label{kin}
\sum_{b=1,b\neq a}^n s_{ab} = 0 ~~{\rm for }~~ a\in \{1,2,\ldots ,n\}.
\ee
Once the solutions are found one computes the Jacobian matrix
\be
\Phi_{ab} =\left\{
             \begin{array}{cc}
               \frac{s_{ab}}{\sigma_{ab}^2} & a\neq b, \\
               -\sum_{c=1,c\neq a}^n \frac{s_{ac}}{\sigma_{ac}^2} & a=b. \\
             \end{array}
           \right.
\ee
Defining $\Phi_{123}^{123}$ as the $(n-3)\times (n-3)$ minor of $\Phi$ obtained by deleting rows and columns $1,2,3$, one has that
\be
m(\alpha|\beta ) = \sum_{I=1}^{(n-3)!}\left.\frac{|123|^2}{\det \Phi_{123}^{123}}\frac{1}{(\alpha(1)\alpha(2)\cdots \alpha(n))}\,\frac{1}{(\beta(1)\beta(2)\cdots \beta(n))}\right|_{\sigma_a = \sigma_a^{(I)}},
\ee
where $\sigma_a^{(I)}$ denotes the value of $\sigma_a$ on the $I^{\rm th}$ solution.

Solving the equations $E_a=0$ is a nontrivial task when $n>5$ as for generic values of $s_{ab}$ and after finding a Groebner basis one is faced with an irreducible polynomial of degree $(n-3)!$\,.

Luckily, it is easy to make a simple proposal for what $m(\alpha|\beta)$ evaluates to and then prove that it is the right answer. This was done in a series of papers \cite{Cachazo:2013iea,Cachazo:2014nsa,Cachazo:2013gna}. Here we simply quote the result and use these integrals as building blocks for generic ones.

%%%%%%%%%%%%%%%%%%%%%%%%%%%%%%%%%%%%%%%%%%%
\subsection{Evaluating $m(\alpha|\beta)$}
%%%%%%%%%%%%%%%%%%%%%%%%%%%%%%%%%%%%%%%%%%%%

Consider any connected tree graph $T$ with $n$ vertices of degree\footnote{The degree of a vertex is defined as the number of edges incident to the vertex.} one and $n-2$ vertices of degree three. Associating a label $\{1,2,\ldots ,n\}$ to the vertices of degree 1 one can assign a rational function of $s_{ab}$ to $T$ as follows. Every internal edge $e$, i.e. not connected to a degree one vertex, divides the graph $T$ into to graphs $T_L$ and $T_R$ if the edge $e$ was removed. Let the subset of vertices of degree one from $\{1,2,\ldots ,n\}$ which lie on $T_L$ be $S_L$ and those on $T_R$ be $S_R$. Then it is easy to show that
\be
\sum_{a,b\in S_L}s_{ab} = \sum_{a,b\in S_R}s_{ab}
\ee
as a consequence of \eqref{kin}. Therefore this is a quantity that can be associated with the edge $e$ and we denote it as
\be
P_e^2\equiv\sum_{a,b\in S_L}s_{ab}.
\ee
The reason for the notation is that in physical applications this is the norm of a Lorentz vector.

The rational function associated with the graph $T$ is then
\be\label{omegaT}
w(T) \equiv \prod_{e\in E_T^{\rm int}} \frac{1}{P_e^2}\,,
\ee
where $E_T^{\rm int}$ is the set of all internal edges of $T$. In physics terminology, $T$ is a Feynman diagram in a massless cubic scalar theory and $w(T)$ is the value of the graph obtained by using Feynman rules.

A given diagram $T$ can be drawn on a plane in a variety of ways. Each way of doing so defines a cyclic ordering of the labels $\{1,2,\ldots ,n\}$. We say that $T$ is consistent with an ordering $\alpha \in S_n/\mathbb{Z}_n$ if $\alpha$ is one of the possible orderings obtained when $T$ is drawn on a plane. Let us denote the set of all graphs $T$ consistent with the ordering $\alpha$ by $\Gamma(\alpha)$.

Now we can state the main result of this subsection. The integral
\be\label{mab}
m(\alpha|\beta) = (-1)^q \sum_{T\in \Gamma(\alpha)\bigcap \Gamma(\beta)} w(T)\,,
\ee
where $q$ was defined in \cite{Cachazo:2013iea} and will not be relevant for our purposes\footnote{For more details see the equation (3.4) in \cite{Cachazo:2013iea}.}.

%%%%%%%%%%%%%%%%%%%%%%%%
\subsection{Examples}
%%%%%%%%%%%%%%%%%%%%%%%%%

Let us consider some simple examples in order to illustrate the use of the general formula \eqref{mab}. The first one is the four-point integral with the canonical ordering
\begin{equation}
m(1234\,|\,1234)=\frac{1}{s_{12}}+\frac{1}{s_{14}},
\end{equation}
which is a trivial computation using \eqref{omegaT} and \eqref{mab}. One can also find different orderings $\alpha$ and $\beta$  such that its result is just one Feynman diagram, for example
\begin{equation}
m(1234\,|\,1243)=\frac{1}{s_{12}},
\end{equation}
which is the diagram
\begin{center}\label{fig21}
\includegraphics[scale=0.35]{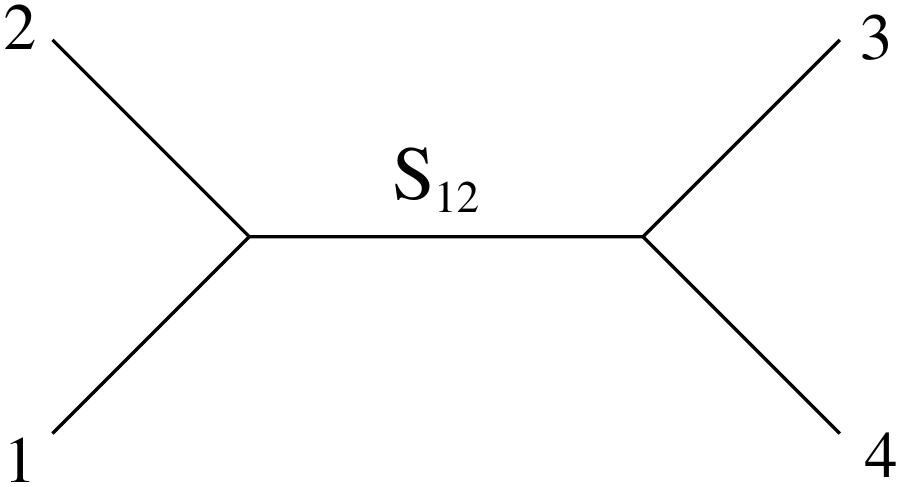}
\begin{center}
({\bf Fig.2.1}) {\small  {\rm Feynman diagram contributing for} $m(1234\,|\,1243)$}.
\end{center}
\end{center}

More interesting examples are the five-point computations. For instance, with $\alpha$ and $\beta$ in the canonical ordering one obtains
\begin{equation}
m(12345\,|\,12345)=\frac{1}{s_{12}s_{45}}+\frac{1}{s_{12}s_{34}}+\frac{1}{s_{23}s_{15}}+\frac{1}{s_{23}s_{45}}+\frac{1}{s_{15}s_{34}}.
\end{equation}
In five points we can also have two different orderings with intersection  on only one Feynman diagram, for example
\begin{equation}
m(12345\,|\,12534)=\frac{1}{s_{12}s_{34}},
\end{equation}
with diagram
\begin{center}\label{fig22}
\includegraphics[scale=0.35]{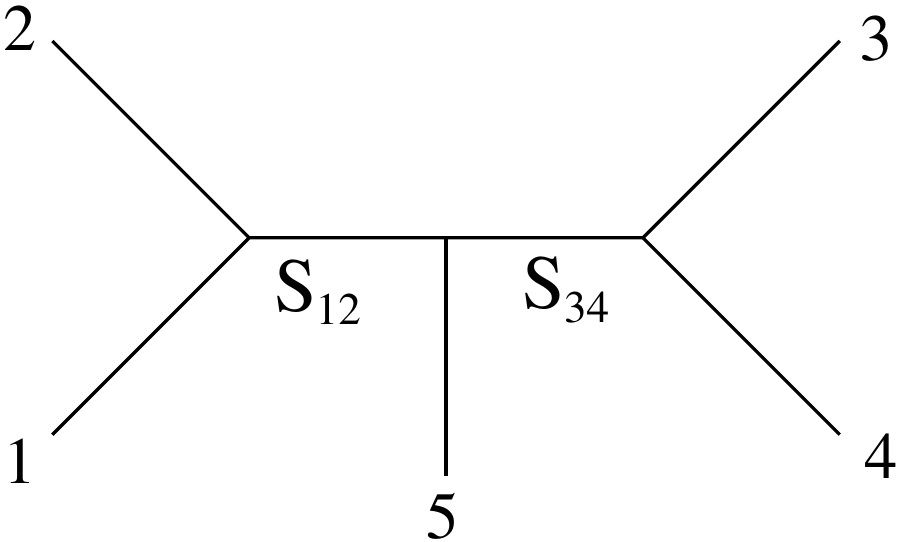}
\begin{center}
({\bf Fig.2.2}) {\small  {\rm Feynman diagram contributing for} $m(12345\,|\,12534)$}.
\end{center}
\end{center}
In addition, the $m(\a | \b)$ five-point integrals have more properties than the $m(\a | \b)$ four point matrix. For example, one can think in a Parke-Taylor with an $\a$ ordering  as a vector and the matrix element $m(\a | \b)$ as the inner product among two of them. So, a natural and interesting question arises, given a  Parke-Taylor with a particular ordering, what is its orthogonal space? for instance, at four points, there does not exist two orthogonal orderings. But, in five points, every Parke-Taylor with an $\alpha$ ordering has a 1-dimensional orthogonal space. For example, let us consider the canonical ordering $(12345)$, its orthogonal space is generated by the Park-Taylor $(14253)$, i.e
\begin{equation}
m(12345\,|\,14253)=0.
\end{equation}

%%%%%%%%%%%%%%%%%%%%%%%%%%%%%%%%%%%%%%%%%%%
\section{Generalized KLT}\label{kltgen}
%%%%%%%%%%%%%%%%%%%%%%%%%%%%%%%%%%%%%%%%%%%

In this section we introduce the first result needed for the computation of general integrals. In the 80's Kawai, Lewellen and Tye (KLT) found a relation connecting scattering amplitudes of closed strings to the sum of products of open strings amplitudes \cite{Kawai:1985xq,Berends:1988,Bern:1998sv}. While closed string amplitudes are computed on the sphere and hence are permutation invariant, open strings are defined as sums over partial amplitudes. Each partial amplitude is computed on a disk where the external states are inserted on the boundary and therefore possess an ordering. Thanks to the Bern-Carrasco-Johansson relations (BCJ) \cite{Bern:2008qj}, the modern version of the KLT formula can be written as (for more details see \cite{Vanhove,Feng:2,Feng:3,Feng:1,Stieberger:2009hq})
\be\label{klt}
M^{\rm closed}_n = \sum_{\hat\alpha,\hat\beta\in S_{n-3}} M^{\rm open}_n(1,\hat\alpha ,n,n-1) {\cal S}^{\rm string}(\hat\alpha |\hat\beta) M^{\rm open}_n(1,\hat\beta ,n-1,n)\,,
\ee
where permutations of $\{2,3,\ldots ,n-2\}$ are denoted $\hat\alpha$ and $\hat\beta$. In the formula above, ${\cal S}^{\rm string}(\hat\alpha|\hat\beta)$ is called the KLT momentum kernel \cite{Vanhove:2} and it is a somewhat complicated function of the variables $s_{ab}$ whose explicit form is not relevant at this point.

The string theory formula \eqref{klt} has a field theoretic analog obtained by taking the infinite tension limit and it relates amplitudes of gravitons to that of gluons. More explicitly \cite{Bern:2008qj,Bern:1998sv,Feng:1,Berends:1988,Vanhove:2},
\be\label{QFTklt}
M^{\rm gravitons}_n = \sum_{\hat\alpha,\hat\beta\in S_{n-3}} M^{\rm gluos}_n(1,\hat\alpha ,n,n-1) {\cal S}(\hat\alpha|\hat\beta) M^{\rm gluons}_n(1,\hat\beta ,n-1,n).
\ee
Here ${\cal S}(\hat\alpha|\hat\beta)$ is the infinite tension limit of ${\cal S}^{\rm string}(\hat\alpha |\hat\beta)$.

In \cite{Cachazo:2014xea}, it was realized that the field theory KLT relation \eqref{QFTklt} and many generalizations naturally follow from the CHY representation of amplitudes.

Let us summarize the construction with special emphasis on the structures needed in section \ref{mainalgo}.

Consider any contour integral as a starting point (sp), it plays the role of $M_n$ in \eqref{klt},
\be\label{inew}
I_{\rm sp} = \int d\mu_n {\cal I}(\sigma),
\ee
with an integrand that can be separated into two parts
\be
{\cal I}(\sigma) = {\cal I}_{L}(\sigma){\cal I}_R(\sigma),
\ee
where each ``half-integrand" has half the $PSL(2,\mathbb{C})$ weight of the full integrand. Examples of such integrands and half-integrands were studied in the previous section with Parke-Taylor factors being the half-integrands.

The evaluation of \eqref{inew} is given by
\be\label{reint}
\sum_{I=1}^{(n-3)!}\left.\frac{|123|^2}{\det \Phi_{123}^{123}}{\cal I}_{L}(\sigma){\cal I}_R(\sigma)\right|_{\sigma_a = \sigma_a^{(I)}}.
\ee
Let us denote the combination $\det \Phi_{123}^{123}/|123|^2$ as $\det '\Phi$.

Introducing an $(n-3)!\times (n-3)!$ diagonal matrix in solution space $D_{IJ} = \det '\Phi(\sigma_a^{(I)})\delta_{IJ}$ and $(n-3)!$-dimensional vectors
$\vec{{\cal I}_L}$ and $\vec{{\cal I}_R}$ one finds a matrix form of \eqref{reint}
\be\label{spone}
I_{\rm sp} = \vec{{\cal I}_L}^T \, D^{-1} \, \vec{{\cal I}_R}.
\ee
The next step is to find an alternative representation of the matrix $D$ in terms of a $(n-3)!\times (n-3)!$ matrix but this time in the ordering space. A natural candidate is to consider a submatrix  of the $(n-1)!\times (n-1)!$ matrix whose entries are given by $m(\alpha|\beta)$ with $\alpha,\beta\in S_{n}/\mathbb{Z}_n$.

More explicitly, the definition in \eqref{mdef} gives
\be
m(\alpha|\beta ) = \sum_{I=1}^{(n-3)!}\left.\frac{1}{\det '\Phi}\frac{1}{(\alpha(1)\alpha(2)\cdots \alpha(n))}\,\frac{1}{(\beta(1)\beta(2)\cdots \beta(n))}\right|_{\sigma_a = \sigma_a^{(I)}}.
\ee
This time it is convenient to introduce a rectangular $(n-1)!\times (n-3)!$ matrix $Q$ with entries
\be
Q_{\alpha}^I = \left.\frac{1}{(\alpha(1)\alpha(2)\cdots \alpha(n))}\times \frac{1}{\det '\Phi}\right|_{\sigma_a = \sigma_a^{(I)}}
\ee
so that
\be
m(\alpha|\beta ) = \sum_{I,J=1}^{(n-3)!} Q_{\alpha}^I D_{IJ} Q_{\beta}^J.
\ee

Next we divide the discussion into the derivation of the standard KLT result \eqref{QFTklt} and then its most general form.

%%%%%%%%%%%%%%%%%%%%%%%%%%%%%%
\subsection{Standard KLT}\label{standardklt}
%%%%%%%%%%%%%%%%%%%%%%%%%%%%%%

Let us first discuss how to recover the standard KLT formula in its modern version \cite{Kawai:1985xq,Bern:1998sv,Bern:2008qj,Feng:1,Berends:1988} before proceeding to the more general discussion. As discussed above one has to select an $(n-3)!\times (n-3)!$ submatrix of the matrix $m(\alpha|\beta)$. The choice that leads to the modern version of the KLT formula is
\be
m(1,\hat\alpha ,n-1,n|1,\hat\beta ,n,n-1) \equiv m^{\rm KLT}(\hat\alpha|\hat\beta)\,,
\ee
with $\hat\alpha,\hat\beta\in S_{n-3}$ permutations of the remaining $n-3$ labels. Once the choice has been made two square matrices can be defined
\be
\hat U_{\hat\alpha}^I \equiv Q_{1,\hat\alpha,n,n-1}^I, \quad \hat V_{\hat\beta}^I \equiv Q_{1,\hat\beta,n-1,n}^I.
\ee
Finally, it is possible find a representation for $D$
\be
m^{\rm KLT} = \hat U^{\rm T} D \hat V \quad \Rightarrow \quad D = (\hat V)^{-1} m^{\rm KLT}(\hat U^{\rm T})^{-1}.
\ee
Using this formula in \eqref{spone}
\be\label{sKLT}
I_{\rm sp} = \vec{{\cal I}_L}^T \, D^{-1} \, \vec{{\cal I}_R} = (\hat U\vec{{\cal I}_L})^T \, (m^{\rm KLT})^{-1} \, \hat V\vec{{\cal I}_R}.
\ee
It is easy to recognize that
\begin{align}\label{UVklt}
(\hat U\vec{{\cal I}_L})(1,\hat\alpha,n,n-1) &= \int d\mu_n \frac{{\cal I}_L(\sigma )}{(1,\hat\alpha,n,n-1)},\\
(\hat V\vec{{\cal I}_R})(1,\hat\beta,n-1,n) &= \int d\mu_n \frac{{\cal I}_R(\s)}{(1,\hat\beta,n-1,n)}\nonumber.
\end{align}
When applied to gravity and Yang-Mills amplitudes \eqref{sKLT} becomes the standard KLT formula \eqref{QFTklt}. Even though it is not used in this work, let us make this more explicit for completeness. In \cite{Cachazo:2013hca} gravity and Yang-Mills amplitudes are computed as follows
\begin{align}\label{grYM}
M^{\rm gravitons}_n &= \int d\mu_n ({\rm Pf}'\Psi (\epsilon,k,\sigma))^2,\\
M^{\rm gluons}_n(\alpha) &= \int d\mu_n \frac{{\rm Pf}'\Psi (\epsilon,k,\sigma)}{(\alpha(1)\alpha(2)\cdots \alpha(n))},\nonumber
\end{align}
where $\Psi (\epsilon,k,\sigma)$ is some $2n\times 2n$ matrix whose precise form can be found in \cite{Cachazo:2013hca}. Now it is clear how this leads to the KLT formula directly. Moreover, it shows that the momentum kernel ${\cal S}(\hat\alpha|\hat\beta) = (m^{\rm KLT})^{-1}(\hat\alpha|\hat\beta)$.

%%%%%%%%%%%%%%%%%%%%%%%%%%%%%%
\subsection{General KLT}\label{gklt}
%%%%%%%%%%%%%%%%%%%%%%%%%%%%%%

Let us now discuss the more general construction. It is clear that the same steps can be followed as in the standard KLT construction if one chooses a general $(n-3)!\times (n-3)!$ sub-matrix of the matrix $m(\alpha|\beta)$. However, not all sub-matrices are allowed as the construction requires the computation of its inverse. It turns out that $m(\alpha|\beta)$ has vanishing determinant and so do some of its $(n-3)!\times (n-3)!$ sub-matrices.

Let ${\cal L}$ and ${\cal R}$ be both subsets of permutations $S_n/\mathbb{Z}_n$ with $(n-3)!$ elements. We say that ${\cal L}$ and ${\cal R}$ are {\it independent} if the matrix with entries
\be
m^{{\cal L}|{\cal R}} \equiv \{ \, m(\alpha|\beta) ~ : ~ \alpha\in {\cal L}, ~~ \beta\in {\cal R} \}
\ee
has non-vanishing determinant.

Provided ${\cal L}$ and ${\cal R}$ are independent one can obtained a formula for $I_{\rm sp}$ of the form
\be\label{genKLT}
I_{\rm sp} = \sum_{\alpha\in {\cal L},\beta\in {\cal R}} ({\cal U I^L})_\a\, (m^{{\cal L},{\cal R}})^{-1}_{\alpha,\beta} \, ({\cal V I^R})_\b,
\ee
where we have defined
\begin{align}
{\cal  U}_{\alpha}^I &\equiv Q_{\alpha}^I\,\,, \qquad \a\in{\cal L}, \\
{\cal V}_{\beta}^I &\equiv Q_{\beta}^I\,\,, \qquad \b\in{\cal R}\nonumber,
\end{align}
and
\begin{align}
({\cal U I^L})_\a &=\int d\mu_n \frac{{\cal I}_L(\sigma )}{(\alpha(1)\alpha(2)\cdots \alpha(n))}\,,\qquad \alpha\in{\cal L},\\
({\cal V I^R})_\b &=\int d\mu_n \frac{{\cal I}_R(\sigma)}{(\beta(1)\beta(2)\cdots \beta(n))}\,,\qquad \beta\in{\cal R}\,\nonumber.
\end{align}

This is the most general form of the KLT relation that is needed in the algorithm presented in section \ref{mainalgo}.

%%%%%%%%%%%%%%%%%%%%%%%%%
\subsection{Examples}\label{kltexample}
%%%%%%%%%%%%%%%%%%%%%%%%%

In this subsection we give a simple example to show how the standard KLT construction can be used in the computation of residue integrals. After that, we formulate an example where the standard KLT is not enough.

Let us start with an example where the standard KLT construction suffices. Consider the five point integral
\be\label{ispf}
I_{\rm sp} = \int d\mu_5 \frac{1}{(12345)}\frac{1}{(12)(345)}.
\ee
Recall the general definition given in the introduction which applied to this case implies $(12)=\sigma_{12}\sigma_{21}$ and $(345)=\sigma_{34}\sigma_{45}\sigma_{53}$.
This integral is not of the form studied in section 2. The idea is then to find a way of writing it in terms of the building blocks of section 2.

Let the KLT basis be given by permutations $(1,\hat\alpha, 32)$ and $(1,\hat\a, 23)$ with $\hat\alpha$ permutations of $\{4,5\}$.

Using the KLT formula one has
\be
I_{\rm sp} = \sum_{\hat\alpha,\hat\beta \in {\rm perm}(4,5)}m(12345|1,\hat\alpha,32)(m^{\rm KLT})^{-1}_{\hat\alpha,\hat\beta}\int d\mu_5\frac{1}{(12)(345)(1,\hat\beta ,23)}.
\ee
At first sight it seems that the problem has been made worse as one has to now deal with two new integrals
\be
\int d\mu_5\frac{1}{(12)(345)(14523)}\,,\qquad \int d\mu_5\frac{1}{(12)(345)(15423)}.
\ee
However, it is simple to check that
\be
(12)(345)(14523) = (12354)(12543)\,, \qquad (12)(345)(15423) = (13542)(12345).
\ee
This means that we have succeeded in expressing $I_{\rm sp}$ in terms of the building blocks.

Using the explicit form of the building blocks one finds that
\begin{align}\label{resultfivepts}
\int d\mu_5 \frac{1}{(12345)}\frac{1}{(12)(345)} &=
\sum_{\hat\alpha,\hat\beta \in {\rm perm}(4,5)}
(\hat U\vec{{\cal I}_L})_{\hat \a} \, (m^{\rm KLT})^{-1}_{\hat\a,\hat\b} \, (\hat V\vec{{\cal I}_R})_{\hat\b}\\
 &=\frac{1}{s_{12}\,s_{34}}+\frac{1}{s_{12}\,s_{45}}+\frac{s_{15}}{s_{12}^2\,s_{34}}+\frac{s_{14}}{s_{12}^2\,s_{45}}+\frac{s_{15}}{s_{12}^2\,s_{45}}\nonumber,
\end{align}
where the $m^{\rm KLT}(1,\hat{\a},32|1,\hat{\b},23)$ matrix is given by
\begin{equation}\label{mfive}
m^{\rm KLT}(1,\hat{\a},32\,|\,1,\hat{\b},23)=\left(
\begin{matrix}
 -\frac{1}{s_{23}\,s_{14}}-\frac{1}{s_{23}\,s_{45}}\,\, &\,\, \frac{1}{s_{23}\,s_{45}}\\
\frac{1}{s_{23}\,s_{45}} \,\,&\,\, -\frac{1}{s_{23}\,s_{15}}-\frac{1}{s_{23}\,s_{45}}\\
\end{matrix}
\right)
\end{equation}
and the vectors
\begin{align}\label{uvfive}
(\hat U\vec{{\cal I}_L})(1,\hat\alpha,32) &= \left(
-\frac{1}{s_{12}\,s_{45}}-\frac{1}{s_{23}\,s_{45}},\frac{1}{s_{12}s_{45}}+\frac{1}{s_{12}s_{34}}+\frac{1}{s_{23}s_{15}}+\frac{1}{s_{23}s_{45}}+\frac{1}{s_{15}s_{34}}
\right)\,,\nonumber\\
(\hat V\vec{{\cal I}_R})(1,\hat\beta,23) &= \left(
\frac{1}{s_{12}\,s_{45}}\,,\,-\frac{1}{s_{12}\,s_{45}}
\right).
\end{align}
In this simple example we have solved a non-trivial integrand just using the building blocks
%given by $m(1,\hat\alpha,32\,|\,1,\hat\beta,23)$
and the standard KLT approach. Note that the matrix \eqref{mfive} and the vectors \eqref{uvfive} are not simple. Very nicely, one can also use the \textit{original} KLT approach in order to obtain a simpler matrix and vectors \cite{Kawai:1985xq,Bern:1998sv,Berends:1988}. For example, let us consider the following decomposition
\be\label{classicklt}
I_{\rm sp} = \sum_{\hat\alpha\in L,\hat\beta \in  R} ({\hat U \vec{{\cal I}}_L})_{\hat\a} \,(m^{\rm KLT})^{-1}_{\hat\alpha,\hat\beta}\,({\hat V \vec{{\cal I}}_R})_{\hat\b},
\ee
where
\begin{align}
L=\{(1,\hat\a(4),3,\hat\a(5),2)\}&=\{ (14352),\, (15342)\},\\
R=\{(1,\hat\b(4),\hat\b(5),2,3)\} &=\{ (14523),\, (15423)\},
\end{align}
and
\begin{align}
(\hat U\vec{{\cal I}_L})_{\hat\alpha} &= m(12345|1,\hat\a(4),3,\hat\a(5),2)\,,\\
(\hat V\vec{{\cal I}_R})_{\hat\b} &= \int d\mu_5\frac{1}{(12)(345)\,\,(1,\hat\b,23)}\,.
\end{align}
It is trivial to show that the $(m^{\rm KLT})_{\hat\alpha,\hat\beta}$ matrix is given by
\begin{equation}\label{mfiveg}
m^{\rm KLT}(\hat\a|\hat\b)=\left(
\begin{matrix}
 \frac{1}{s_{14}\,s_{25}}\,\, &\,\, 0\\
0 \,\,&\,\, \frac{1}{s_{15}\,s_{24}}\\
\end{matrix}
\right)
\end{equation}
and the vectors
\begin{align}\label{uvfiveg}
(\hat U\vec{{\cal I}_L})_{\hat\alpha}&= \left(
\frac{1}{s_{12}\,s_{34}}\,,\,\frac{1}{s_{12}s_{34}}+\frac{1}{s_{15}s_{34}}
\right),\nonumber\\
(\hat V\vec{{\cal I}_R})_{\hat\b} &=
\left(
\frac{1}{s_{12}\,s_{45}}\,,\,\frac{-1}{s_{12}\,s_{45}}
\right),
\end{align}
so the computation of \eqref{ispf} becomes trivial. One can easily check that
\begin{align}
\int d\mu_5 \frac{1}{(12345)}\frac{1}{(12)(345)} &=  
 \sum_{\hat\alpha\in L,\hat\beta \in  R} ({\hat U \vec{{\cal I}}_L})_{\hat\a} \,(m^{\rm KLT})^{-1}_{\hat\alpha,\hat\beta}\,({\hat V \vec{{\cal I}}_R})_{\hat\b}\\
&=
\frac{1}{s_{12}s_{45}}\left( \frac{s_{14}s_{25}}{s_{12}s_{34}}-\frac{s_{15}s_{24}}{s_{34}}\left(\frac{1}{s_{12}}+\frac{1}{s_{15}}\right) \right)\nonumber\,,
\end{align}
which 
%the formula \eqref{classicklt} obtained directly in this basis 
agrees with \eqref{resultfivepts}.

Finally, consider the following six-point integral
\be\label{benzene}
\int d\mu_6 \frac{1}{(123456)}\frac{1}{(12)(34)(56)}.
\ee
It turns out that choosing a KLT basis of $(n-3)! =6$ permutations where three labels are fixed to some particular locations is not enough in this case. One can show that it is not possible to find such a set so that when multiplied with $(12)(34)(56)$ always gives a product of two six-point Parke-Taylor factors. In section \ref{allsix} we present an explicit set of six permutations that do give rise to two Parke-Taylor factors in all six cases and hence a building block.

%%%%%%%%%%%%%%%%%%%%%%%%%%%%%%%%%%%%%%%%%%%%
\section{Petersen's Theorem}\label{peter}
%%%%%%%%%%%%%%%%%%%%%%%%%%%%%%%%%%%%%%%%%%%%%

In this section we present the second key result needed for the algorithm in section \ref{mainalgo}. Consider a special class of integrands $F(\sigma)$ defined by rational functions with no zeroes. This means that $F(\sigma)$ has $2n$ factors $\sigma_{ab}$ in the denominator and a trivial numerator which can be taken to be unity. We denote this kind of integrands as $F_d(\s)$. The required $PSL(2,\mathbb{C})$ transformation of $F_d(\sigma)$ implies that each label $a$ must appear exactly in four factors. In this section all integrands $F$  satisfy these special properties, i.e $F(\s)=F_d(\s)$.

An integrand with these properties is uniquely determined by a graph ($G_F$) with $n$ vertices constructed by including an (non-oriented) edge connecting $a$ and $b$ for every factor of $\sigma_{ab}$ in the denominator of $F_d(\s)$. Note that the same factor can appear more than once and hence the graph is not simple in general\footnote{Graphs with multiple edges connecting two vertices are also called multigraphs.}. To illustrate this  morphism we give a simple example. Let us  consider the rational  function
\begin{equation}\label{Pexample1}
F_d(\s)=\frac{1}{(123)(345)(561)(246)},
\end{equation}
Using the previous rules, the graph associated with the function $F_d(\s)$, which we have called $G_F$, is given by the pair $G_F=(V_F,E_F)$, where $V_F$ and $E_F$ are the vertex and edge sets, respectively
\begin{align}
V_F &=\{1,2,3,4,5,6\}\,,\\
E_F&=\{[1;2],[2;3],[1;3],[3;4],[4;5],[3;5],[5;6],[1;6],[1;5],[2;4],[4;6],[2;6] \},\nonumber
\end{align}
where $[a;b]=[b;a]$ (non-oriented graph). The line drawing of the graph $G_F$ is
\begin{center}\label{fig41}
\includegraphics[scale=0.5]{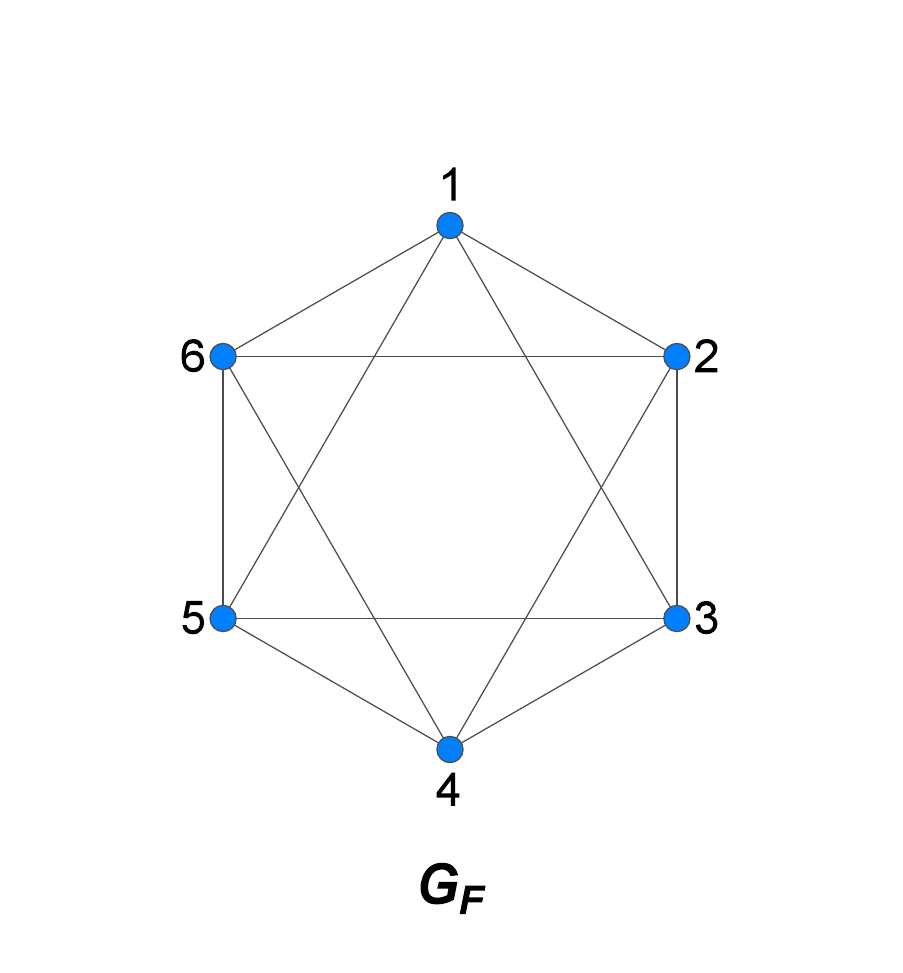}\quad .
\begin{center}
({\bf Fig.4.1}) {\small {\rm Graph $G_F$}}
\end{center}
\end{center}

The graph associated with any integrand $F_d(\s)$ has the property that every vertex has degree exactly four. These are called 4-regular. In general a graph where each vertex has degree $k$ is called $k$-regular.

A result of Petersen dating back to 1891 states that every 4-regular graph $G$ with $n$ vertices contains a 2-factor. A 2-factor is a 2-regular subgraph of $G$ with $n$ vertices. Of course, after removing all the edges in $G$ from such a 2-factor one is left with another 2-regular graph with $n$ vertices \cite{graph1,graph2}.

It should now be clear why this theorem is useful in our construction. Given any function $F_d(\s)$ and its associated graph $G_F$, find the two 2-regular subgraphs implied by Pertersen's theorem\footnote{This decomposition is not unique.} and denote them by $G^L_{F}$ and $G^R_{F}$. Then it is possible to write
\be
F_d = F^L_d \times F^R_d\,,
\ee
where
\be
F^L_d = \frac{1}{\prod_{e\in G^L_F}\sigma_{v_e,u_e}}\,,\qquad
F^R_d = \frac{1}{\prod_{e\in G^R_F}\sigma_{v_e,u_e}}\,,
\ee
with $v_e$ and $u_e$ the end vertices of the $e$ edge. Given that $G^L_{F}$ and $G^R_{F}$ are 2-regular, $F^L_d$ and $F^R_d$ both transform as half integrands and the generalized KLT construction can be used to decompose
\be
\int d\mu_n F_d
\ee
as a sum over product of simpler integrals. The concept of ``simpler" will be made precise in sections \ref{HD} and \ref{mainalgo}.

For example, in the graph represented by (Fig.4.1) it is simple to see the following two 2-factors
\\
\begin{center}\label{fig42}
\includegraphics[scale=0.45]{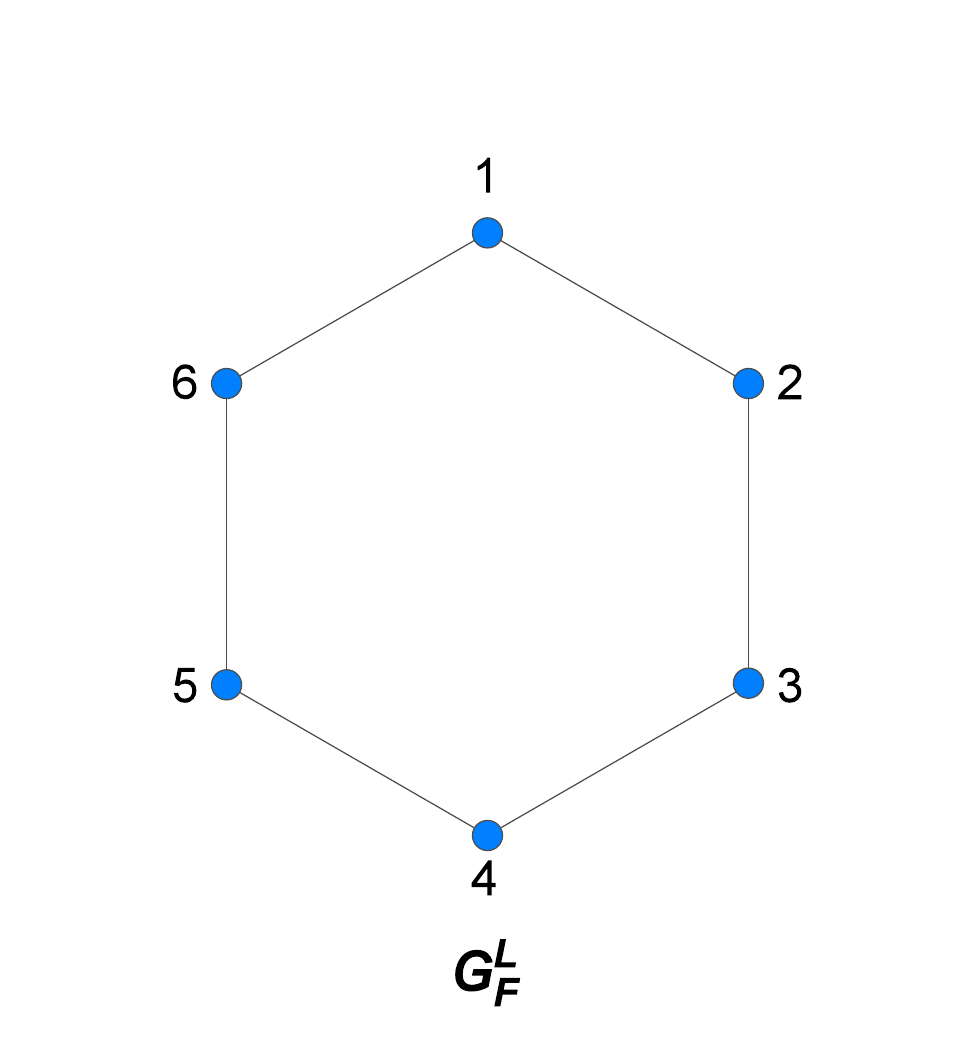},\quad \includegraphics[scale=0.45]{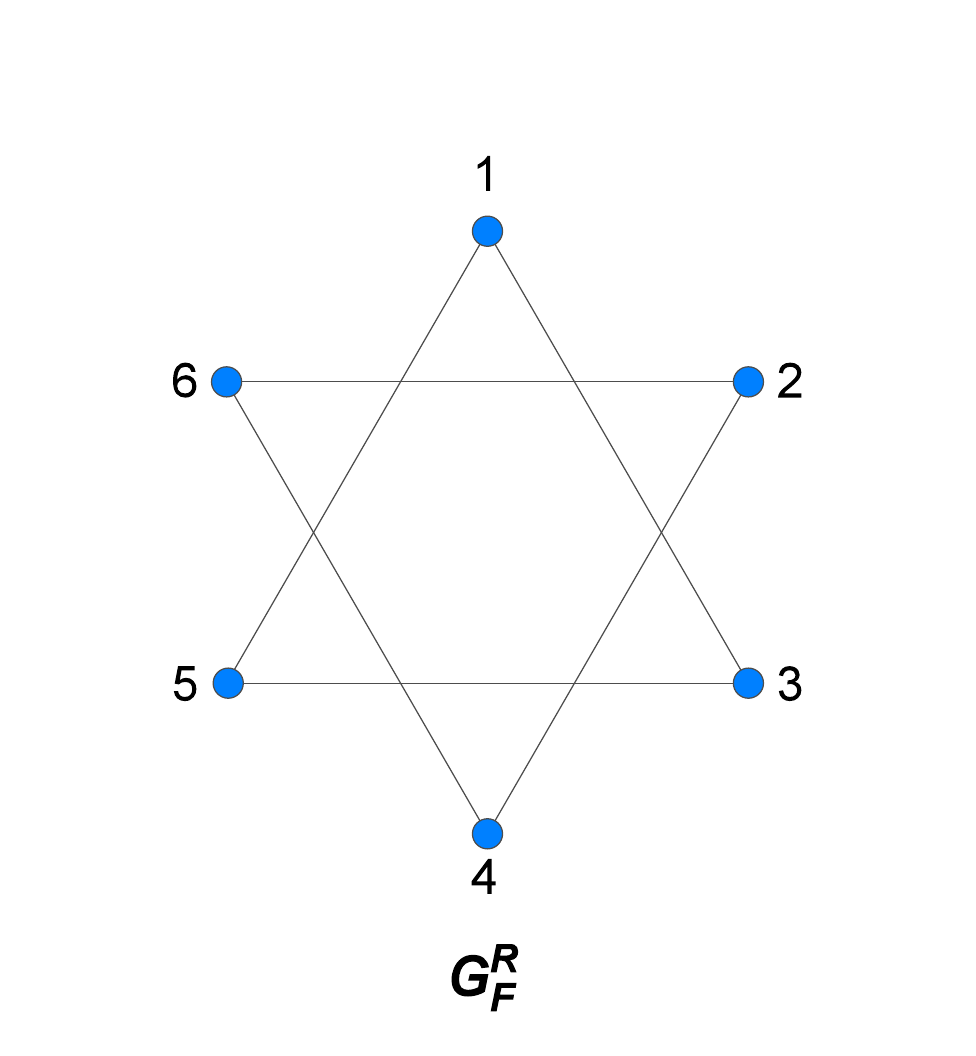}
\begin{center}
({\bf Fig.4.2}) {\small {\rm 2-Factor decomposition of $G_F$.}}
\end{center}
\end{center}
and therefore
$$
F^L_d=\frac{1}{(123456)},\qquad F^R_d=\frac{1}{(135)(246)}.
$$
Although this example is very simple, finding the decomposition into two 2-factors of a general 4-regular graph can be a daunting exercise. In appendix A we provide an explicit algorithm for finding two 2-factors $G^L_{F}$ and $G^R_{F}$ and a sketch of the proof of Pertersen's theorem.

%%%%%%%%%%%%%%%%%%%%%%%%%%%%%%%%%%%%%%%%%%%%%%%%%
\section{Hamiltonian Decomposition}\label{HD}
%%%%%%%%%%%%%%%%%%%%%%%%%%%%%%%%%%%%%%%%%%%%%%%%%%

In this section we present the third and final ingredient needed for the general algorithm. Recall the example presented in section \ref{kltexample}
\be
\int d\mu_5 \frac{1}{(12345)(12)(345)}.
\ee
The KLT procedure worked because we were able to find Parke-Taylor factors such that
\begin{align}\label{decom}
(12)(345)\times (14523) &= (12354)\times (12543),\\
(12)(345)\times (15423) &= (13542)\times (12345)\nonumber.
\end{align}
It is clear that by combining Petersen's theorem with the KLT construction one would be able to compute any integral with a trivial numerator if a decomposition of the form \eqref{decom} was always possible. This is the main subject of this section and we start by introducing some terminology standard in the graph theory literature.

A 4-regular graph $G$ is said to have a Hamiltonian decomposition if $G$ has two edge-disjoint Hamiltonian cycles $H_1$ and $H_2$ \cite{graph1}.

Let us remind the reader what a Hamiltonian cycle is. In any connected graph with $n$ vertices one can ask if there is a connected closed path that visits all vertices exactly once. Such a closed path, i.e., collection of edges, is called a Hamiltonian cycle.

We are going to assume that we are given a generic 2-regular graph $G^R$ with $n$ vertices, for example when $n=5$ we can consider the graph associated to the denominator $(12)(345)$. If the 2-regular graph $G^R$ is connected then we do not have to continue since it is already of the form needed. Assume that the 2-regular graph is made out of $m$ disconnected 2-regular graphs, each with the corresponding Parke-Taylor factor, $(r_1)(r_2)\cdots (r_m)$.

We say that an $n$-point Parke-Taylor factor $(\alpha(1)\alpha(2)\cdots \alpha(n))$ is {\it compatible} with a general combination $(r_1)(r_2)\cdots (r_m)$ if the union of both graphs, which is obviously a 4-regular graph, has a Hamiltonian decomposition. In our more physical terminology, $(\alpha(1)\alpha(2)\cdots \alpha(n))$ is compatible with $(r_1)(r_2)\cdots (r_m)$ if
\be
(r_1)(r_2)\cdots (r_m)\times (\alpha(1)\alpha(2)\cdots \alpha(n)) = (\beta(1)\beta(2)\cdots \beta(n))\times (\gamma(1)\gamma(2)\cdots \gamma(n))
\ee
for some $\beta,\gamma\in S_n/\mathbb{Z}_n$.

How many permutations $\alpha\in S_n/\mathbb{Z}_n$ are compatible with a given form $(r_1)(r_2)\cdots (r_m)$ seems to be a complicated question in general. However we have made an extensive computer search and have found strong evidence that not only the number is always larger than $(n-3)!$ but it becomes much larger than $(n-3)!$ as $n$ increases. In the next subsection we discuss the results obtained in the computer search as well as a conjecture for the precise number when every $(r_a)$ in $(r_1)(r_2)\cdots (r_m)$ contains exactly two elements.

We restrict our search to the cases where all $(r_a)$'s except one have exactly two elements (i.e., are bubbles) and the cases where all $(r_a)$ except one have exactly three element (i.e., are triangles).

%%%%%%%%%%%%%%%%%%%%%%%%%%%%%%%%%%%%%%
\subsection{Bubbles and One Polygon}
%%%%%%%%%%%%%%%%%%%%%%%%%%%%%%%%%%%%%%

Let us start with 2-regular graphs of the form
\be
(12)(34)\ldots (2m-1,2m)(2m+1,2m+2,\ldots ,2m+k).
\ee
The graph is then given by $m$ bubbles and one polygon with $k$ sides. The total number of vertices is $n=2m+k$.

The results from our computer search are presented in table \ref{t1} .

\begin{center}\label{t1}
\includegraphics[scale=0.46]{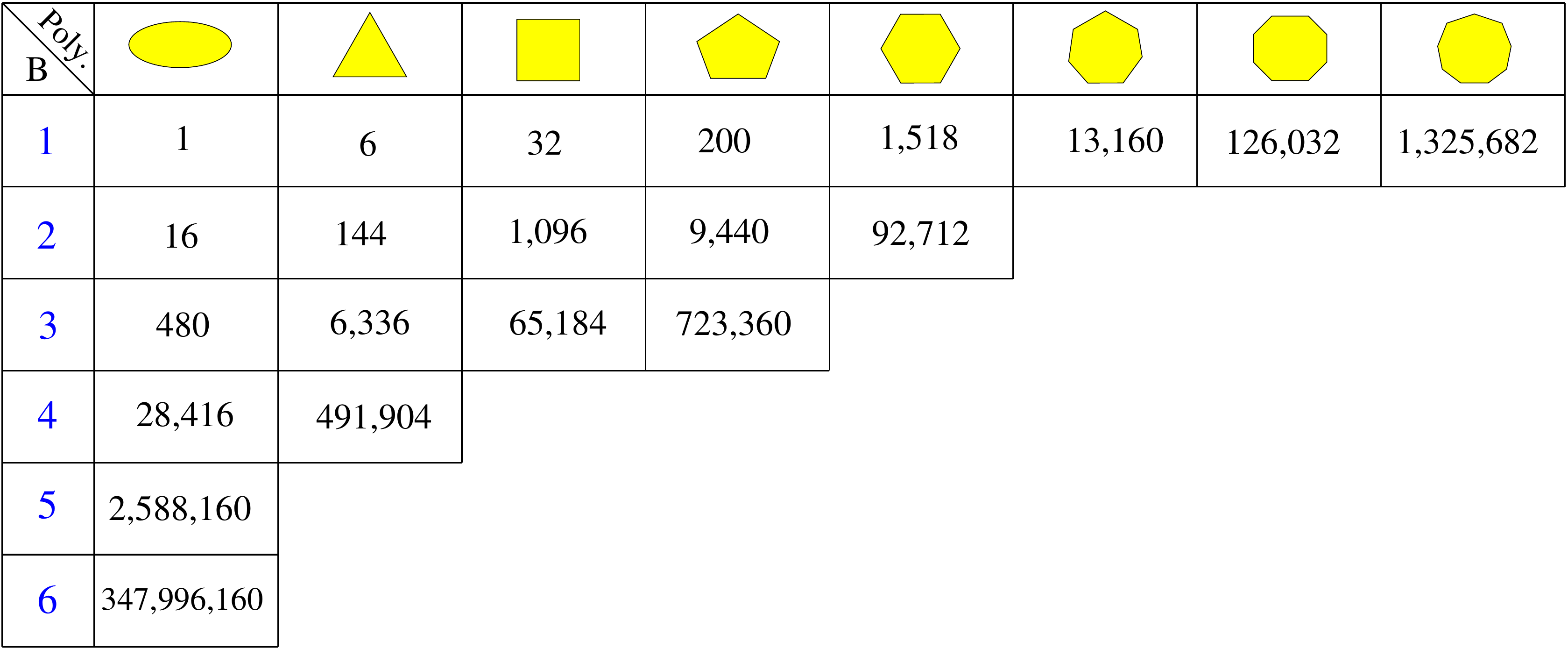}
\begin{center}
{\bf Table 5.1.} {\small {\rm Number of compatible  Parke-Taylor factors to the graph obtained by combining bubbles (vertical-{\bf B}) with various polygons (horizontal-{\bf Poly}).}}
\end{center}
\end{center}

Recall that the numbers presented in the table are the total numbers of permutations that are compatible with the corresponding 2-regular graph. It is easy to see that the number in the table are always greater than $(n-3)! = (2m+k-3)!$ (except for n=4 when it is equal) and that the ratio increases as $n$ gets larger.

Quite nicely, we have been able to find a sequence of numbers that reproduces the case $k=2$ for all $m$ tested and this is why we conjecture the following.

Consider the 2-regular graph with $2s$ vertices which is made out of $s$ bubbles (here $s=m+1$ since the polygon added to the $m$ bubbles in the table is also a bubble). The number of compatible Parke-Taylor factors is given by
\be\label{numb}
2^{s-2}(s-1)!A(s)\,,
\ee
where $A(s)$ is the number of types of sequential s-swaps moves for the travelling salesman problem which has a closed formula presented as sequence A001171 in the OEIS webpage \cite{math:2,math}.

It is interesting to note that the case of only bubbles is the one with the least number of compatible permutations for a given number $n$ of vertices. Using the closed formula, it is possible to find the ratio of \eqref{numb} to $(n-3)!=(2s-3)!$ as $s$ goes to infinity,
\be
r(s) = \frac{2^{s-2}(s-1)!A(s)}{(2s-3)!} \,\,\sim \,\,\frac{\pi}{2}\, s\,.
\ee

\subsection{Triangles and One Polygon}

Finally, we consider the case in which we have several triangles and only one polygon with $k$ vertices. Clearly the case of one triangle and $k=2$ is already in the table \ref{t1}. The results for this new computer search are presented in table \ref{t2}.

\begin{center}\label{t2}
\includegraphics[scale=0.5]{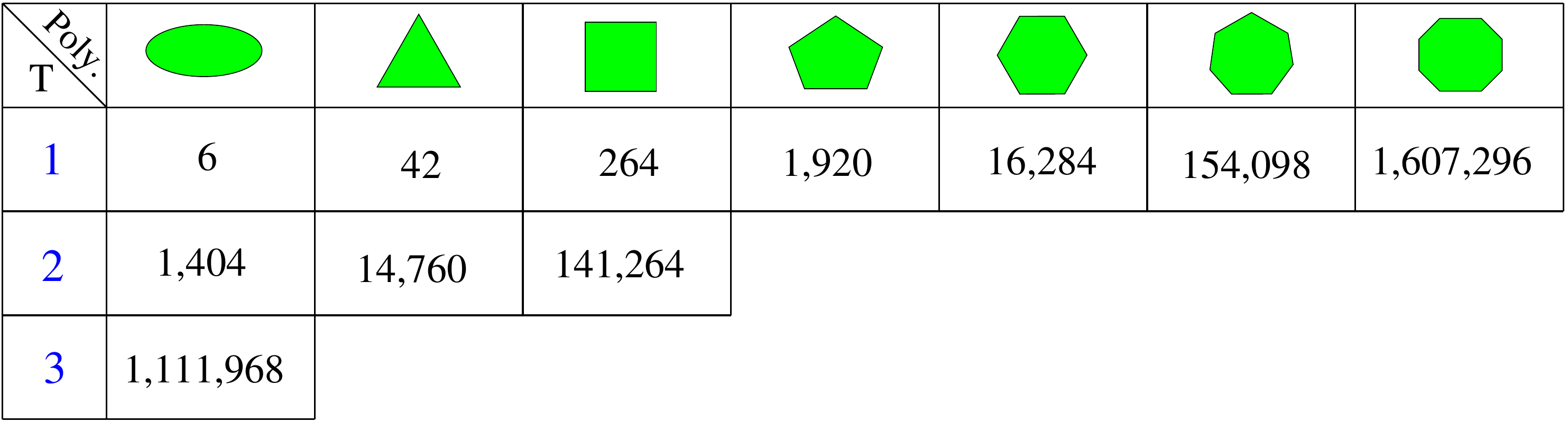}
\begin{center}
{\bf Table 5.2.} {\small {\rm Number of compatible Parke-Taylor factors to the graph obtained by combining triangles (vertical-{\bf T}) with various polygons (horizontal-{\bf Poly}).}}
\end{center}
\end{center}

Once again, it is clear that the number of compatible Parke-Taylors is larger than $(n-3)!$ and the ratio grows with the number of vertices.

%%%%%%%%%%%%%%%%%%%%%%%%%%%%%%%%%%%%%%%%%%
\section{Main Algorithm}\label{mainalgo}
%%%%%%%%%%%%%%%%%%%%%%%%%%%%%%%%%%%%%%%%%%%

In this section we present the main algorithm for the decomposition of an integral of a general rational function $F$ in terms of the building blocks $m(\alpha |\beta)$.

%%%%%%%%%%%%%%%%%%%%%%%%%%%%%%%%%%%%%%%%%
\subsection{Reduction of the Numerator}
%%%%%%%%%%%%%%%%%%%%%%%%%%%%%%%%%%%%%%%%%

A general function $F$ can have a numerator which can be taken, without loosing of generality, to be a monomial in the variables $\sigma_{ab}$. Let us define cross ratios to be the basic $PSL(2,\mathbb{C})$ invariant functions
\be
r_{abcd} \equiv \frac{\sigma_{ab}\sigma_{cd}}{\sigma_{ad}\sigma_{bc}}.
\ee
It is clear that by choosing any pair of Parke-Taylor factors, say
\be
\{ (\gamma(1)\gamma(2)\cdots \gamma(n)), (123\cdots n) \},
\ee 
to multiply $F$, one gets an $SL(2,\mathbb{C})$ invariant function which can be expressed in terms of cross ratios, i.e., 
\be
(\gamma(1)\gamma(2)\cdots \gamma(n))(123\cdots n)F = \prod_{I=1}^m r_{a_I,b_I,c_I,d_I}.
\ee
for some value of $m\geq 0$.

Using this we choose as our starting point a representation for $F$ of the form
\be\label{iterSP}
F = \frac{1}{(\gamma(1)\gamma(2)\cdots \gamma(n))(123\cdots n)}\prod_{I=1}^m r_{a_I,b_I,c_I,d_I}.
\ee
The following is a procedure which might not be the most efficient in particular cases but it is general. Let us isolate the $m^{\rm th}$ cross ratio and write
\begin{align}
F &= F_L\times F_R \nonumber\\
&= \left(\frac{1}{(\gamma(1)\gamma(2)\cdots \gamma(n))}\right)\times \left(\frac{1}{(12\cdots n)}\prod_{I=1}^{m-1} r_{a_I,b_I,c_I,d_I}\times \frac{\sigma_{a_m,b_m}\sigma_{c_m,d_m}}{\sigma_{a_m,d_m}\sigma_{b_m,c_m}}\right)\,.
\end{align}
Use the KLT procedure to separate the left and right factors (defined by the parenthesis) by using a basis of $(n-3)!$ Parke-Taylor factors of the form $1/(c_m,d_m,a_m, \alpha)$ where $\alpha$ is some permutation of the $n-3$ left over labels. Applying this procedure one finds that
\be
\int d\mu_n F = \sum_{\alpha,\beta}\,m(\gamma(1)\gamma(2)\cdots \gamma(n)\,|\,c_m,d_m,a_m, \alpha)\,\, {\cal S}(\alpha|\beta) \int d\mu_n \frac{F_R}{(c_m,d_m,a_m, \beta)}\nonumber .
\ee
Now we are left with the computation of new integrals in which the factor $\sigma_{c_m,d_m}$ in then numerator has been canceled by the same factor arising in the expansion of $(c_m,d_m,a_m, \alpha)$. More explicitly we now have to compute
\begin{equation}\label{midstep}
\frac{F_R}{(c_m,d_m,a_m, \beta)} = \left(\frac{1}{(12\cdots n)}\prod_{I=1}^{m-1} r_{a_I,b_I,c_I,d_I}\right)\times
\left( \frac{\sigma_{a_m,b_m}}{\sigma_{a_m,d_m}\sigma_{b_m,c_m}\sigma_{d_m,a_m}\sigma_{a_m,\beta_1}\cdots \sigma_{\beta_{n-3},c_m}}\right).
\end{equation}
Now we can repeat the KLT procedure for each such new integrals. Once again we take the left and right factors as those collected in parenthesis in \eqref{midstep} but this time we use a basis of the form $1/(a_m,b_m,d_m,\alpha)$. This gives rise to integrals on the right of the form
\be
\int d\mu_n \frac{\sigma_{a_m,b_m}}{(\sigma_{a_m,d_m}\sigma_{b_m,c_m}\sigma_{d_m,a_m}\sigma_{a_m,\beta_1}\cdots \sigma_{\beta_{n-3},c_m})}
\frac{1}{(a_m,b_m,d_m,\alpha)}\,,
\ee
which after expanding $(a_m,b_m,d_m,\alpha)$ and canceling the factor $\sigma_{a_m,b_m}$ in the numerator becomes an integral of the special kind discussed in section \ref{peter}. We leave the computation of such integrals to the next subsection.

On the left side of the KLT formula one has
\be
\left(\frac{1}{(a_m,b_m,d_m,\beta)\,(12\cdots n)}\prod_{I=1}^{m-1} r_{a_I,b_I,c_I,d_I}\right).
\ee
But this is identical in structure to the starting point \eqref{iterSP} but with $m-1$ cross ratios. Iterating the procedure one ends up reducing the computation of the original integral
\be
\int d\mu_n\left(\frac{1}{(\gamma(1)\gamma(2)\cdots \gamma(n))}\right)\times \left(\frac{1}{(12\cdots n)}\prod_{I=1}^{m-1} r_{a_I,b_I,c_I,d_I}\times \frac{\sigma_{a_m,b_m}\sigma_{c_m,d_m}}{\sigma_{a_m,d_m}\sigma_{b_m,c_m}}\right)
\ee
to that of integrals of the special form
\be
\int d\mu_n \frac{1}{\prod_{e\in G}\sigma_{v_e,u_e}}\,,
\ee
where $G$ is some 4-regular graph with $n$ vertices.

\subsection{Finding a Compatible KLT Basis}

This is the final step of the algorithm. Consider any integral of the special type as the ones found above
\be\label{one}
\int d\mu_n \frac{1}{\prod_{e\in G}\sigma_{v_e,u_e}}.
\ee
If the integrand can be written as the product of two Parke-Taylor factors then this integral is of the form $m(\alpha|\beta)$ and it is the end of the procedure. This is the ideal point to introduce more of the mathematical terminology corresponding to our physical issue. A Parke-Taylor factor
of the integrand corresponds to a Hamilton cycle of $G$. Therefore, if $G$ has two edge-disjoint Hamilton cycles we stop.

Assuming that $G$ does not have two edge-disjoint Hamilton cycles, using Petersen's theorem as described in section \ref{peter} we write \eqref{one}  as
\be
\int d\mu_n \frac{1}{\prod_{e\in G_L}\sigma_{v_e,u_e}}\times \frac{1}{\prod_{e\in G_R}\sigma_{v_e,u_e}}\,,
\ee
where both $G_L$ and $G_R$ are 2-regular graphs of $n$-vertices. This means that the left and the right factors can be used as the starting points in a KLT decomposition.

The next step is based on the existence of compatible Parke-Taylor factors as discussed in section \ref{HD}. Recall that a Parke-Taylor factor $(\alpha(1)\alpha(2)\cdots \alpha(n))$ is said to be compatible with a 2-regular graph, say $G_L$, if the union of the two graphs, which is a 4-regular graph, admits a Hamiltonian decomposition, i.e., it is the union of two edge-disjoint Hamilton cycles (see section \ref{HD} for definitions and more details).

In section \ref{HD} we gave computer-based evidence for the fact that for any 2-regular graph, the number of compatible Parke-Taylor factors is larger than $(n-3)!$. We now also assume that from the set of compatible graphs of $G_L$ and of $G_R$ it is possible to choose $(n-3)!$ of each, denoted as ${\cal L}$ and ${\cal R}$ such that they are independent in the sense defined in section \ref{gklt}. For the reader's convenience we recall that ${\cal L}$ and ${\cal R}$ are said to be independent of the $(n-3)!\times (n-3)!$ matrix
\be
m^{{\cal L}|{\cal R}} = \{ m(\alpha |\beta )~~ :~~\alpha \in {\cal L},~~ \beta \in {\cal R}\}
\ee
is not singular.

Under these assumptions we take ${\cal L}$ and ${\cal R}$ in order to build a generalized KLT relation that expresses
\be
\int d\mu_n \frac{1}{\prod_{e\in G_L}\sigma_{v_e,u_e}}\times \frac{1}{\prod_{e\in G_R}\sigma_{v_e,u_e}}
\ee
as a sum over product of
\be\label{int1}
\int d\mu_n \frac{1}{(\prod_{e\in G_L}\sigma_{v_e,u_e})(\alpha(1)\alpha(2)\cdots \alpha(n))}
\ee
and
\be\label{int2}
\int d\mu_n \frac{1}{(\prod_{e\in G_L}\sigma_{v_e,u_e})(\beta(1)\beta(2)\cdots \beta(n))}
\ee
where $\alpha$ and $\beta$ define graphs that are $G_L$ and $G_R$ compatible respectively.

By definition of $G_L$ and $G_R$ compatibility both integrals \eqref{int1} and \eqref{int2} can be written as the product of two Parke-Taylor factors, e.g,
\be
(\prod_{e\in G_L}\sigma_{v_e,u_e})(\alpha(1)\alpha(2)\cdots \alpha(n))= (\rho(1)\rho(2)\cdots \rho(n))(\rho'(1)\rho'(2)\cdots \rho'(n))
\ee
for some $\rho,\rho'\in S_n$ and therefore
\be
\int d\mu_n \frac{1}{(\prod_{e\in G_L}\sigma_{v_e,u_e})(\alpha(1)\alpha(2)\cdots \alpha(n))} = m(\rho |\rho')
\ee
as desired.

%%%%%%%%%%%%%%%%%%%%%%%%%%%%%%%%%%%%%%%%%%%%%%%%%%%%
\section{All Six-Point Integrals}\label{allsix}
%%%%%%%%%%%%%%%%%%%%%%%%%%%%%%%%%%%%%%%%%%%%%%%%%%%

In this section we illustrate the techniques introduced in previous sections in the case of six-point integrals. Starting with a general integral
\be
\int d\mu_6 F(\sigma)
\ee
one can use the reduction procedure explained in the previous section to turn it into combinations of the special integrals studied in section \ref{peter} which have a constant numerator. More explicitly, we are left with the computation of integrals of the form
\be
\int d\mu_6\, \frac{1}{\prod_{1\leq a<b\leq 6}\sigma_{ab}^{w_{ab}}}
\ee
with $w_{ab}\geq 0$, $w_{aa}=0$ and $\sum_{b=1}^n w_{ab}=4$ for all $a\in \{1,2,\ldots ,n\}$.

As explained in section 4, the matrix $w_{ab}$, with $w_{ba}=w_{ab}$, is the adjacency matrix of a 4-regular graph $G_F$ \cite{graph1,graph2}. Employing Petersen's theorem any such integrand can then be decomposed as
\be
\int d\mu_6 \,\frac{1}{\prod_{e\in G^L_F}\sigma_{v_e,u_e}}\times\frac{1}{\prod_{e\in G^R_F}\sigma_{v_e,u_e}}
\ee
where $G^L_F$ and $G^R_F$ are two 2-regular graphs that provide one of the possible decompositions of $G_F$.

Each one of the 2-regular graphs $G^L_F$ and $G^R_F$ is one of the following options: a hexagon; a square and a bubble; two triangles; or three bubbles.

The last step is to find two sets of six Parke-Taylor factors ${\cal L}$ and ${\cal R}$ such that all Parke-Taylor factors in ${\cal L}$ (${\cal R}$) are compatible with $G^L_F$ ($G^R_F$). Of course, the two sets ${\cal L}$ and ${\cal R}$ must also be independent. Once this is found, the KLT procedure of section \ref{kltgen} completes the computation in terms of the basic building blocks $m(\alpha|\beta)$.

At this point is clear that our only remaining task in this section is to provide explicit sets of Parke-Taylor factors which are compatible with each one of the possible 2-regular graph. 

Clearly, there is nothing needed when the 2-regular graph is a hexagon as it is by itself a Parke-Taylor factor and therefore compatible with all $5!$ Parke-Taylor factors at six-points.

%%%%%%%%%%%%%%%%%%%%%%%%%%%%%%%%%%%%%%%
\subsection{A Bubble and a Square}
%%%%%%%%%%%%%%%%%%%%%%%%%%%%%%%%%%%%%%

Consider the configuration of labels for a bubble and a square given in figure 7.1 
%%% Figure 7.1 %%%%%
\begin{center}
\includegraphics[scale=0.4]{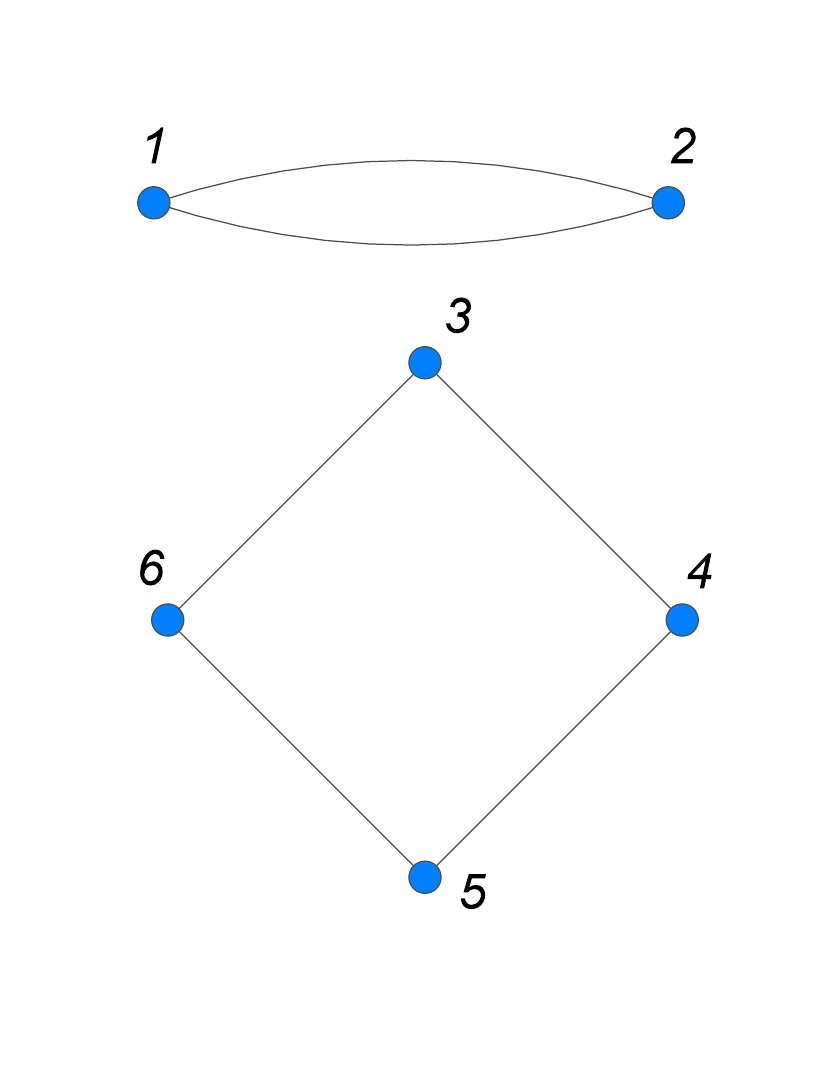}\,\,\,\,\,\, .
\begin{center}
({\bf Fig.7.1}) {\small {\rm A bubble and a square geometry.}}
\end{center}
\end{center}
%%%%%%%%%%%%%%%%%%%%
Clearly, any other assignments of labels can be obtained from this one by a simple relabelling. An independent basis of compatible six Parke-Taylor factors chosen from the 32 possible ones is given in figure 7.2
%%%%%  Figures 7.2  %%%%%%%%%%
\begin{center}
\includegraphics[scale=0.3]{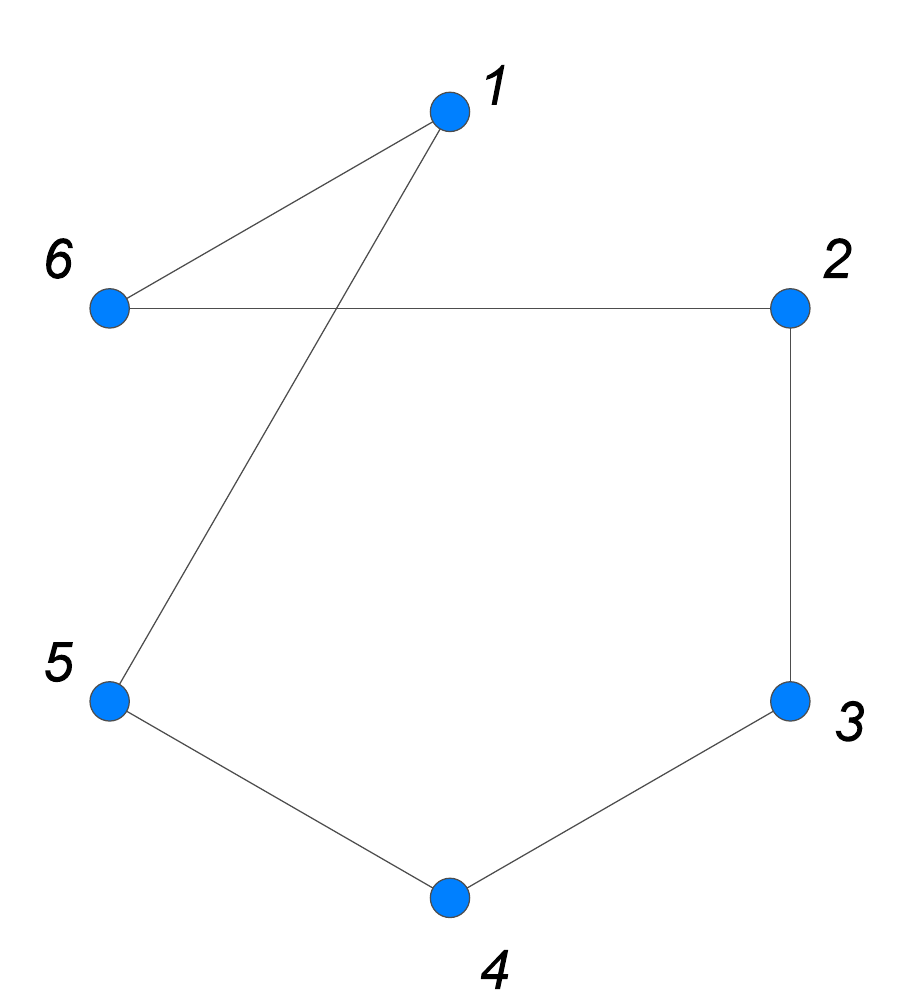},\qquad\,\,\,\,
\includegraphics[scale=0.3]{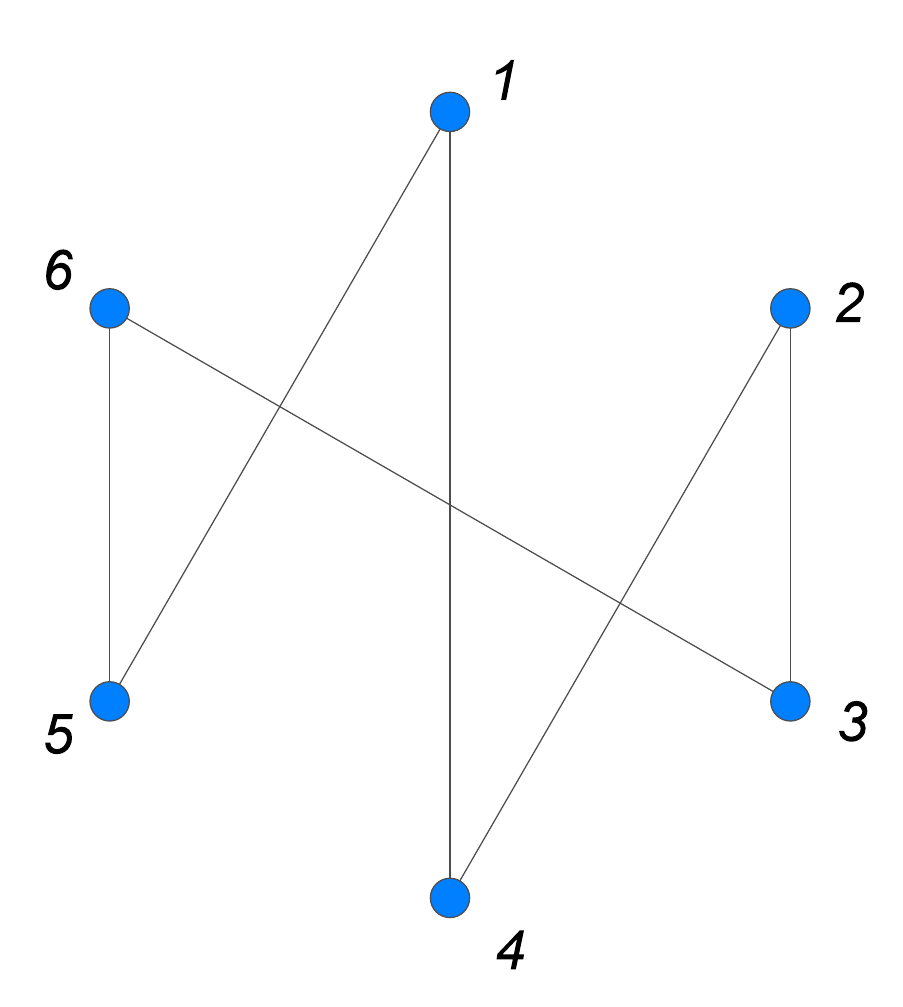},\qquad\,\,\,\,
\includegraphics[scale=0.3]{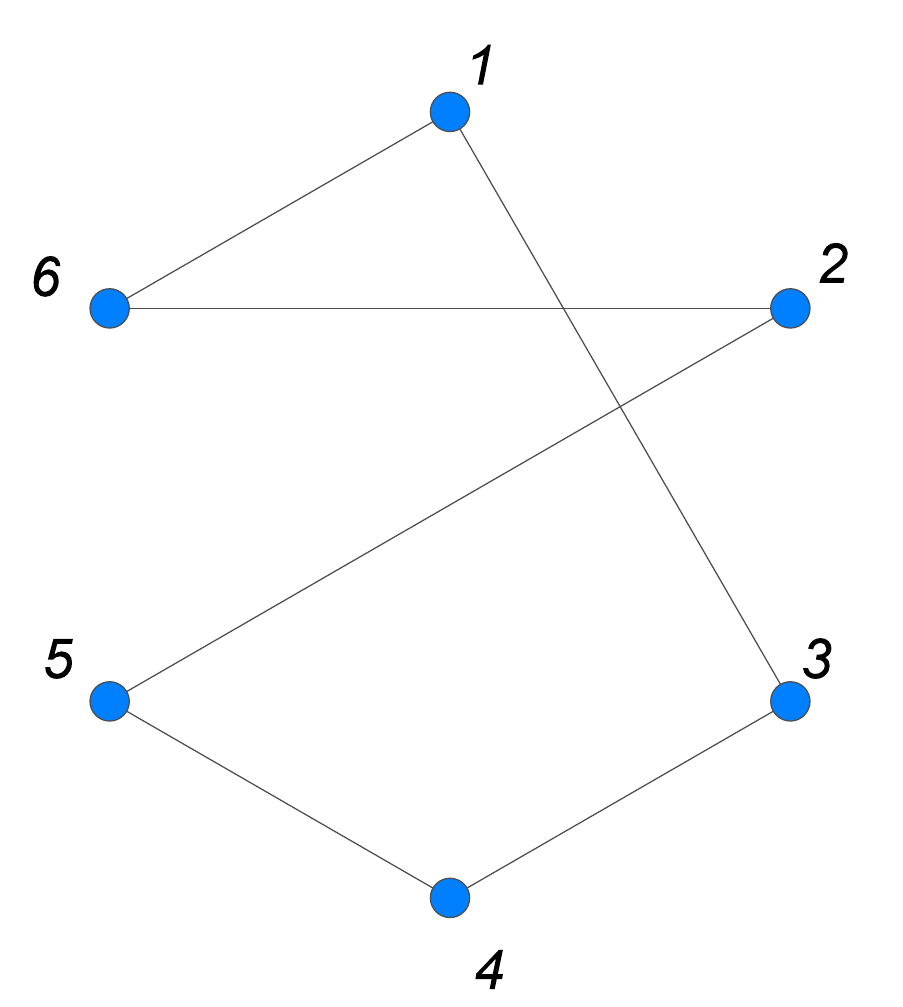},\\
\includegraphics[scale=0.3]{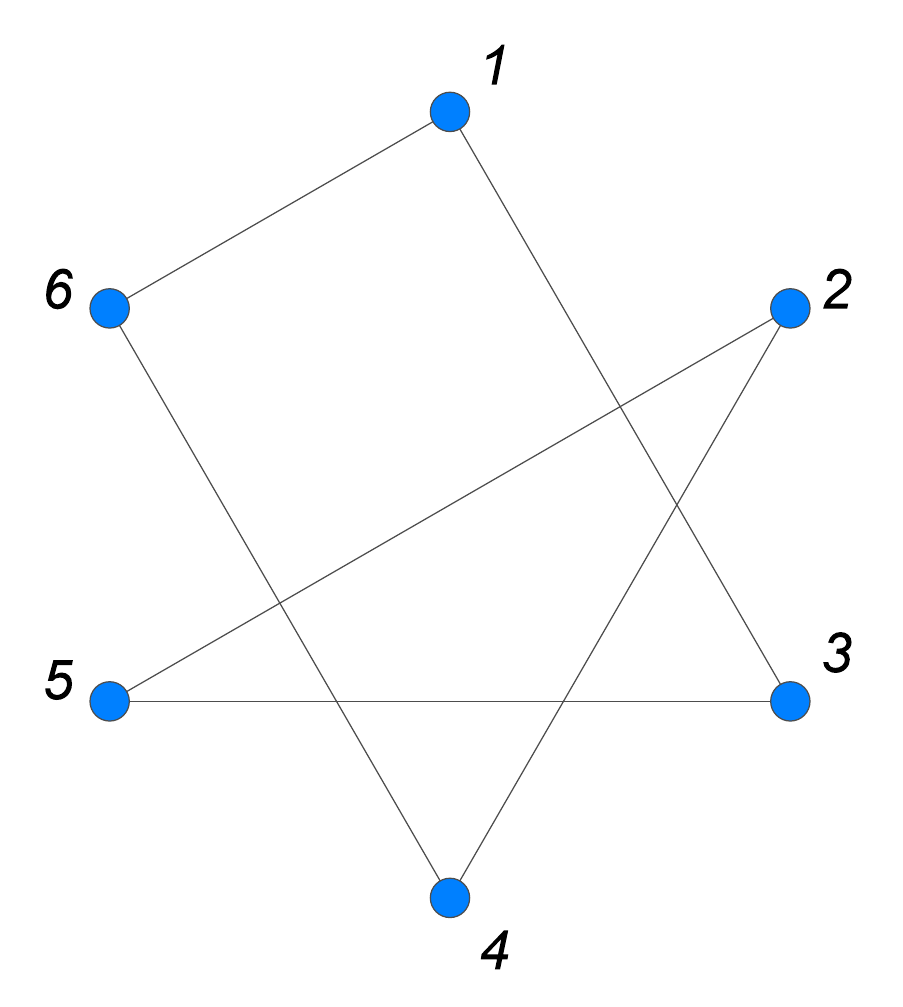},\qquad\,\,\,\,
\includegraphics[scale=0.3]{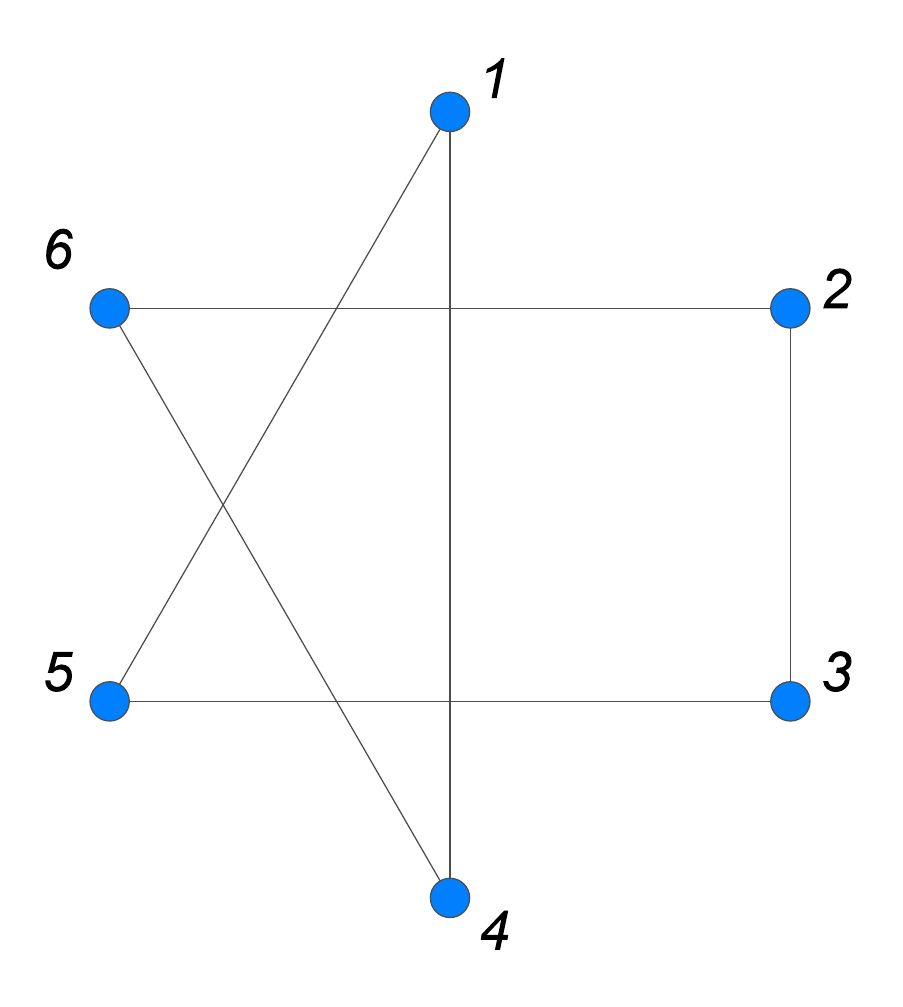},\qquad\,\,\,\,
\includegraphics[scale=0.3]{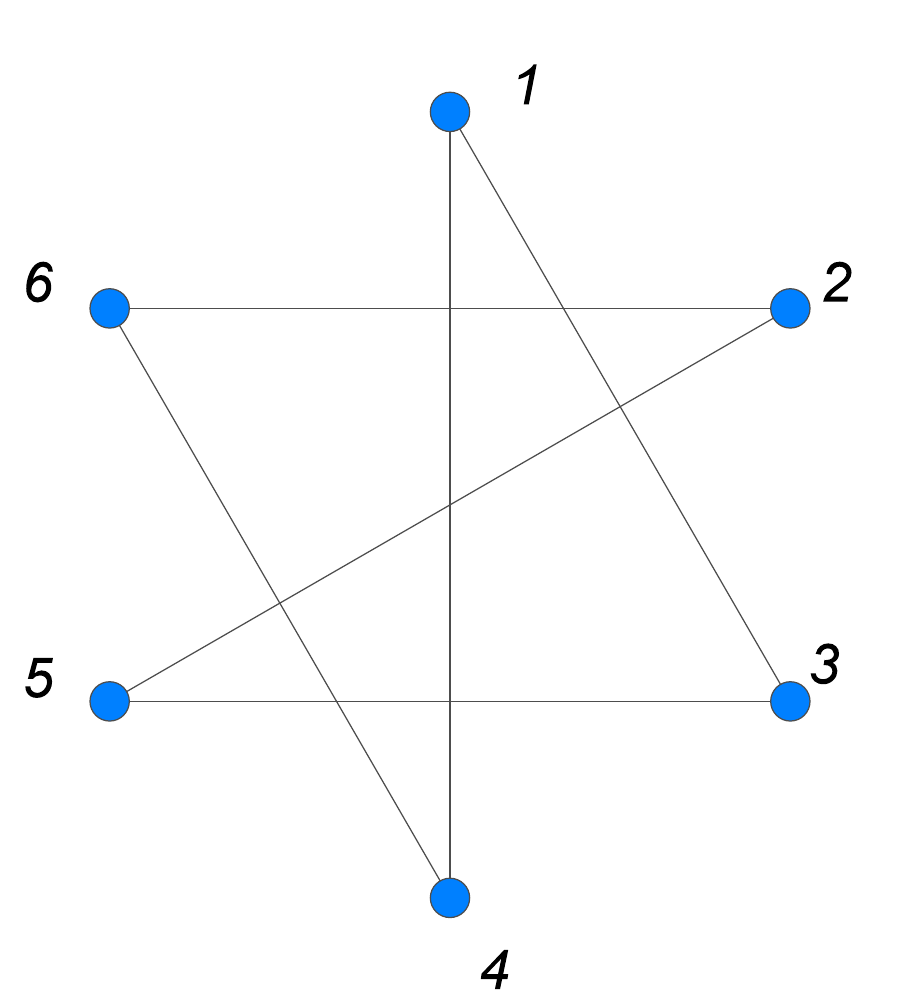}.
\begin{center}
({\bf Fig.7.2}) {\small {\rm Compatible cycles with a bubble and a square.}}
\end{center}
\end{center}
%%%%%%%%%%%%%%%%%%%%%%%%%%%%%%

%%%%%%%%%%%%%%%%%%%%%%%%%%%%%%%%%%
\subsection{Two Triangles} 
%%%%%%%%%%%%%%%%%%%%%%%%%%%%%%%%%

Consider the configuration of labels for two triangles given in figure 7.3
%%% Figure 7.3 %%%%%
\begin{center}
\includegraphics[scale=0.4]{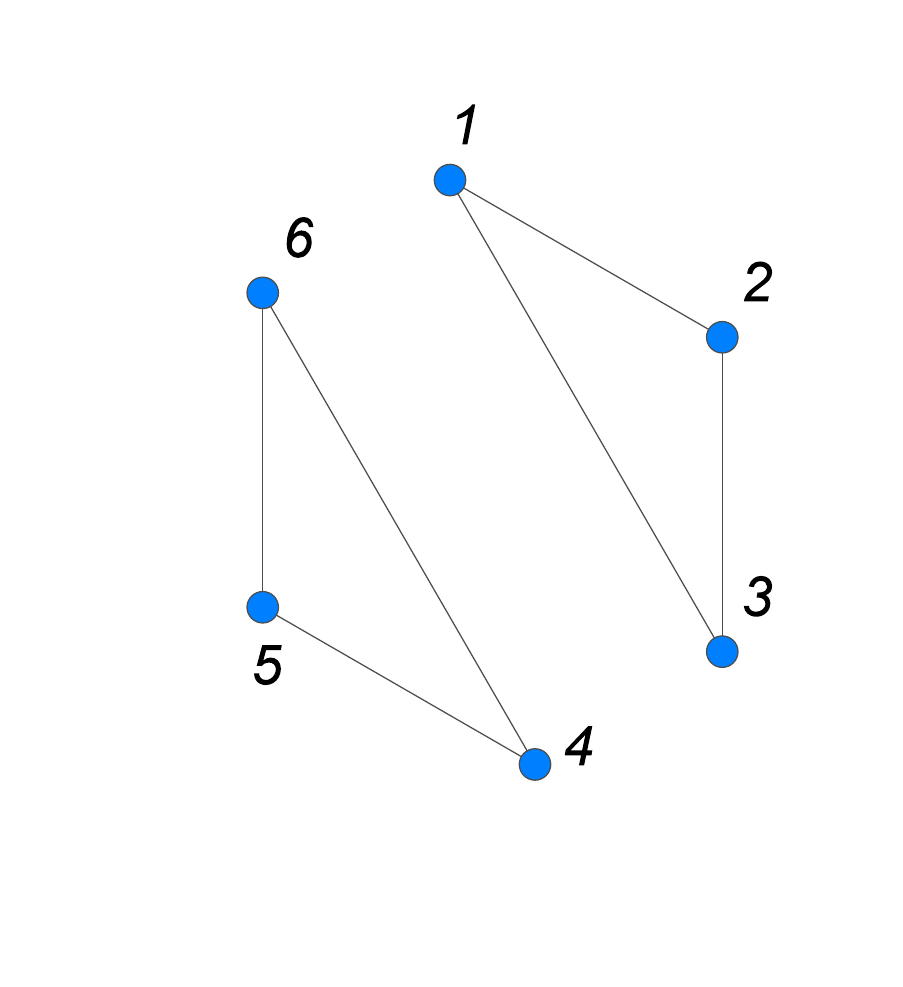}\,\,\,.\,\,\,
\begin{center}
({\bf Fig.7.3}) {\small {\rm Two triangles geometry.}}
\end{center}
\end{center}
%%%%%%%%%%%%%%%%%%%
An independent basis of compatible six Parke-Taylor factors chosen from the 42 possible ones is given in figure 7.4
%%%%%  Figures 7.4 %%%%%%%%%%
\begin{center}
\includegraphics[scale=0.3]{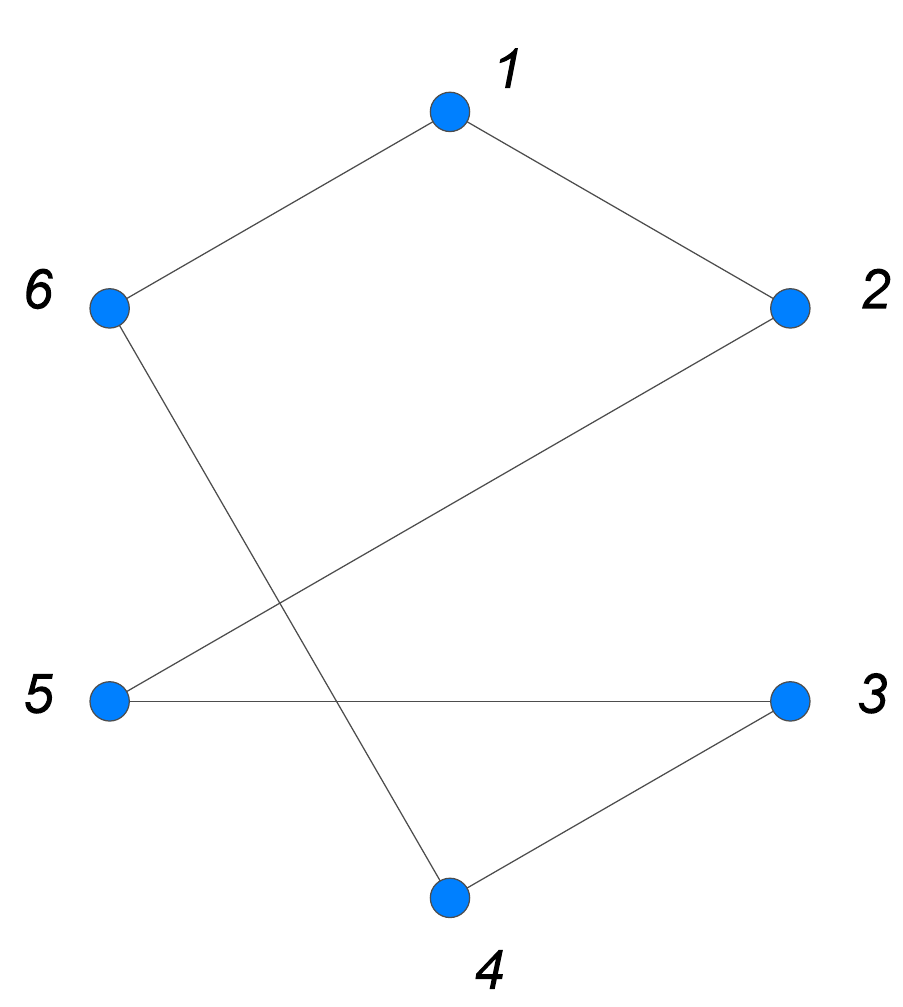},\qquad\,\,\,\,
\includegraphics[scale=0.3]{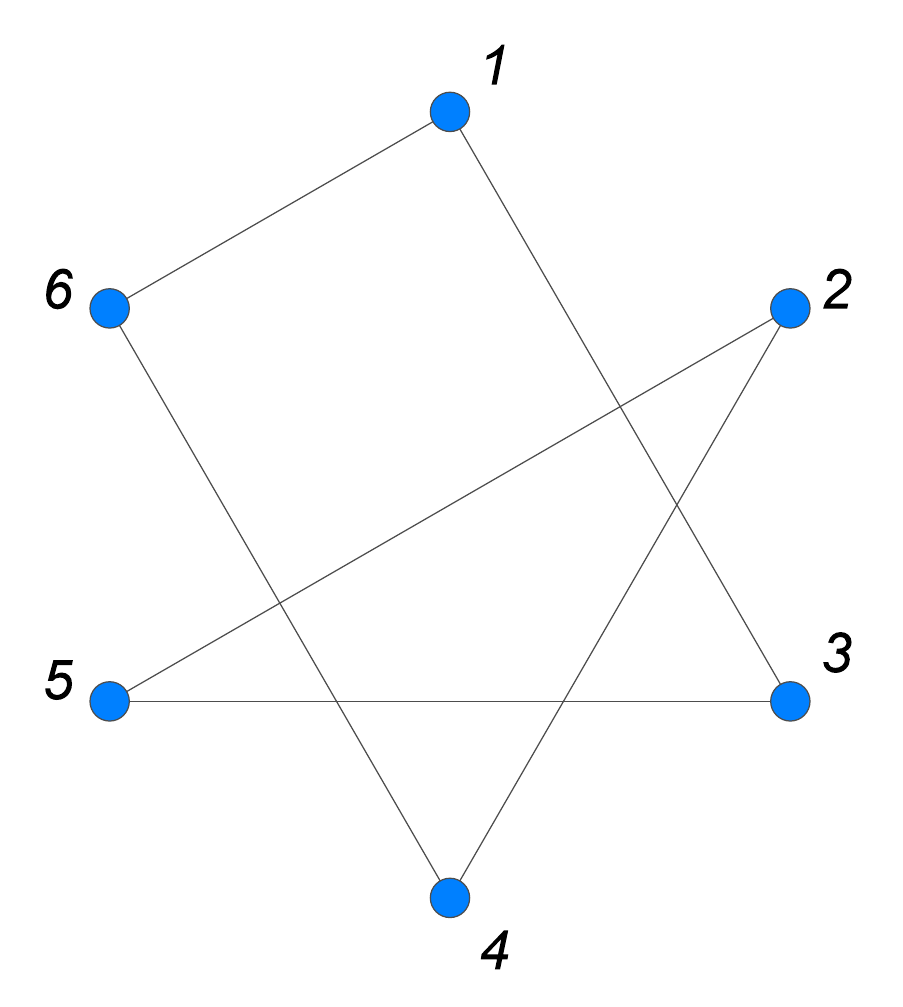},\qquad\,\,\,\,
\includegraphics[scale=0.3]{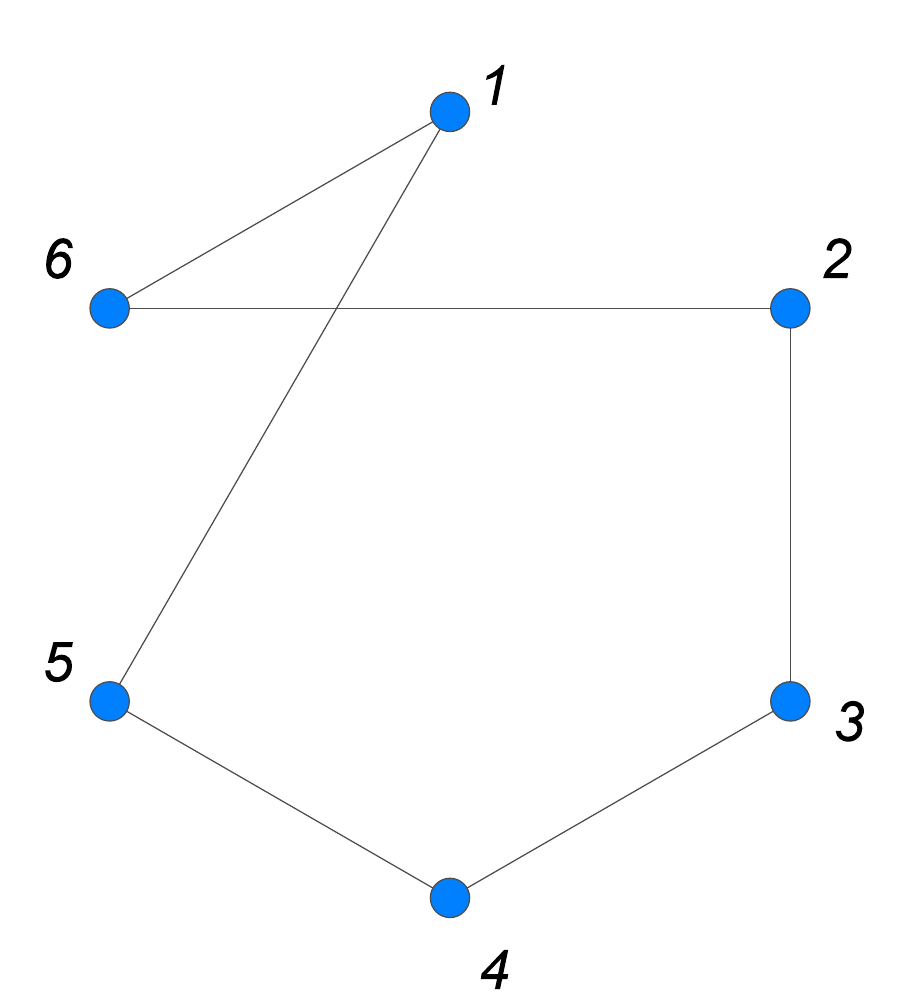},\\
\includegraphics[scale=0.3]{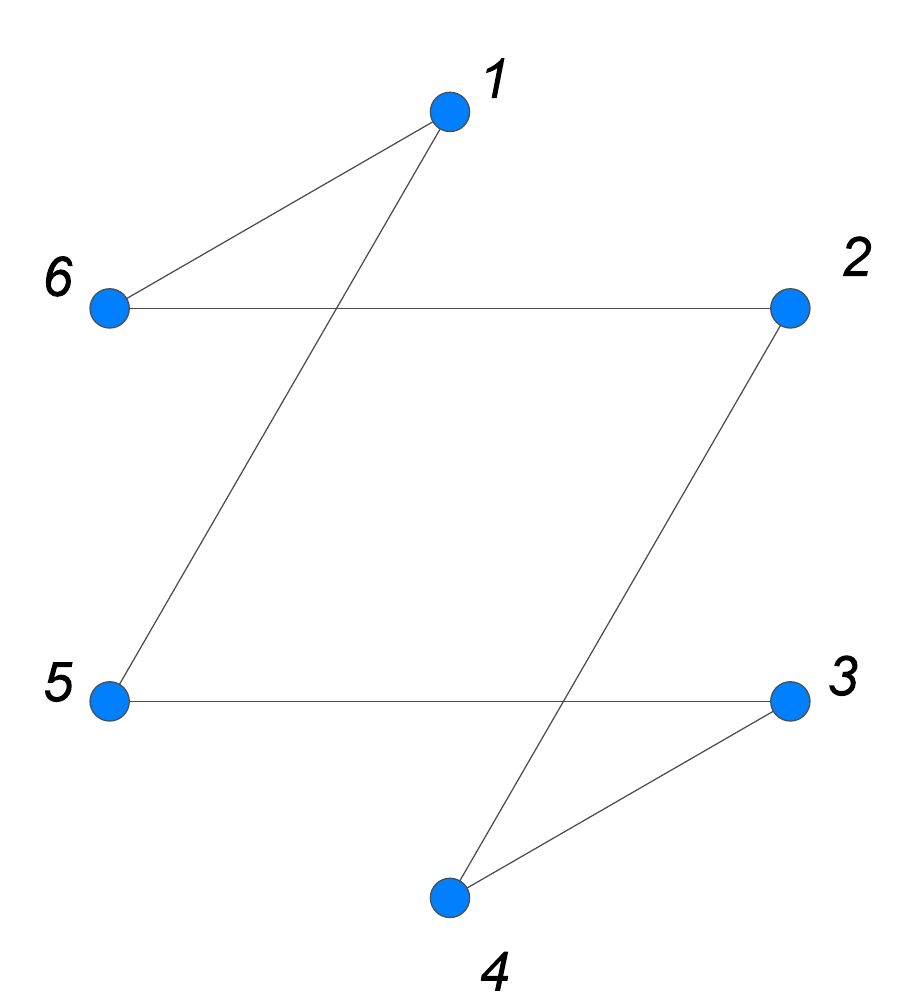},\qquad\,\,\,\,
\includegraphics[scale=0.3]{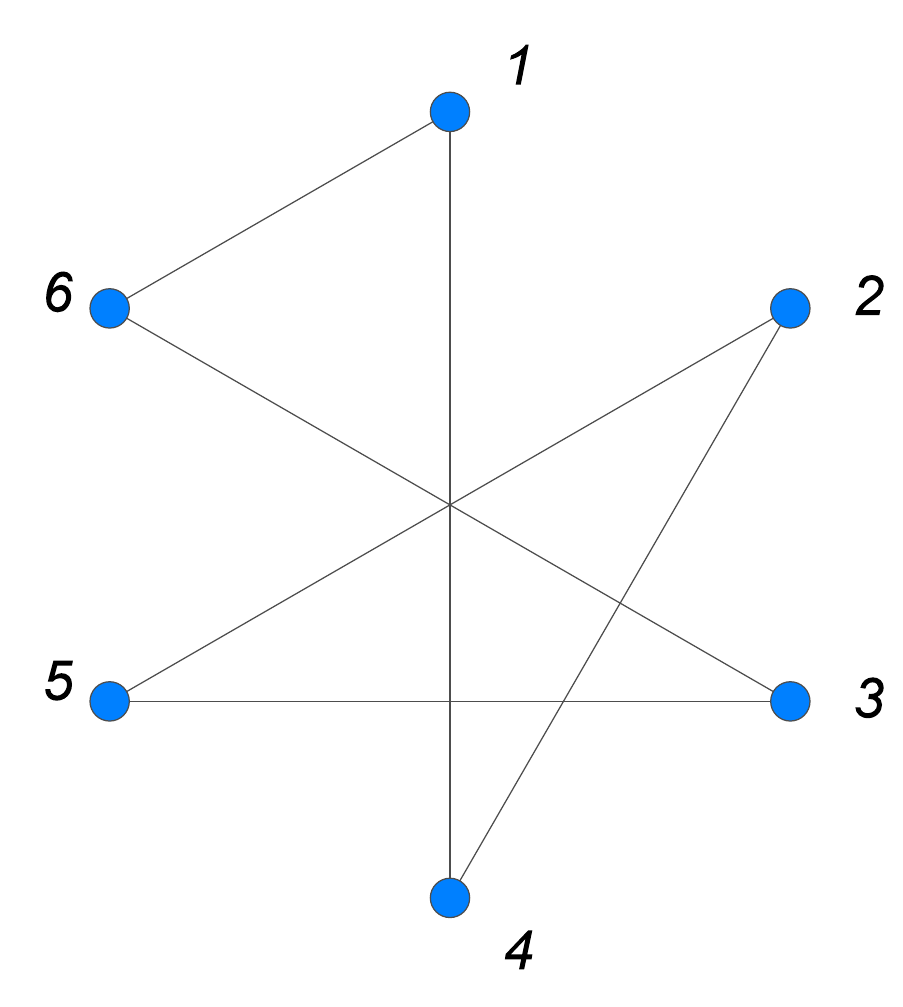},\qquad\,\,\,\,
\includegraphics[scale=0.3]{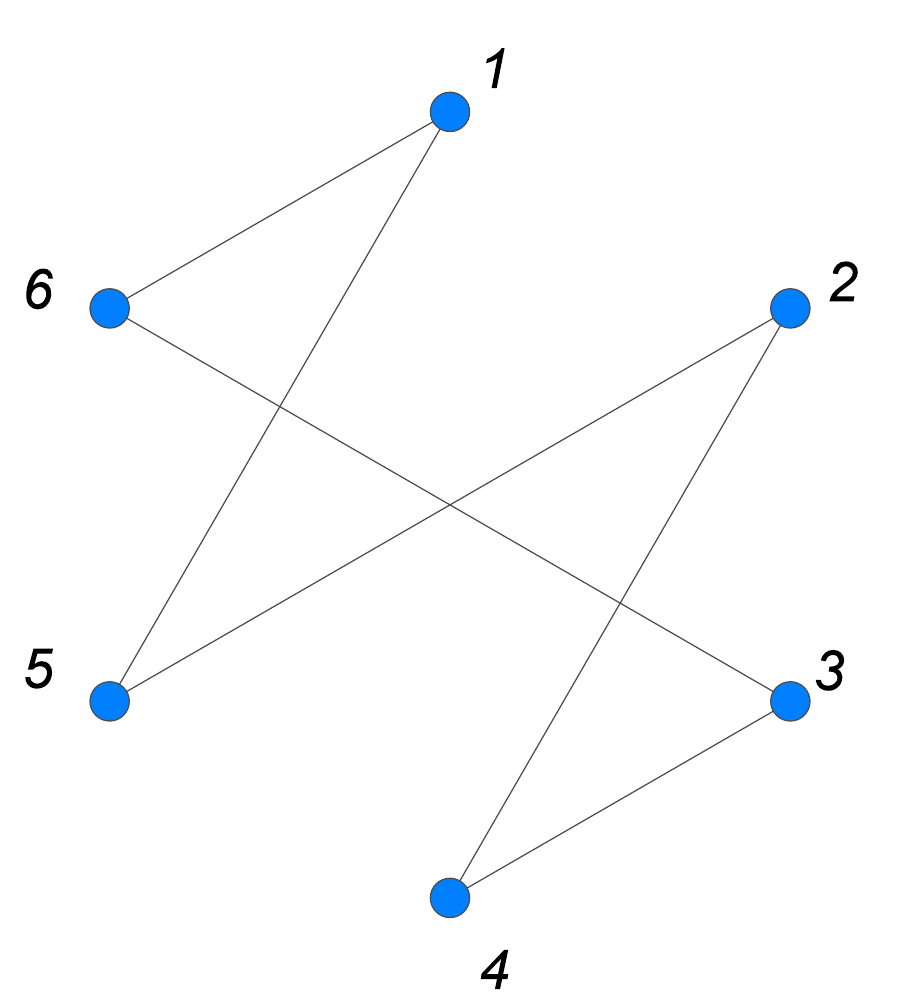},
\begin{center}
({\bf Fig.7.4}) {\small {\rm Compatible cycles with two triangles.}}
\end{center}
\end{center}
%%%%%%%%%%%%%%%%%%%%%%%%%%%%%%%

%%%%%%%%%%%%%%%%%%%%%%%%%%%%%%%
\subsection{Three Bubbles}
%%%%%%%%%%%%%%%%%%%%%%%%%%%%%%%

Finally, consider the configuration of labels for three bubbles given in figure 7.5
%%% Figure 7.5 %%%%%
\begin{center}
\includegraphics[scale=0.5]{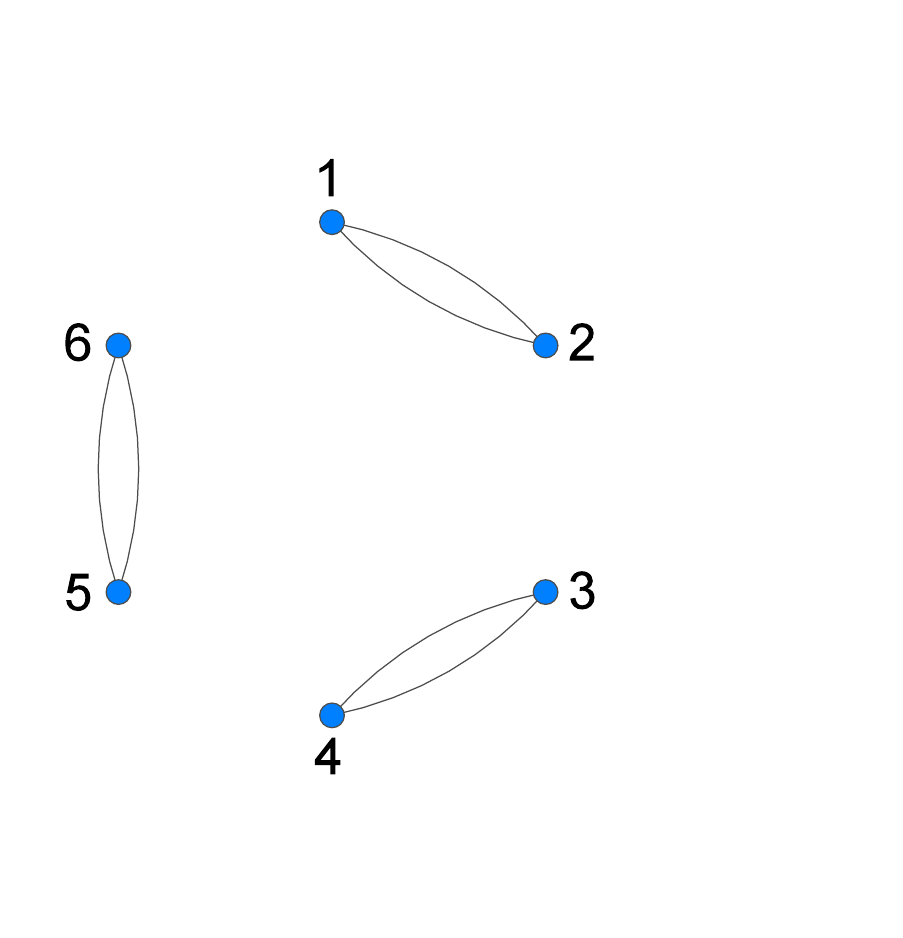}.
\begin{center}
({\bf Fig.7.5}) {\small {\rm Three Bubbles.}}
\end{center}
\end{center}
%%%%%%%%%%%%%%%%%%%%%%%%%%%%%

From the table \ref{t1}, we know that the total number of  2-regular compatible graphs with three bubble is 16. These 16 Parke-Taylor factors are given in figure 7.6
%%%%%  Figures   7.6 %%%%%%%%%%
\begin{center}
\includegraphics[scale=0.3]{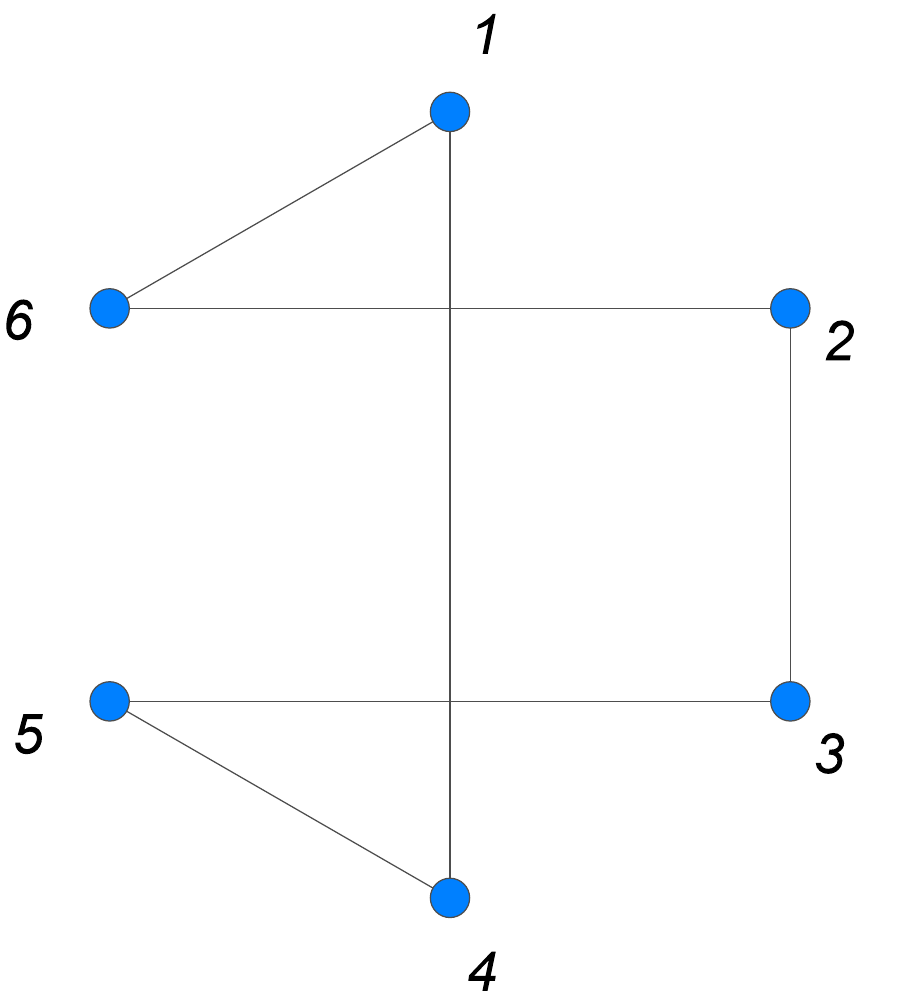},\qquad\,\,\,\,
\includegraphics[scale=0.3]{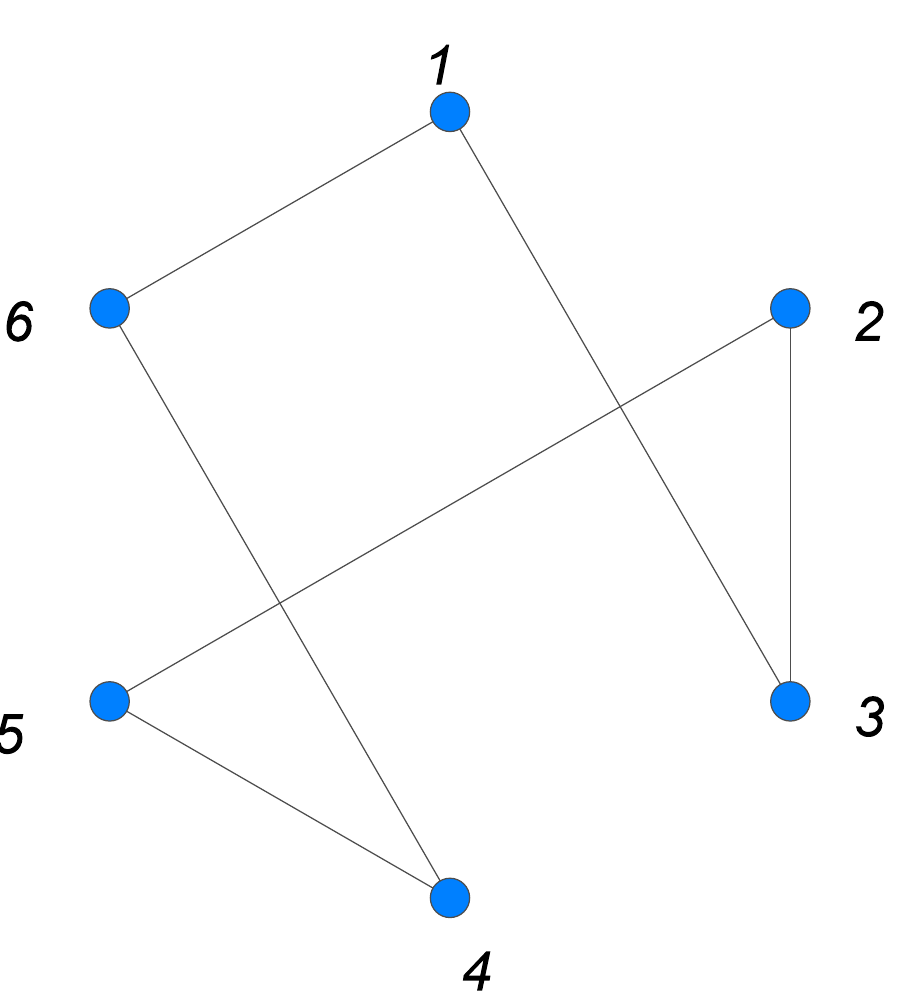},\qquad\,\,\,\,
\includegraphics[scale=0.3]{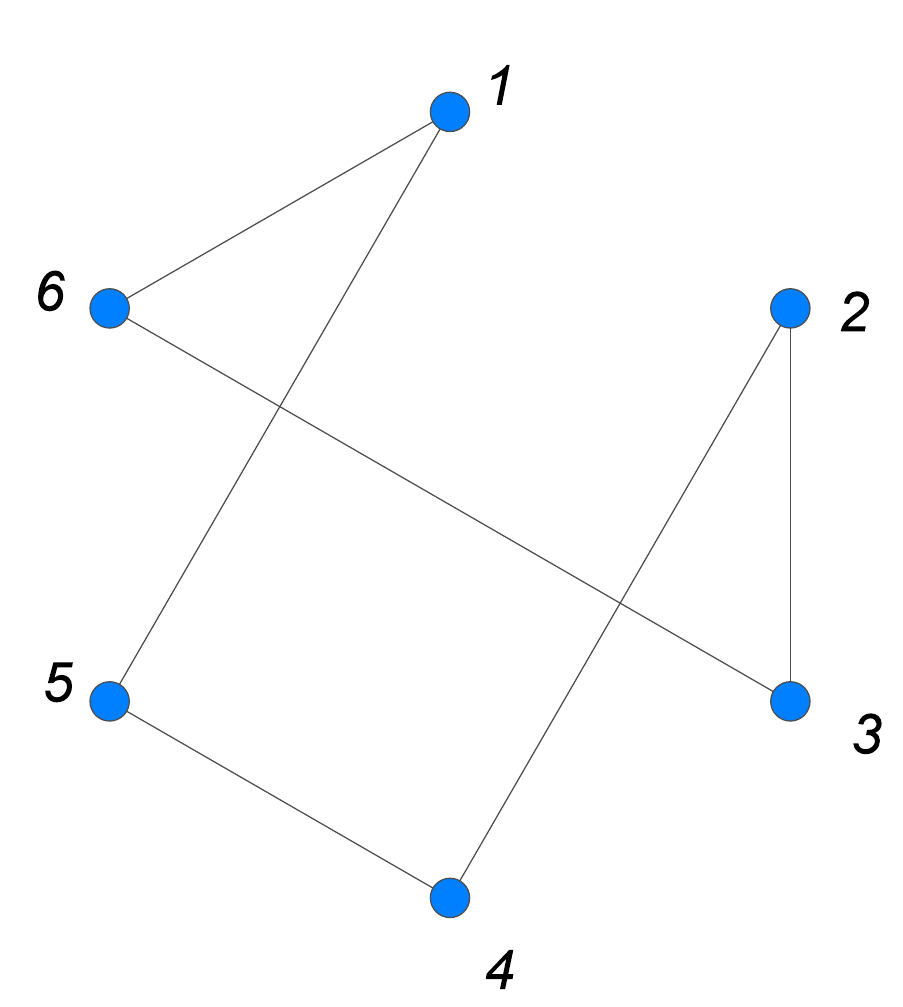},\,\\
\includegraphics[scale=0.3]{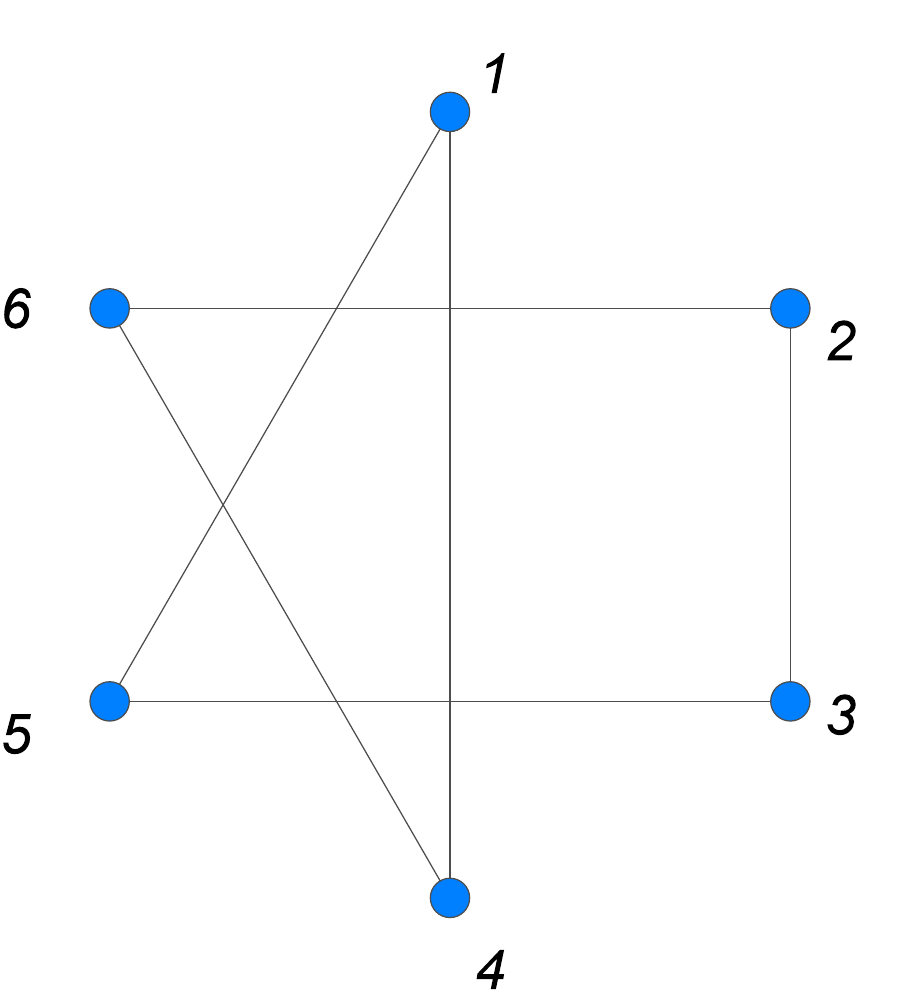},\qquad\,\,\,\,
\includegraphics[scale=0.3]{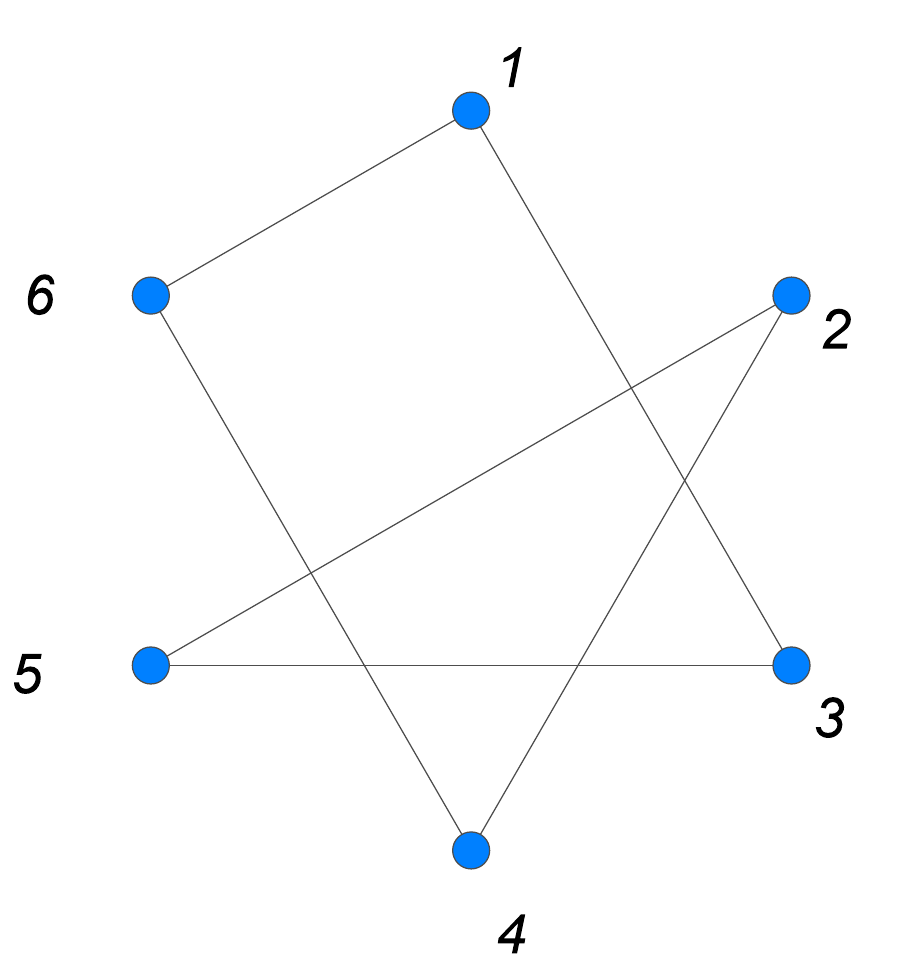},\qquad\,\,\,\,
\includegraphics[scale=0.3]{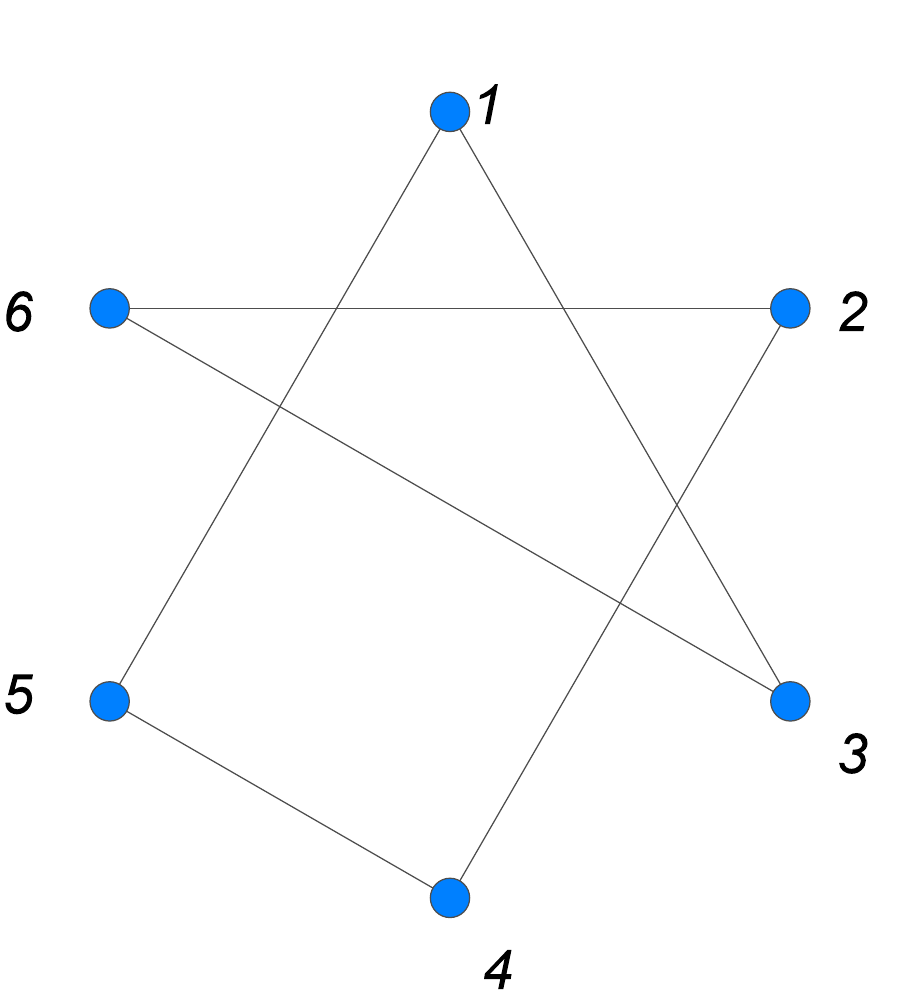},\,\\
\includegraphics[scale=0.3]{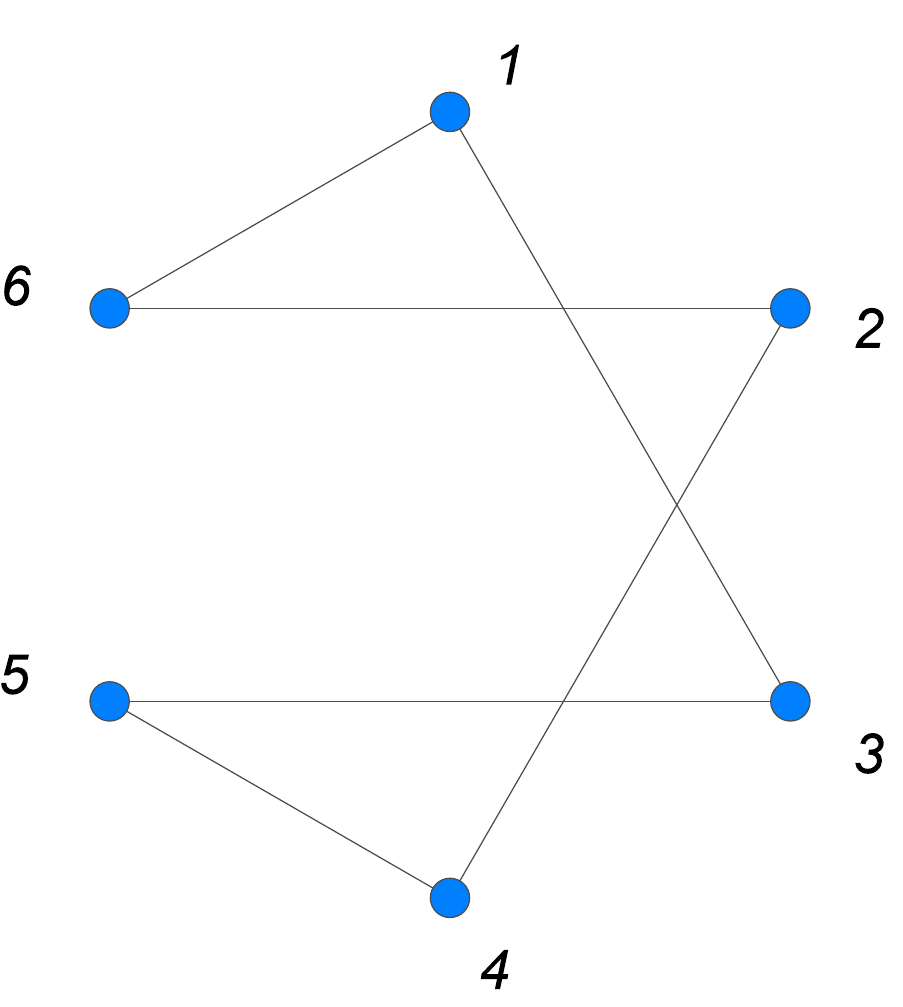},\qquad\,\,\,\,
\includegraphics[scale=0.3]{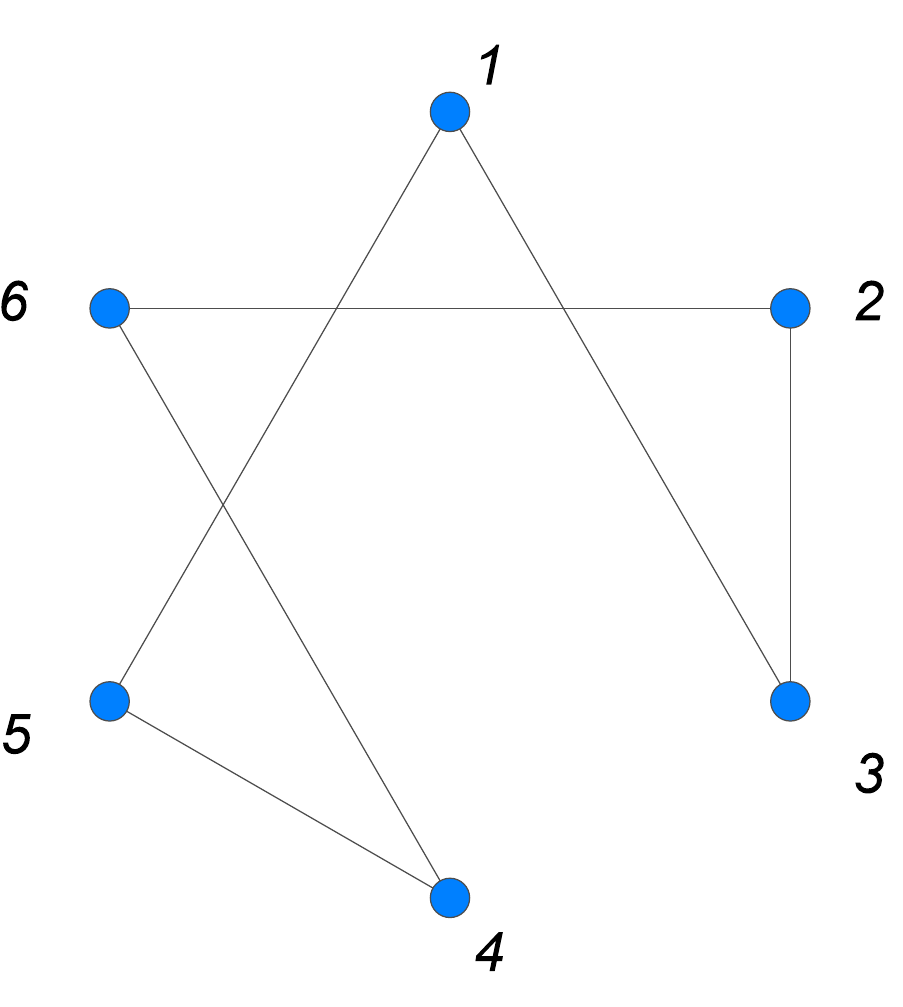},\qquad\,\,\,\,
\includegraphics[scale=0.3]{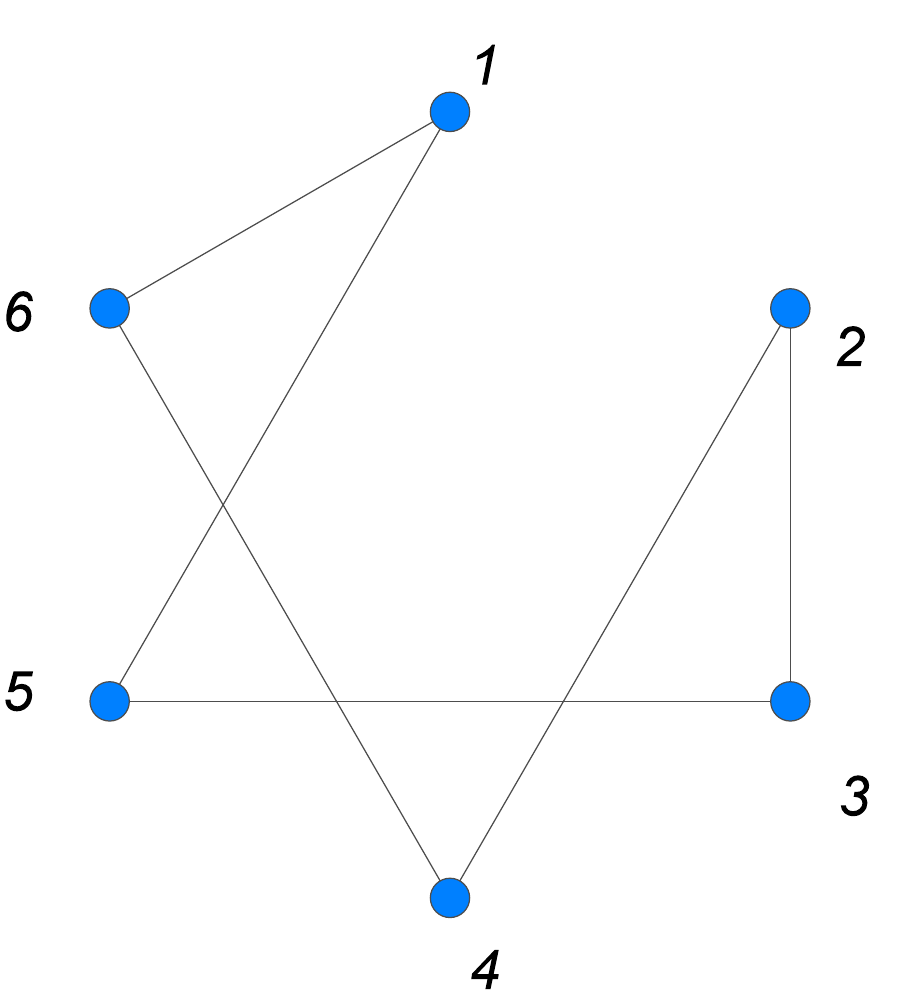},\,\\
\includegraphics[scale=0.3]{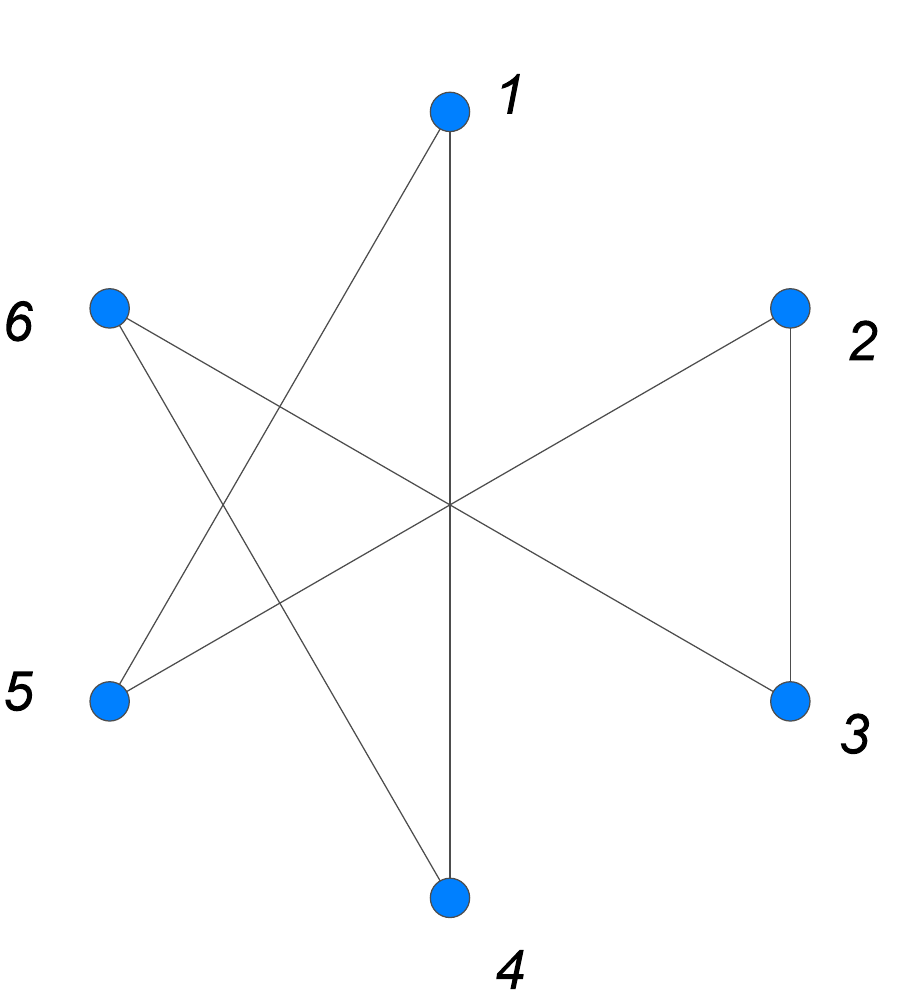},\qquad\,\,\,\,
\includegraphics[scale=0.3]{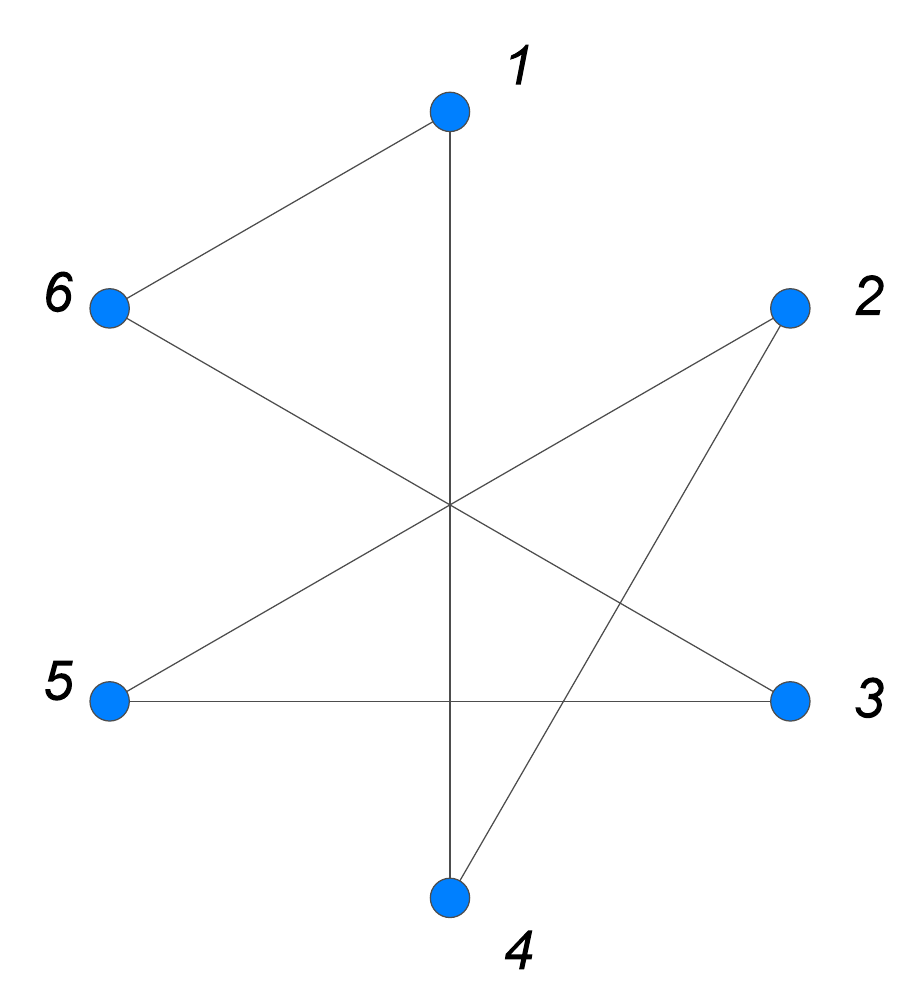},\qquad\,\,\,\,
\includegraphics[scale=0.3]{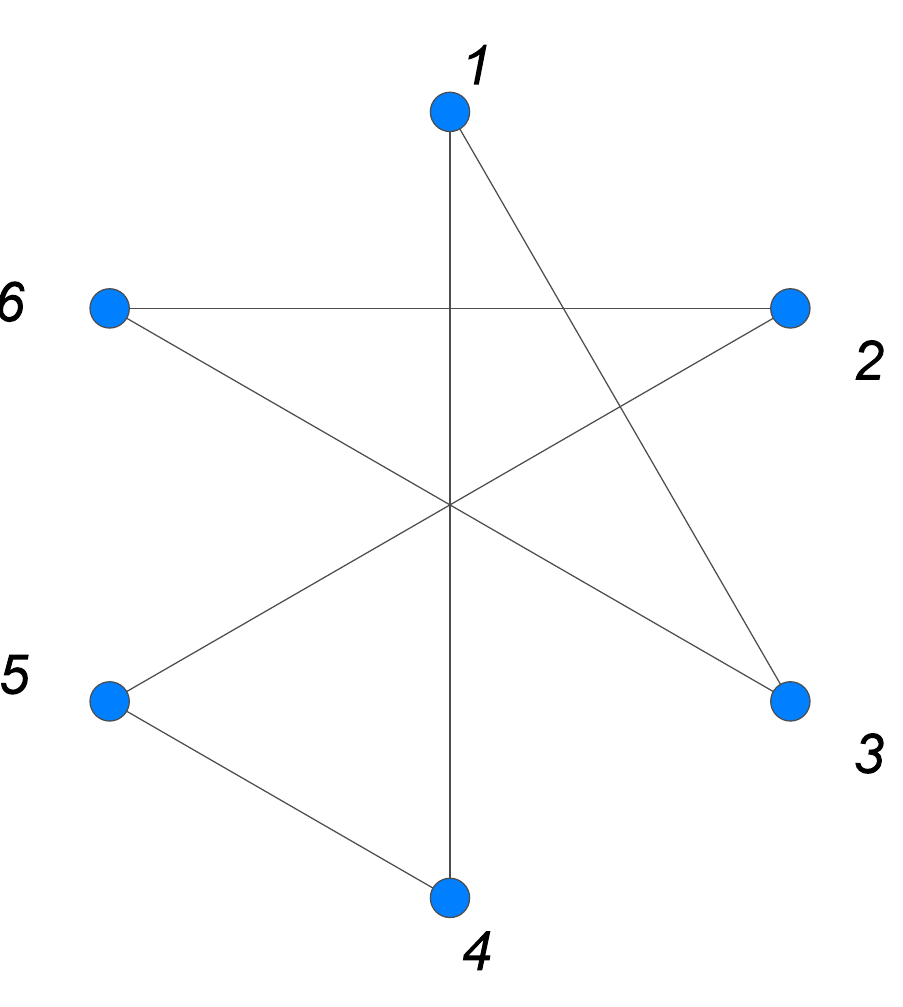},\,\\
\includegraphics[scale=0.3]{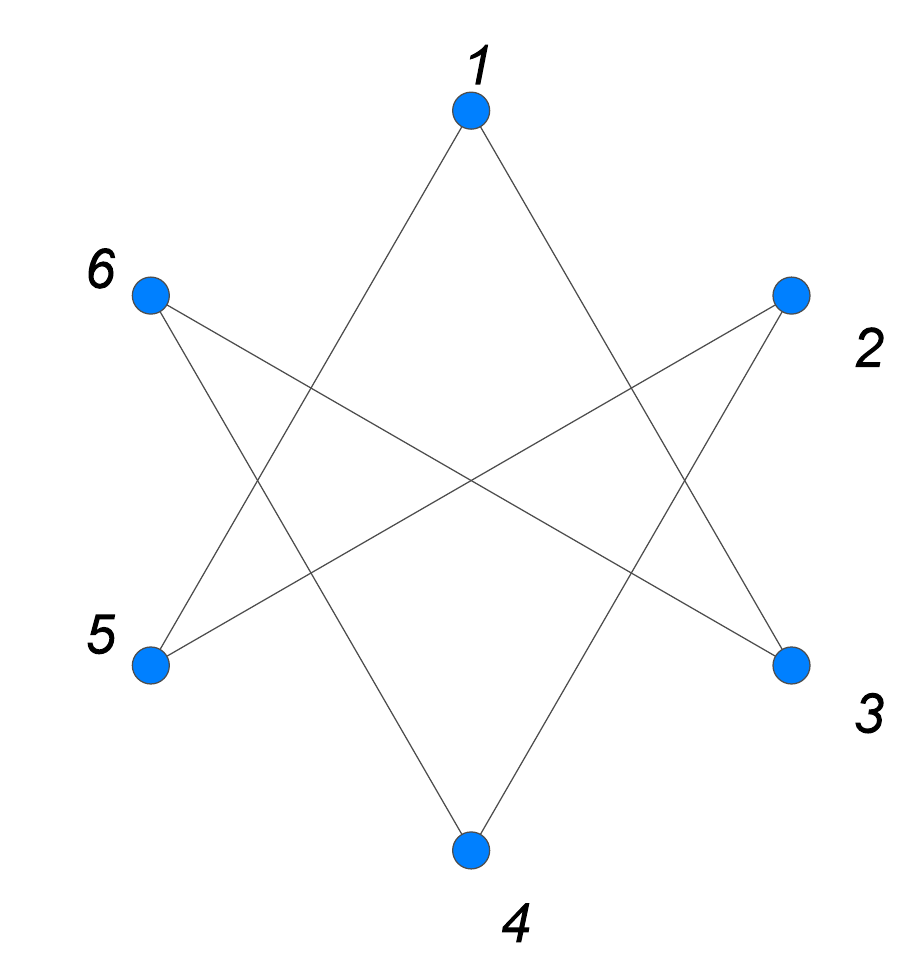},\qquad\,\,\,\,
\includegraphics[scale=0.3]{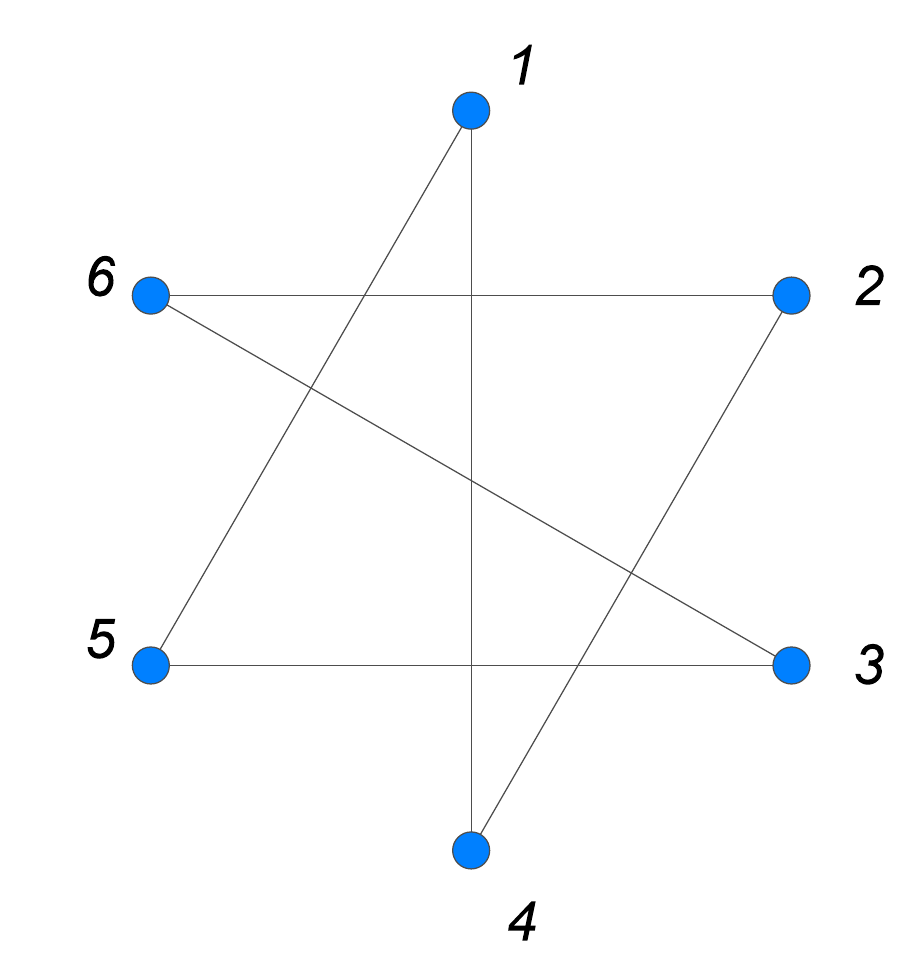},\qquad\,\,\,\,
\includegraphics[scale=0.3]{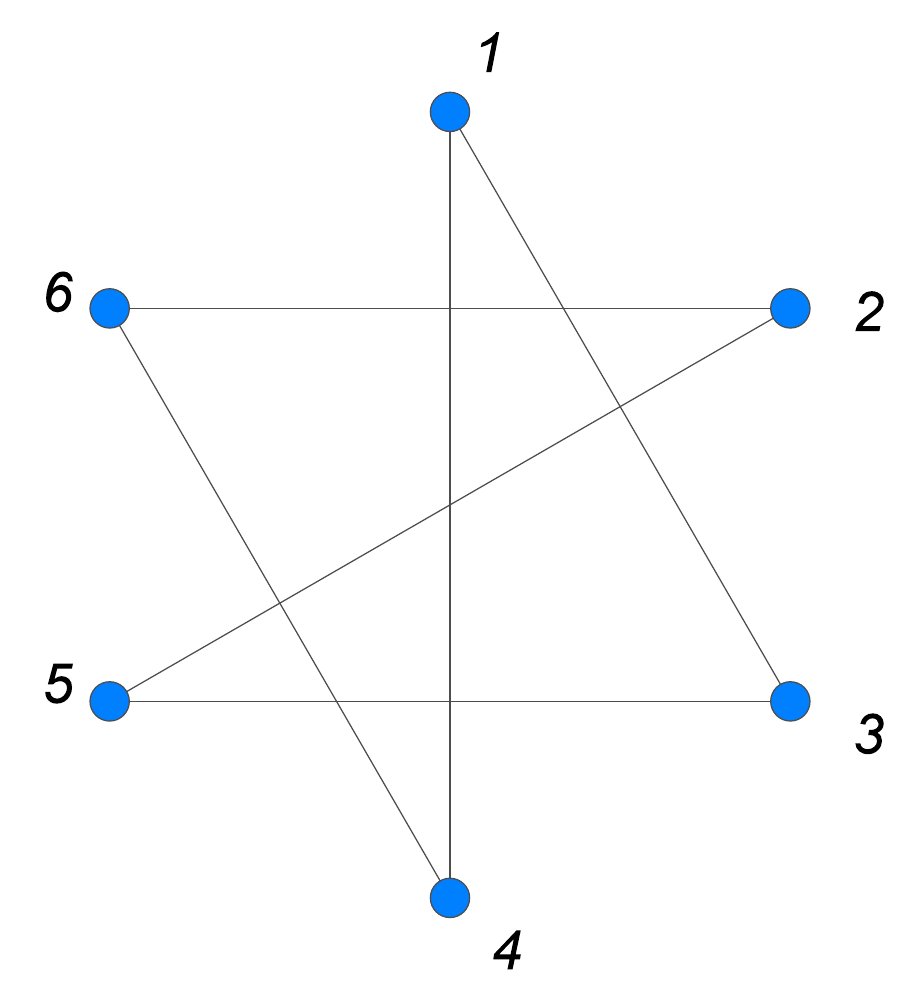},\,\\
\includegraphics[scale=0.3]{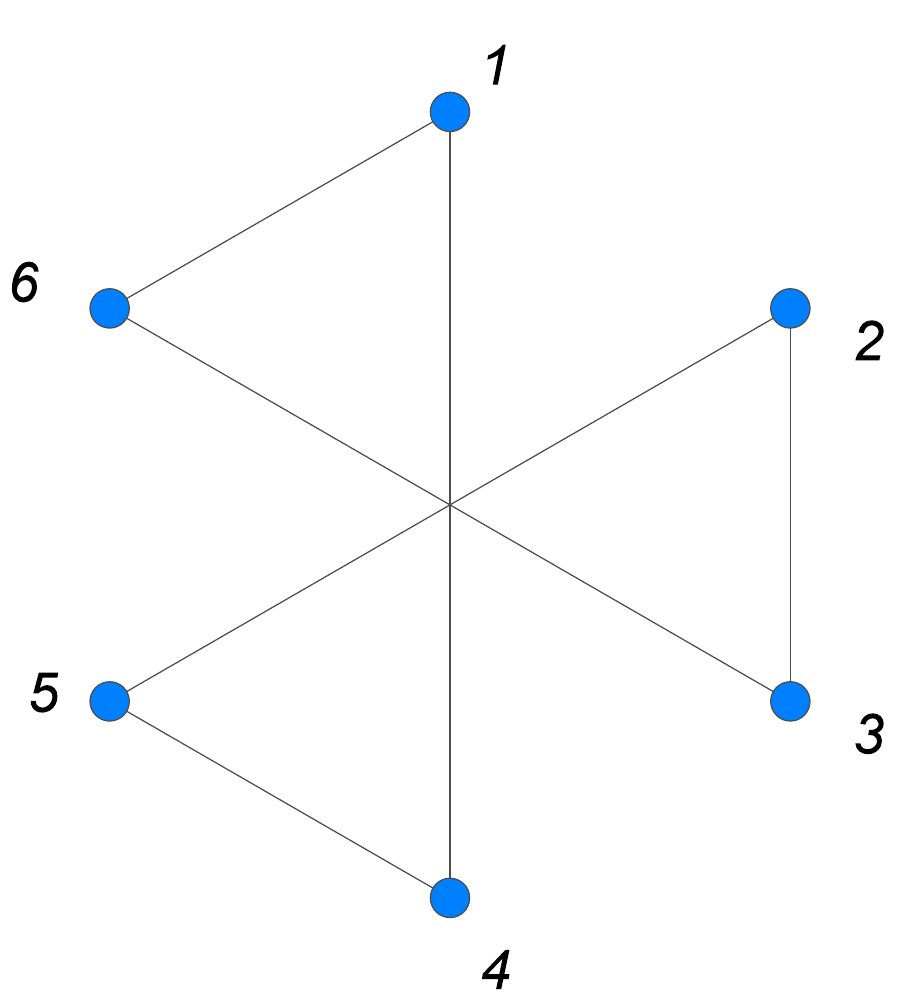}.
\begin{center}
({\bf Fig.7.6}) {\small {\rm Compatible cycles with three Bubbles. }}
\end{center}
\end{center}
In the next section we choose an independent basis of compatible six Parke-Taylor factors from the 16 possibilities given in figure 7.6, in order to solve the non-trivial example \eqref{benzene}.

%%%%%%%%%%%%%%%%%%%%%%%%%%%%%%%%
\subsection{Explicit Example}
%%%%%%%%%%%%%%%%%%%%%%%%%%%%%%%%

As an explicit illustration of the full procedure we end this section with the computation of the integral we left open at the end of section \ref{kltexample}
\be\label{bencene}
I_{\rm sp}=\int d\mu_6 \,\frac{1}{(123456)}\,\frac{1}{(12)(34)(56)}\,\,\,,
\ee
with its associated graph
\begin{center}
\includegraphics[scale=0.4]{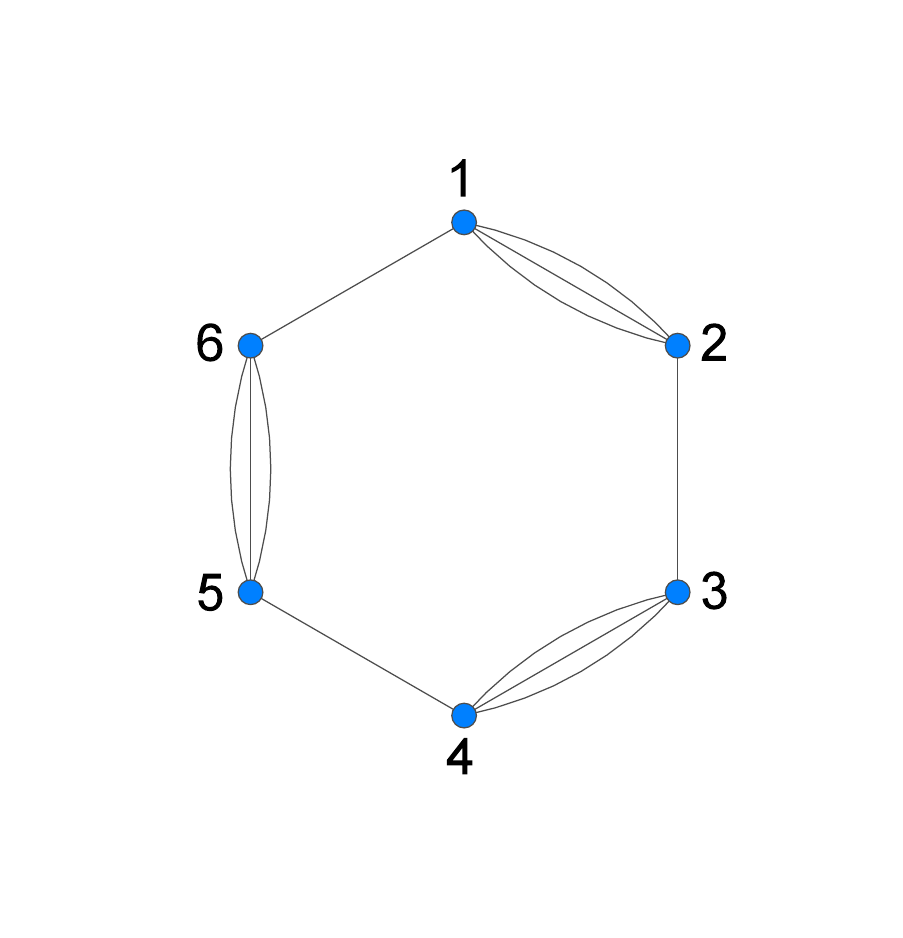}
\includegraphics[scale=0.4]{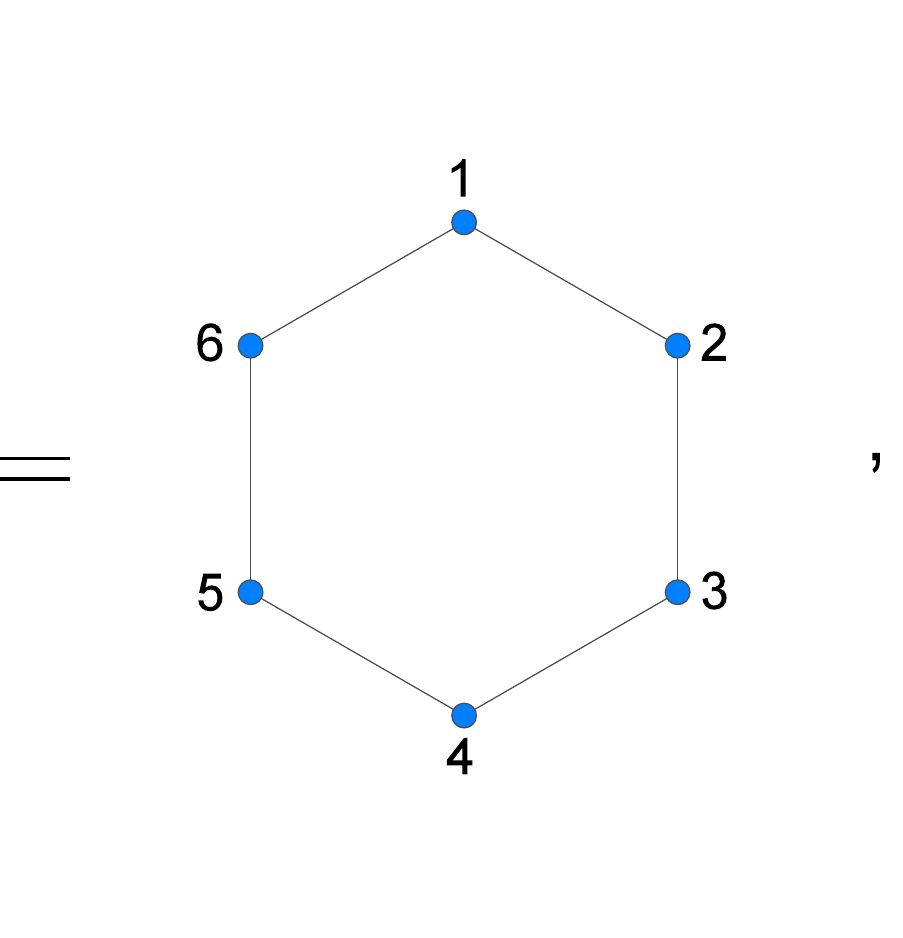}
\includegraphics[scale=0.4]{three_b-eps-converted-to.pdf}
\begin{center}
({\bf Fig.7.7}) {\small {\rm Graph representation of the integrand ${\cal I}(\s)={\cal I}_L(\s){\cal I}_R(\s)$, where ${\cal I}_L(\s)=\frac{1}{(123456)},\,\,{\cal I}_R(\s)=\frac{1}{(12)(34)(56)}$}}
\end{center}
\end{center}
On the r.h.s the canonical Parke-Taylor (${\cal I}_L(\s)$) does not have any incompatibility with the standard KLT approach, but, for the three bubbles (${\cal I}_R(\s)$) the standard KLT construction is not enough. The idea is to find a set of six cycles (six Parke-Taylor) such that the union  of each one of them with the three bubbles graph can be decomposed in two-disjoint Hamiltonian cycles (i.e. in two Parke-Taylor).

From the figure  7.6 we choose the following set
\be\label{rbase}
{\cal R} = \{(145326),(145236),(164523),(153246),(154623),(154236)\}.
\ee
In order to check that these six elements are compatibles with the three bubbles we, in figure 7.8, give an Hamiltonian decomposition 
\begin{center}
\includegraphics[scale=0.41]{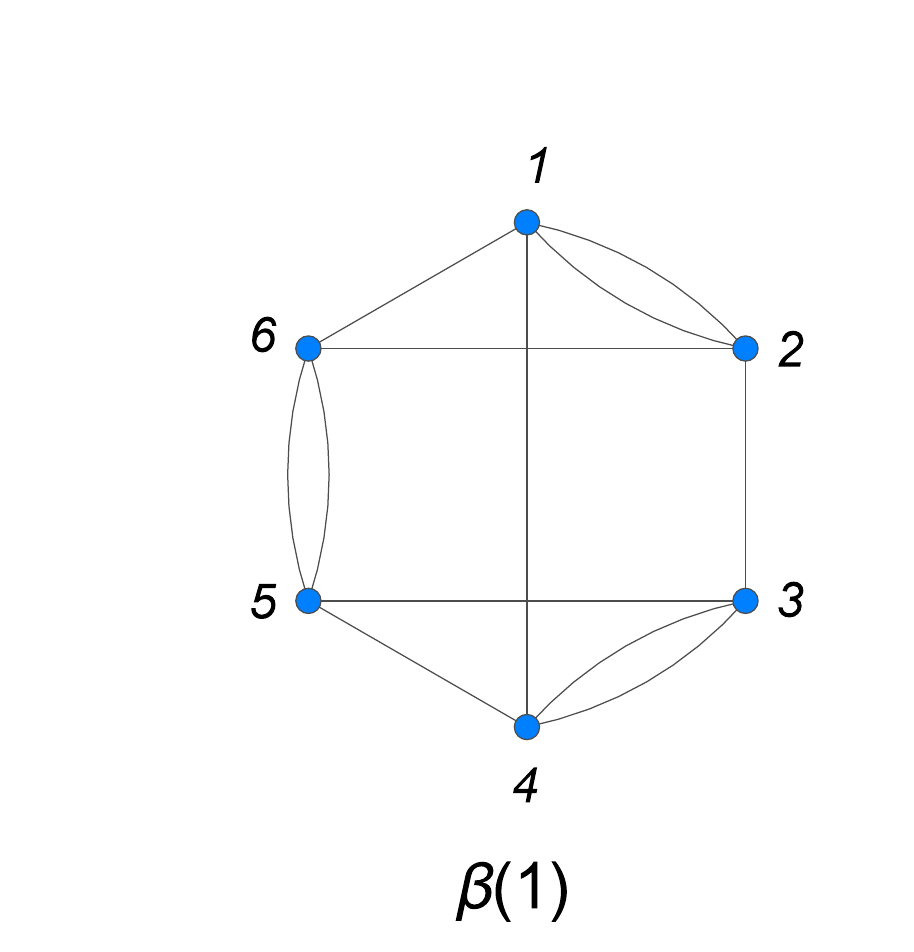}
\includegraphics[scale=0.41]{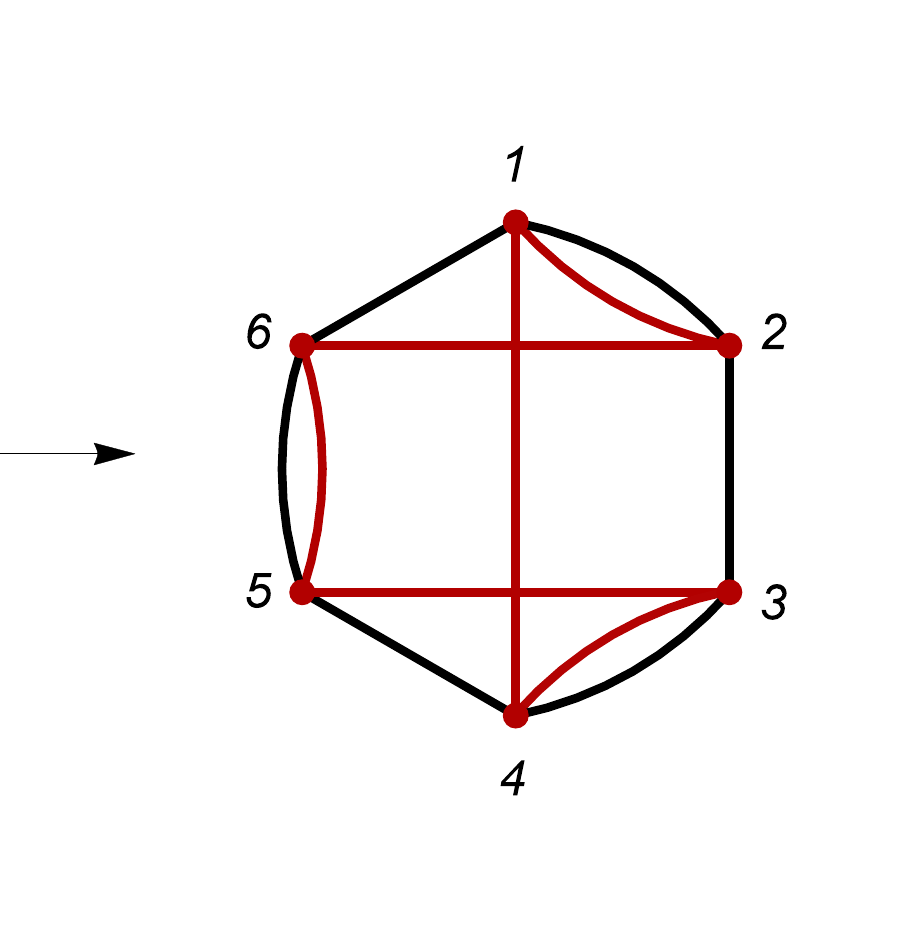}
\quad \includegraphics[scale=0.42]{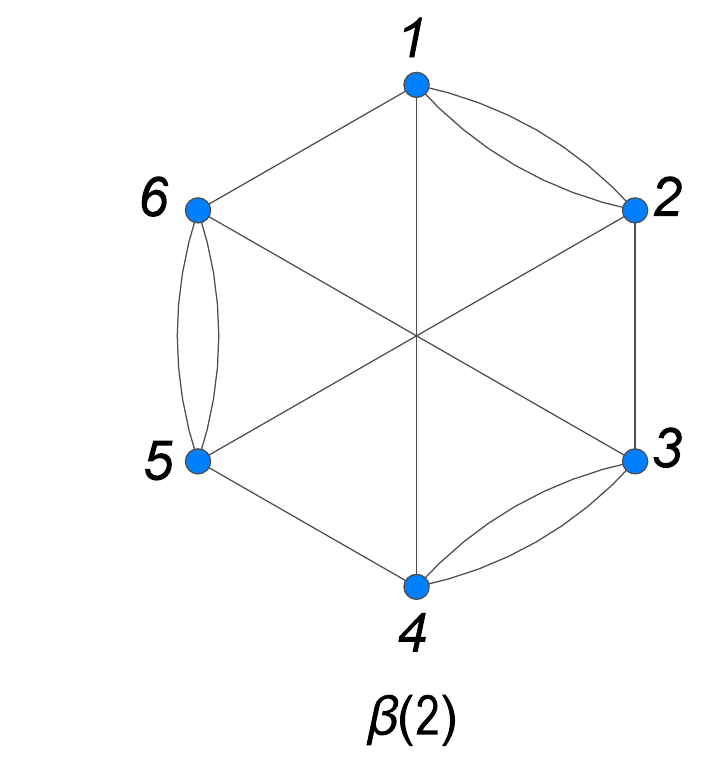}
\includegraphics[scale=0.41]{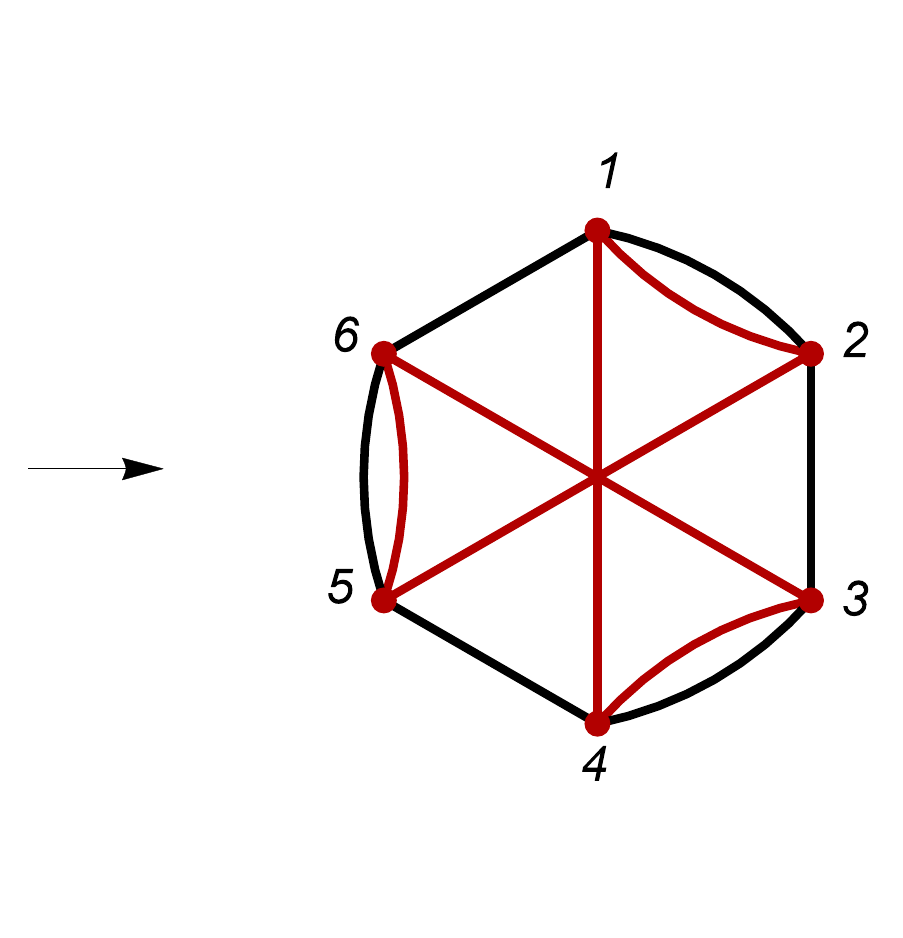}\\
\,\,\,\includegraphics[scale=0.41]{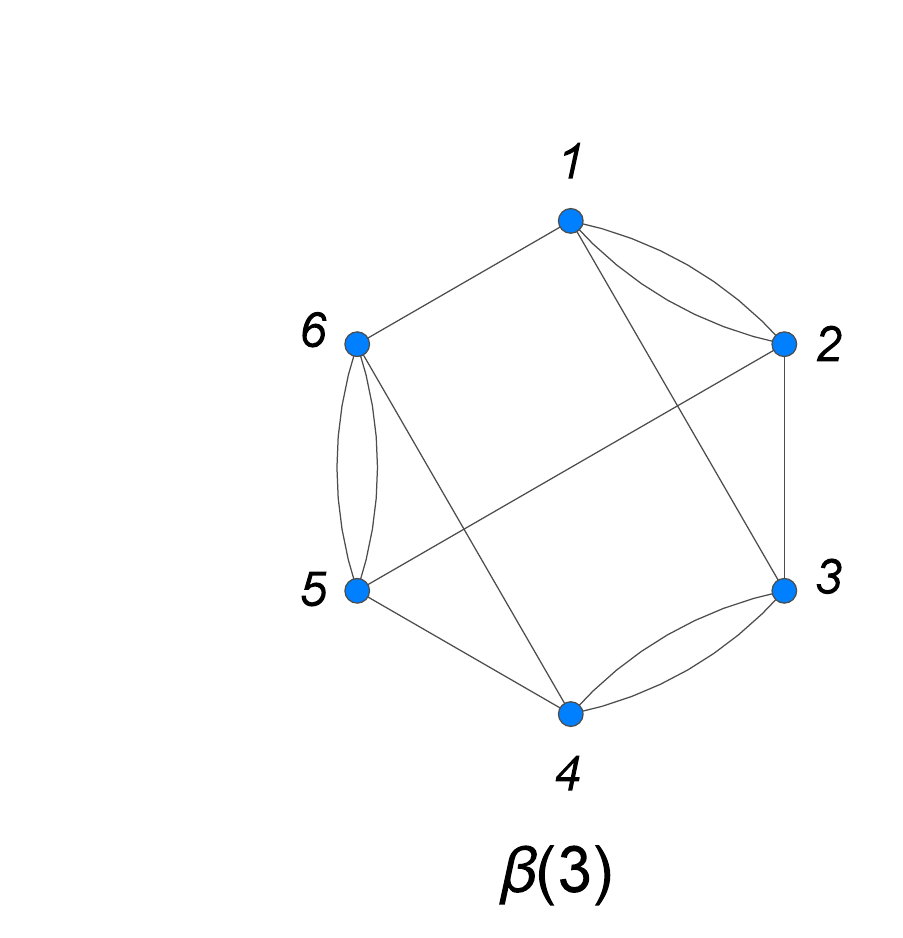}
\includegraphics[scale=0.41]{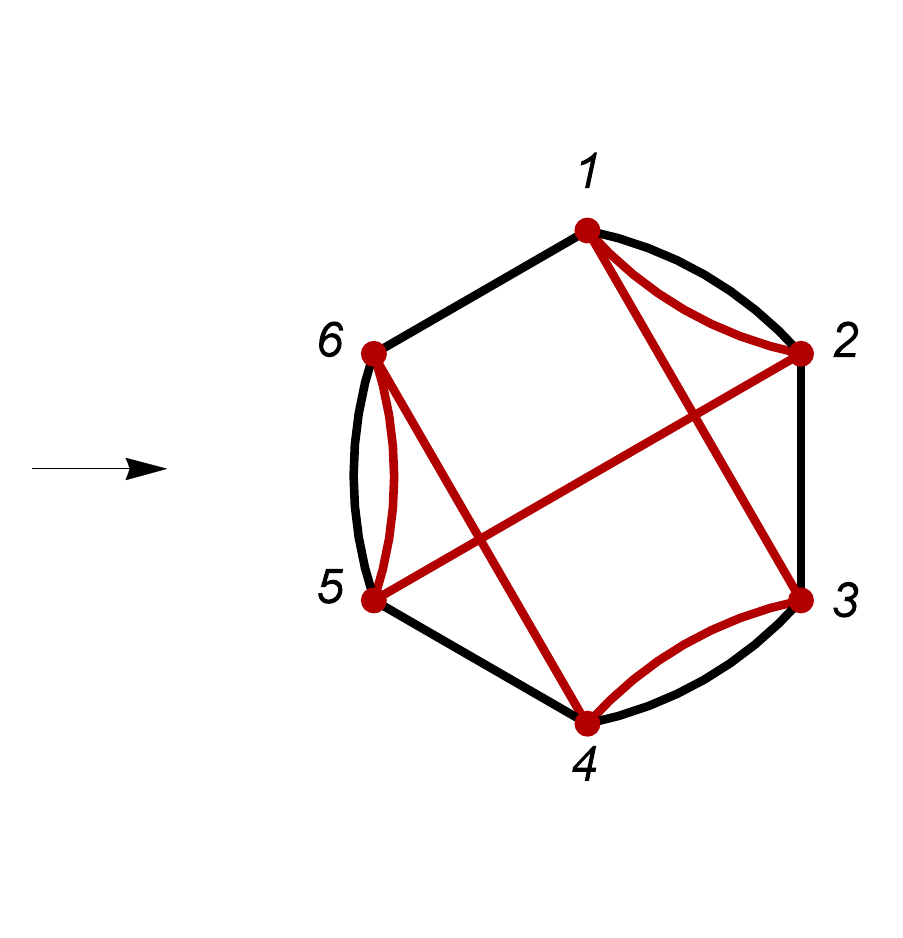}
\includegraphics[scale=0.41]{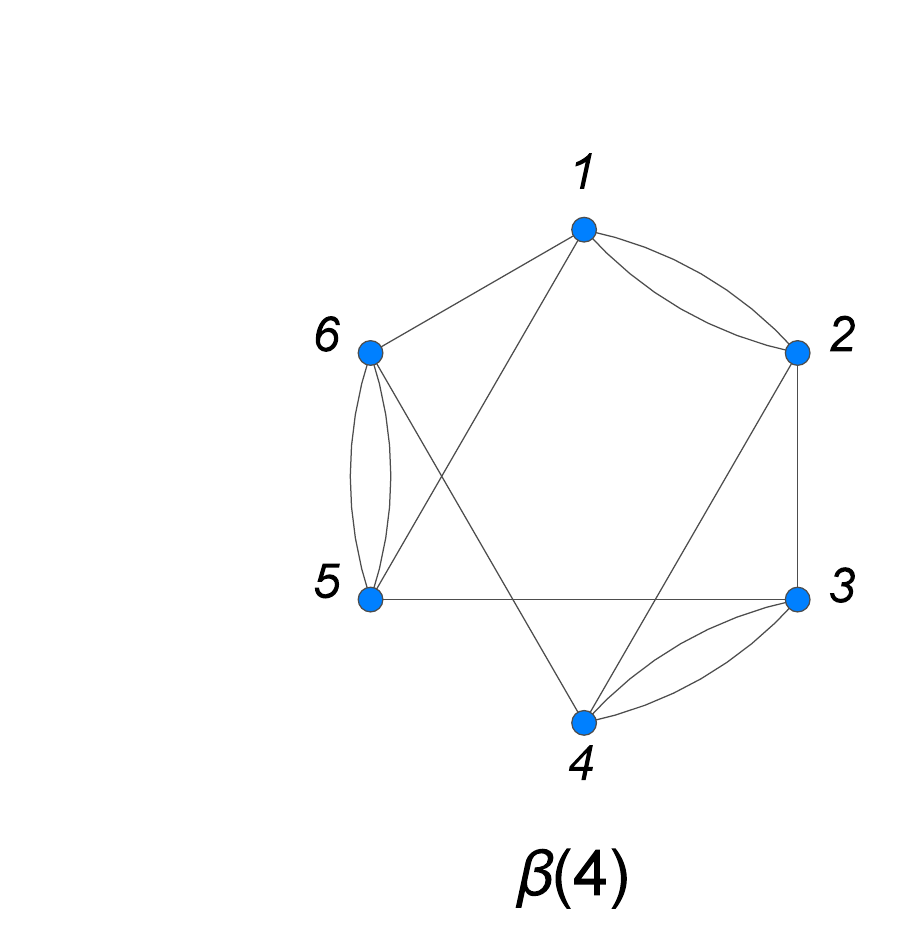}
\includegraphics[scale=0.41]{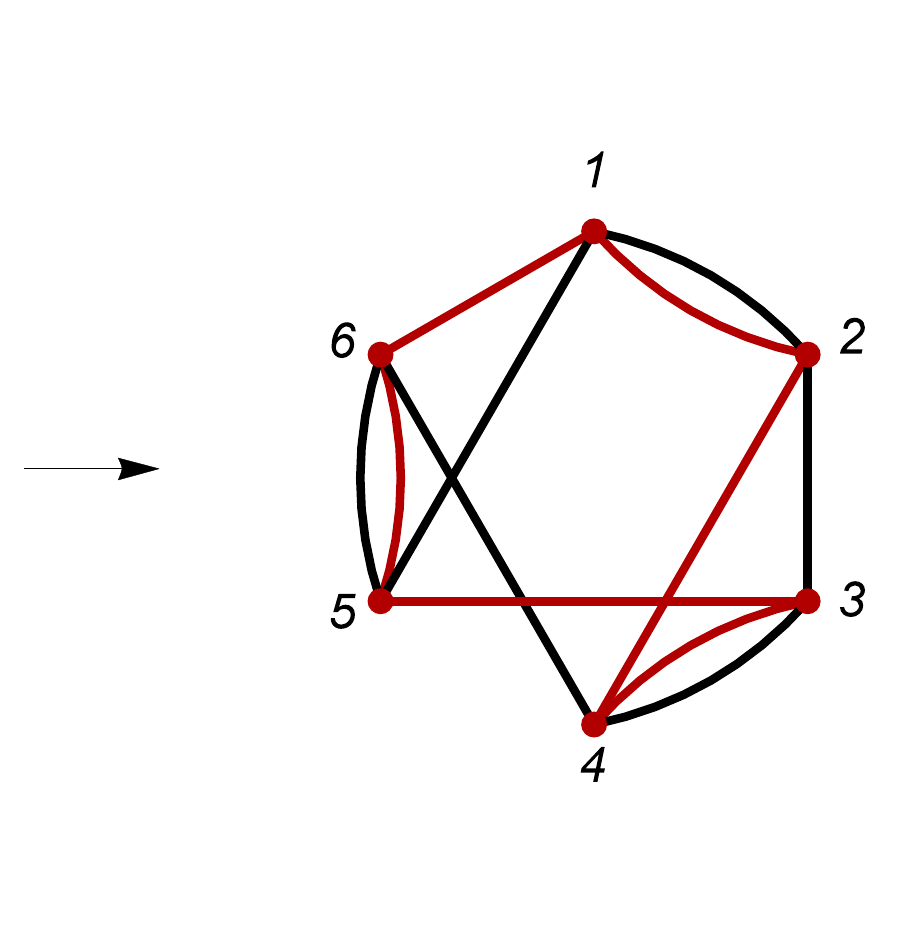}\\
\includegraphics[scale=0.41]{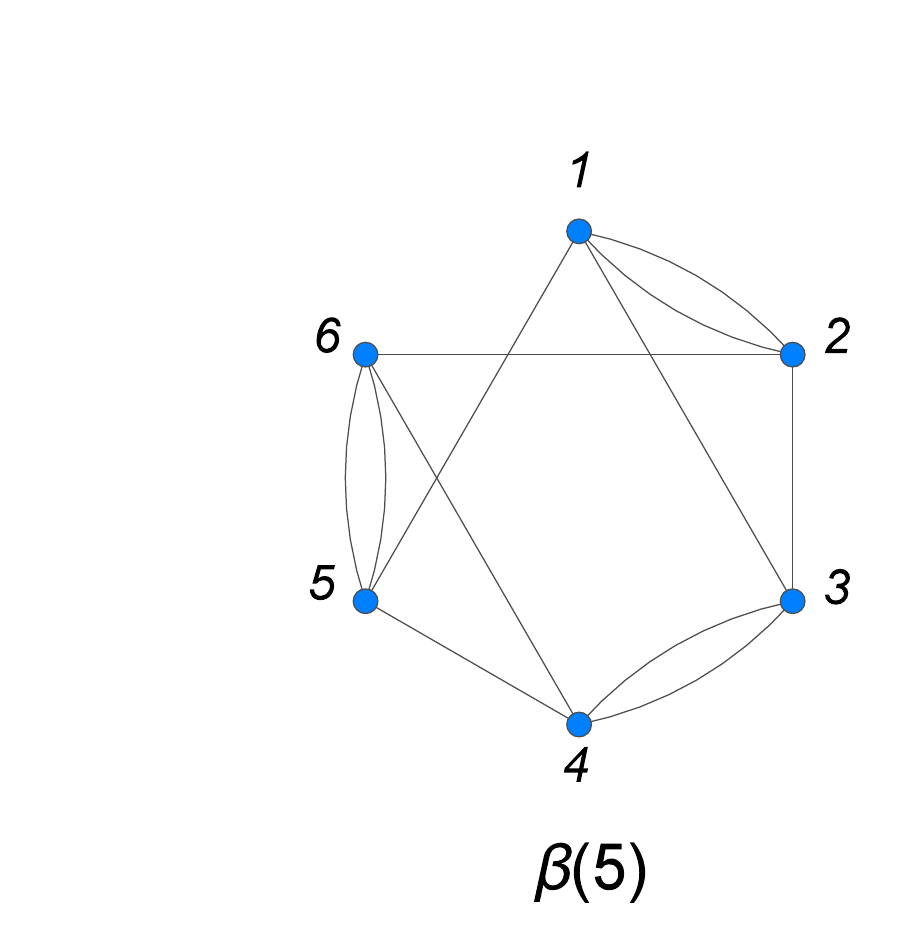}
\includegraphics[scale=0.41]{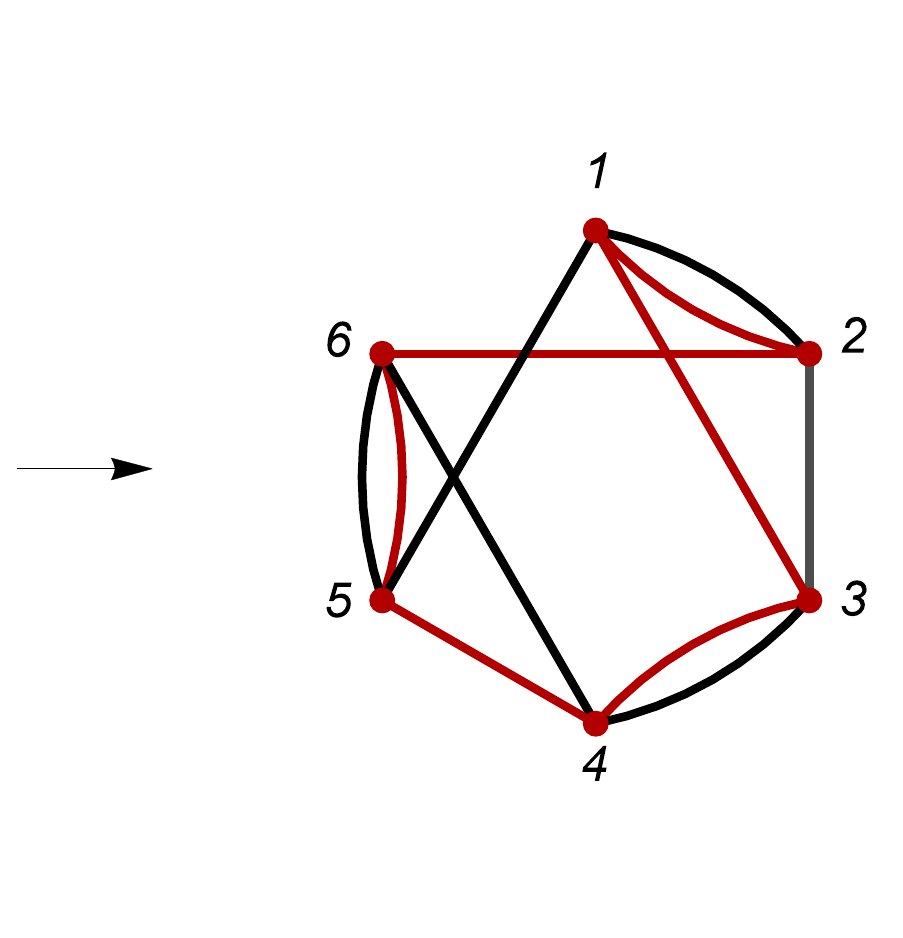}
\qquad \includegraphics[scale=0.44]{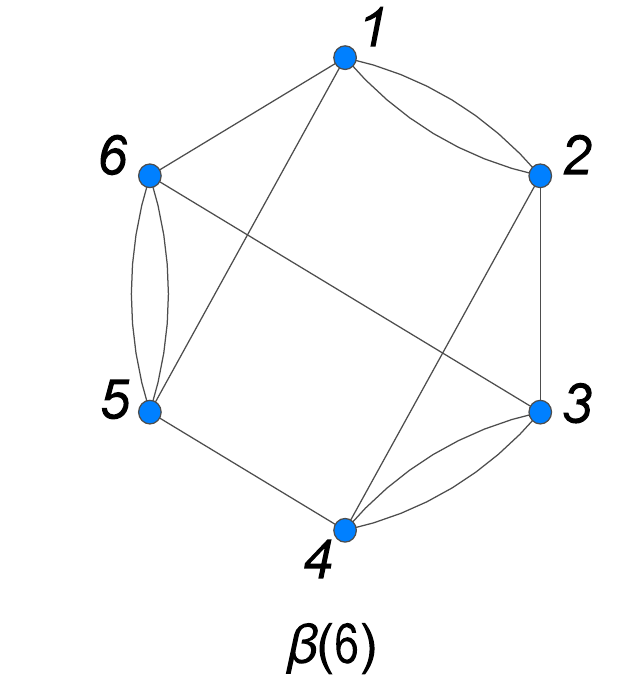}
\includegraphics[scale=0.41]{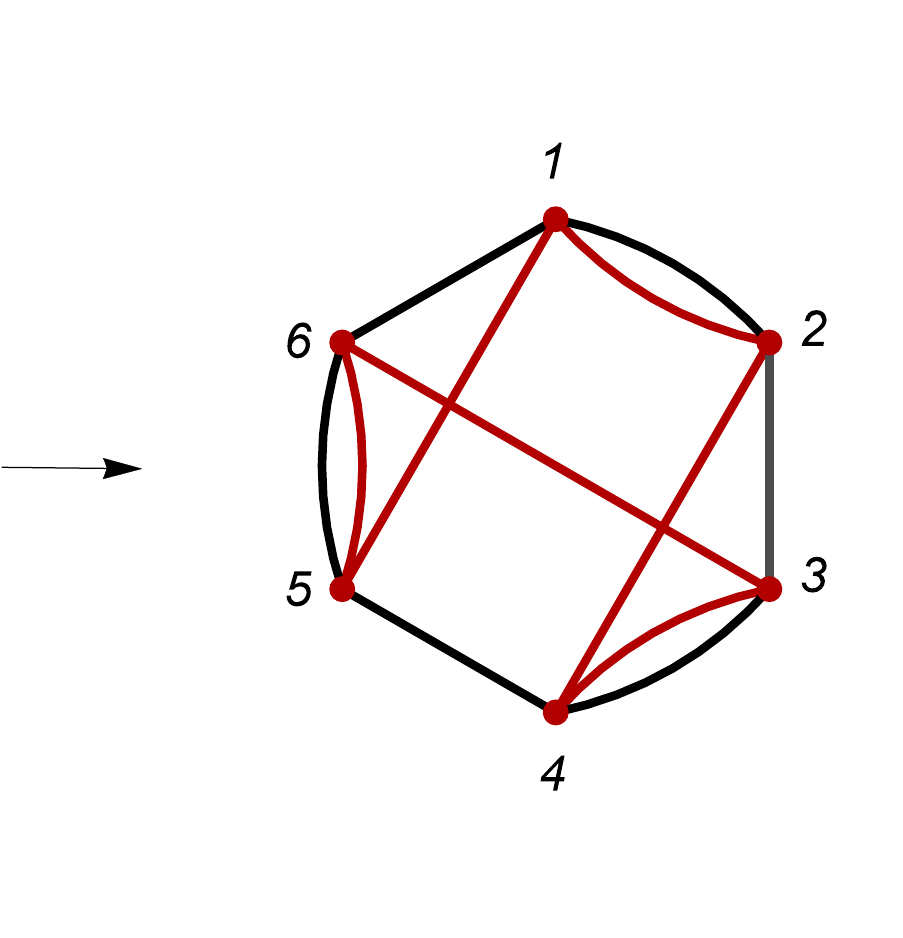}
\begin{center}
({\bf Fig.7.8}) {\small {\rm Hamiltonian decomposition  of $(12)(34)(56)(\b)$, $\b\in {\cal R}$. The red lines give one Hamiltonian cycle and the black lines form the other one (disjoint cycles). }}
\end{center}
\end{center}
Therefore, the integrand
$$
\frac{1}{(12)(34)(56)(\b)},
$$
with $\b\in {\cal R}$, can be written as the product of two six-point Parke-Taylor factors and hence their integrals are part of the building blocks.

So far, we have just found a ``right base'', i.e. a linearly independent\footnote{Remember, the set is linearly independent using the $m(\a|\b)$ inner product.} ${\cal R}$ set given in \eqref{rbase}. Now, it is necessary to find a ``left base'', i.e. a ${\cal L}$ set which must satisfy the following two conditions:\\
(1) the matrix defined by the inner product $m^{\cal L,R}(\a|\b)$,  with $\alpha \in {\cal L}$ and  $\beta \in {\cal R}$, is not singular;\\
(2) the $\a'$s elements, $\a\in {\cal L}$, must be compatibles  with the left integrand ${\cal I}_L(\s)$, i.e. they admit a Hamiltonian decomposition (see section \ref{HD}), such as it happened on the ``right side'', (see Fig.7.8).\\
In our example \eqref{bencene}, the left integrand, ${\cal I}_L(\s)=\frac{1}{(123456)}$, is the canonical Parke-Taylor, therefore the second condition is automatically satisfied (product of two Parke-Taylor). If we defines the left set (${\cal L}$) with the same elements of the right set, i.e.   ${\cal L}\equiv {\cal R}$, then  one can show that the matrix $m^{\cal L,R}(\a|\b)$ is not singular. However, note that this is a huge matrix since that the diagonal elements are Parke-Taylor squared \footnote{${\cal L}\equiv {\cal R}$ is a good left base  if the left integrand is also given by three bubbles, i.e. for a total integrand given by the graph
\begin{center}
\includegraphics[scale=0.35]{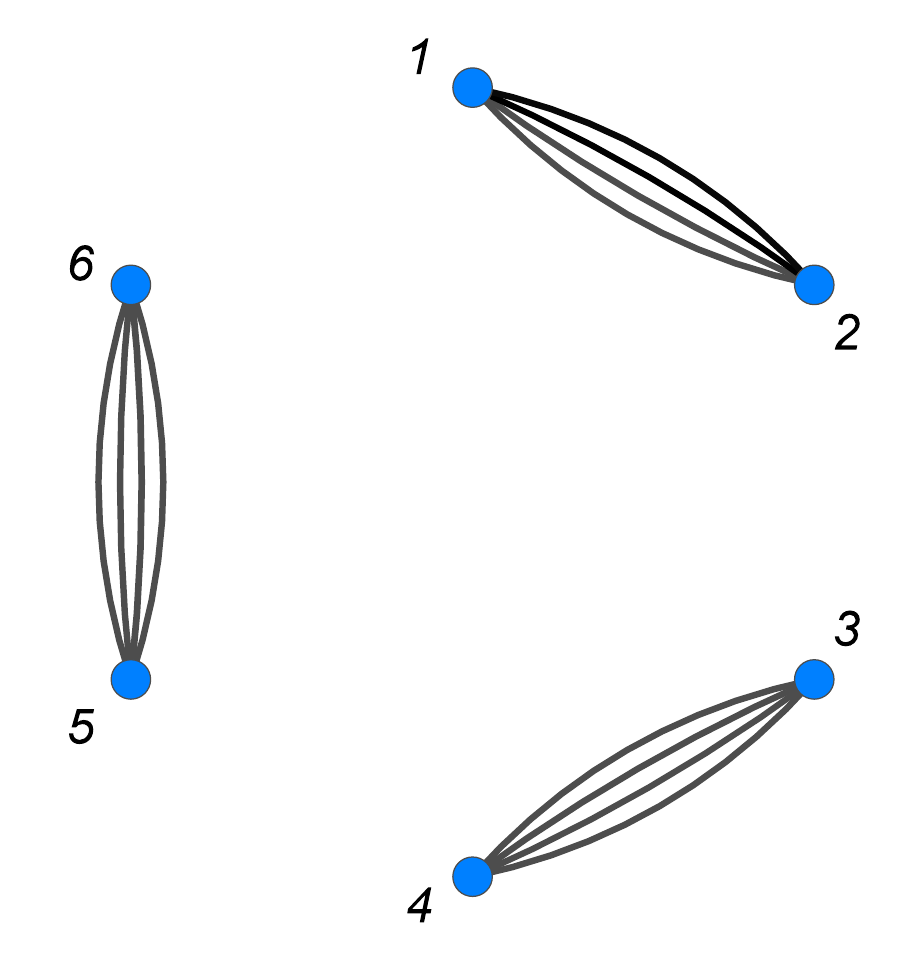}
\end{center}
}. We define the left set such that its intersection with the right base is disjoint $({\cal L}\cap{\cal R} = \emptyset)$, in order to avoid terms  with Parke-Taylor squared in the matrix. For example, choosing the linearly independent set
\be\label{lbase}
{\cal L} = \{(126534),(125634),(125643),(124365),(124653),(136524)\},
\ee
one can check that the matrix $m^{\cal L,R}_{\a|\b}$, $\a\in {\cal L}$ and $\b\in{\cal R}$,  given by
\begin{equation}\label{msix}
\left(
\begin{matrix}
\frac{B[14\,:\,345]}{s_{35}\,s_{26}}&0 & 0 & 0 & 0 & 0\\
0 & -(s_{14}\,s_{25}\,s_{36})^{-1}& 0 &  0& 0 & \frac{B[14\,:\,136]}{s_{25}\,s_{36}}\\
0 & 0 &  \frac{B[25\,:\,123]}{s_{13}\,s_{46}} & 0 & (s_{13}\,s_{46}\,s_{123})^{-1} & (s_{13}\,s_{25}\,s_{245})^{-1}\\
0 &0 &  0 & (s_{24}\,s_{15}\,s_{234})^{-1}& -\frac{B[46\,:\,24]}{s_{35}\,s_{135}} & 0\\
0 & 0 & (s_{13}\,s_{46}\,s_{123})^{-1} & 0 & \frac{B[123\,:\,246]}{s_{13}\,s_{46}} & 0\\
0 & 0 & 0 & \frac{B[36\,:\,234]}{s_{15}\,s_{24}} & 0 & (s_{24}\,s_{36}\,s_{245})^{-1}
\end{matrix}
\right)\nonumber
\end{equation}

with $s_{i_1 i_2\cdots i_j}\equiv (k_{i_1}+k_{i_2}+\cdots +k_{i_j})^2$ and 
\begin{equation}
B[i_1\cdots i_k\,:\,j_1\cdots j_m]\equiv\frac{1}{s_{i_1\cdots i_k}}+\frac{1}{s_{j_1\cdots j_m}}
\end{equation}
is not singular. Moreover, the vectors defined as
\begin{align}
({\cal UI}^{\cal L})_\a&=\int d\mu_6 \frac{1}{(123456)(\a)},\qquad\,\,\,\a\in {\cal L},\\
({\cal VI}^{\cal R})_\b&=\int d\mu_6 \frac{1}{(12)(34)(56)(\b)},\quad\b\in {\cal R},
\end{align}
are given by
%\fontsize{9pt}{8pt}\selectfont
\begin{align}\label{UIa}
({\cal UI}^{\cal L})_\a=&\left(
\begin{matrix}
\frac{B[56\,:\,345]}{s_{12}\,s_{34}} ,  & -(s_{12}\,s_{34}\,s_{56})^{-1} ,& \frac{B[34\,:\,123]}{s_{12}\,s_{56}}, & \frac{B[12\,:\,234]}{s_{34}\,s_{56}}, & (s_{12}\,s_{56}\,s_{123})^{-1} ,& 0\\
\end{matrix}
\right),\nonumber\\
({\cal V I}^{\cal R})_\b=&\left(
\begin{matrix}
\frac{B[56\,:\,345]}{s_{12}\,s_{34}},& -(s_{12}\,s_{34}\,s_{56})^{-1},
& \frac{B[34\,:\,456]}{s_{12}\,s_{56}}, &
\frac{B[12\,:\,234]}{s_{34}\,s_{56}}, & \frac{B[34\,:\,123]}{s_{12}\,s_{56}} ,&
\frac{B[12\,:\,234]}{s_{34}\,s_{56}}
\end{matrix}
\right).\nonumber
\end{align}
%\normalsize
Finally, we can  write the answer of the  \eqref{bencene} integral  as a rational function of the $m(\a|\b)$ building blocks
\begin{equation}\label{answerb}
I_{\rm sp}=\int d\mu_6 \frac{1}{(123456)}\frac{1}{(12)(34)(56)}\,\,=\,\sum_{\a\in{\cal L},\,\b \in{\cal R}}
({\cal U I}^{\cal L})_\a\,\, (m^{\cal L,R})^{-1}_{\a|\b}\,\, ({\cal V I}^{\cal R})_\b.
\end{equation}

We have been able to solve  a non-trivial integral using the algorithm described  in this paper. In addition, one can note, such as it was done in the Hamiltonian decomposition section, that it is possible to split the geometry of the 4-regular graph (initial integrand) and to  study them separately as two 2-regular graphs.

%%%%%%%%%%%%%%%%%%%%%%%%%%%%%%%%%%%%%
\section{Discussions}\label{disc} 
%%%%%%%%%%%%%%%%%%%%%%%%%%%%%%%%%%%%%%

In this paper we provided an algorithm for the computation of contour integrals of the form
\be
\int d\mu_n F(\sigma )
\ee
where $F(\sigma )$ is a general rational function with the $SL(2,\mathbb{C})$ transformations of two Parke-Taylor factors and therefore can be written, without loss of generality, as
\be
F(\sigma ) =\frac{1}{(12\cdots n)(\gamma(1)\gamma(2)\cdots \gamma(n))}\prod_{i=1}^m r^{(i)}
\ee
where $r^{(i)}$ are cross ratios of the positions of four punctures and $m$ is an arbitrary positive integer.
These integrals appear in many physical applications and hence having an algorithm for their computation is important for the study of a variety of theories.
The contour and measure are defined using the critical points of a Morse function on ${\cal M}_{0,n}$ \cite{morse,Gross:1987ar,Ohmori:2015}
\be\label{morse}
\phi(\sigma,\bar{\sigma} )=\frac{1}{2}\sum_{1\leq a<b \leq n}s_{ab}\,\ln |\sigma_{a}-\sigma_b|^2.
\ee
This function is a universal part of the ``action'' that controls correlation functions that compute scattering amplitudes in string theory. One of the most pressing issues is to find a direct connection with string theory computations. It is well known that in the Gross-Mende limit of string amplitudes (which is a certain tensionless limit and thus opposite from field theory) correlations functions also localize to the critical points of \eqref{morse} \cite{Gross:1987ar}. Constructions based on ambitwistor space have been very successful but still do not provide a direct link with a limit of string theory \cite{Mason:2013sva,Adamo:2013tsa,Geyer:2014fka,Ohmori:2015}. The approach of Berkovits that uses an infinite tension limit of string theory in the pure spinor formalism is clearly connected to string theory but is not directly connected to the CHY formulas for gravity and Yang-Mills \cite{Berkovits:2013xba,Gomez:2013wza,Adamo:2015}. It is reasonable to hope that a direct connection of CHY formulas to a limit of string theory may teach us new lessons on how string theory is connected to field theory and how the tools developed in this work can extend to applications in string theory.\\
A simple byproduct of this work is its application to produce a variety of field theoretic relations. One of the simplest examples is a relation that expresses double-trace amplitudes of gluons in the Einstein-Yang-Mills theory discussed in \cite{Cachazo:2014nsa,Cachazo:2014xea} to single trace amplitudes and scalar amplitudes \cite{Cachazo:2013hca,Cachazo:2013iea}. To see the relation consider the CHY formula for a double-trace amplitude with gluons $1,2,\ldots ,m$ in the first trace and gluons $m+1,m+2,\ldots , n$ in the second,
\be
A^{(2)}(1,2\ldots, m:m+1,m+2,\ldots ,n) = \int d\mu_n \frac{s_{12\cdots m}}{(12\cdots m)\,(m+1\,\,m+2\,\,\cdots \,\,n)}{\rm Pf}'\Psi\nonumber\,.
\ee
The graph of the integrand is simply the product of two polygons and hence one can find a basis of $(n-3)!$ permutations (Parke-Taylor factors) that are compatible with that graph. Recall that by being compatible we mean that the union of the graph associated to a permutation and the two polygons admits an edge-disjoint Hamiltonian decomposition. In other words, it can be expressed as the union of two Parke-Taylor factors. Denoting such as a basis as a left basis ${\cal L}$ and choosing any other convenient basis as a right basis ${\cal R}$ one has
\be\label{doubleR}
A^{(2)}(1,2\ldots ,m:m+1,m+2,\ldots, n) = s_{12\cdots m}\sum_{\alpha\in {\cal L},\beta\in {\cal R}} m(\alpha'|\alpha'') \left(m^{{\cal L}|{\cal R}}\right)^{-1}_{\a|\b}A^{(1)}(\beta),
\ee
with
\be
m(\alpha'|\alpha'') \equiv \int d\mu_n \frac{1}{(\alpha )(12\cdots m)\,(m+1\,\,m+2\,\,\cdots\,\, n)} = \int d\mu_n \frac{1}{(\alpha')(\alpha'')}
\ee
and where $\alpha'$ and $\alpha''$ are the two edge-disjoint Hamilton cycles that decompose
$$(\alpha )\,(12\cdots m)\,(m+1\,\,m+2\,\,\cdots \,\,n).$$
It would be interesting to explicitly construct the bases ${\cal L}$ and ${\cal R}$ such that the relation \eqref{doubleR} takes its simplest possible form. Also interesting is to find a possible string theoretic origin for this relation. 
It is clear that there are plenty of relations such as \eqref{doubleR} which connect very different kinds of objects. In fact, these new kind of relations greatly extend the large class already found using KLT in \cite{Cachazo:2014xea} which linked theories such as the special Galileon with the $U(N)$ nonlinear $\sigma$ model.
The algorithm presented in section \ref{mainalgo} is completely general but it is not the most efficient one in particular cases. It would be interesting to select certain families of integrals that appear in particular theories and refine the algorithm to make it the most efficient possible. One way to improve the efficiency is by selecting basis of permutations ${\cal L}$ and ${\cal R}$ such that the matrix $m^{{\cal L}|{\cal R}}$ is as sparse as possible.
Finally, finding direct mathematical application of contour integrals over ${\cal M}_{0,n}$ could lead to yet another link between the elegant math of Riemann surfaces and that of graph theory. One closely related link was established by using Strebel differentials \cite{strebel} which provide a connection between the decorated moduli space, ${\cal M}_{0,n}\times (\mathbb{R}^+)^n$ and ribbon graphs \cite{ribbon}. It would be fascinating to find connections with this or other constructions.

\acknowledgments
FC would like to thank M. Mosca and A. Postnikov for useful discussion. Both authors thank  S. He, C. Mafra, O. Schlotterer and E. Yuan for useful discussions. This work is supported by the Perimeter Institute for Theoretical Physics. HG would like to thank the hospitality of Perimeter Institute where most of this work was done. Research at Perimeter Institute is supported by the Government of Canada through Industry Canada and by the Province of Ontario through the Ministry of Research \& Innovation. HG thanks to C. Cardona for his carefully reading and comments. HG is supported by FAPESP grant 2011/13013-8.

%%%%%%% Appendix %%%%%%%

\appendix

%%%%%%%%%%%%%%%%%%%%%%%%%%%%%%%%%%%%%%%%%%%%%
\section{Algorithm and Petersen's Theorem}
%%%%%%%%%%%%%%%%%%%%%%%%%%%%%%%%%%%%%%%%%%%%%

Before stating the Petersen's Theorem it is useful to give  an algorithm to find a decomposition of any 4-regular graph in  two edge-disjoint 2-factors, in order to elucidate the Petersen's Theorem.

%%%%%%%%%%%%%%%%%%%%%%%%%%%
\subsection{Algorithm}
%%%%%%%%%%%%%%%%%%%%%%%%%%

The main aim of this algorithm is to show in a simple way how to obtain two edge-disjoint 2-regular spanning\footnote{A Spanning graph is a subgraph such that its vertex set is the same as the original graph.} graphs from one 4-regular graph, i.e.
$$
G_F = G^L_F\,\cup\, G^R_F
$$
where the ``$\cup$" symbol means the union of the edge sets \footnote{The vertex sets are the same for the two 2-regular graphs and the 4-regular graph, i.e. $V_{F}=V^L_{F}=V^R_{F }$.}.

The algorithm is divided in four fundamental steps. In order to better understand it we give an example in each step using the graph drew in the (Fig.4.1).

\begin{itemize}

\item {\bf Eulerian Path (EP)}

The first step of the algorithm is to find an Eulerian path. An Eulerian path is a trail in a graph which visits every edge exactly once. Since we are working with $2k$-factor undirected graphs, $k\in \mathbb{N_+}$, then the existence of an Eulerian path is  guaranteed and, in addition, every Eulerian path is closed, i.e. it is a cycle \cite{graph1,graph2}.

An Eulerian orientation of an undirected graph $G_F$ is an assignment of a direction to each edge of $G_F$ such that, at each vertex $a_i$, the indegree of $a_i$ (number of incoming edges) equals the outdegree of $a_i$ (number of outgoing edges).

There are several algorithms to get an Eulerian path but we do not discuss them in this paper \cite{graph1,graph2}.
In order to unify the notation we denote the cycle defined by the trail $\{[a_1;a_2],[a_1;a_2],...,[a_{n-1};a_n],[a_n;a_1]\}$ as
$$
(a_1 a_2... a_n)\equiv \{[a_1;a_2],[a_2;a_3],...,[a_{n-1};a_n],[a_n;a_1]\},
$$
where the orientation of the path is given by the ordering of the edges in the set.

With a view to understand better the Eulerian path concept we have found two of them on the example given in the (Fig.4.1)
%%%%%%%%%%%%%%%%%%
\begin{center}
\includegraphics[scale=0.6]{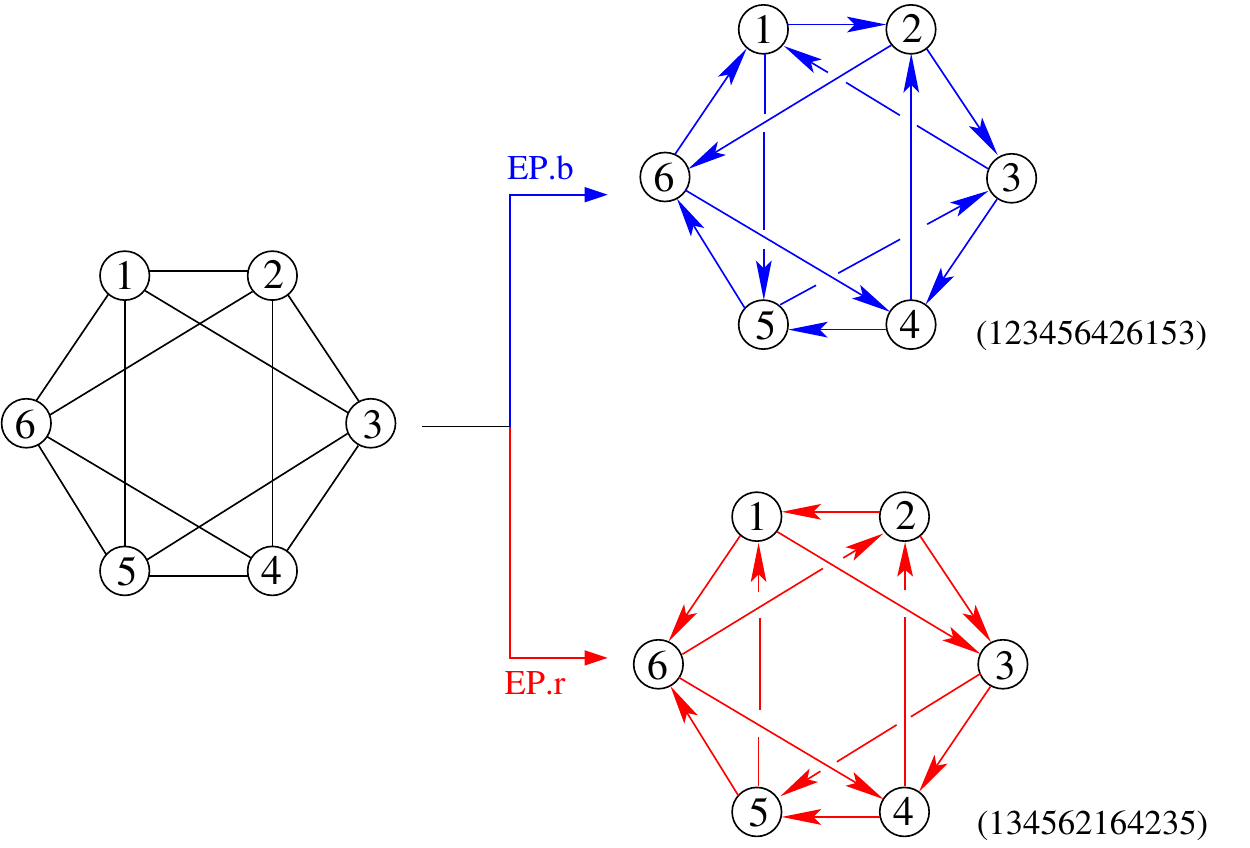},
\begin{center}
({\bf Fig.A.1}) {\small {\rm Eulerian paths.}}
\end{center}
\end{center}
%%%%%%%%%%%%%%%%%%%
Note that on each vertex there are two incoming and two outgoing edges, such as it was said above.

\item {\bf Bipartite Graph}

After finding an Eulerian path, the second step is to construct a bipartite graph from this EP. Thus, we must define what is a bipartite graph.

{\bf Bipartite Graph}\\
A bipartite graph  $G$  is a graph whose vertex-set $V$ can be partitioned into two subsets $U$ and $W$, such that each edge of $G$ has one endpoint in $U$ and  one endpoint in $W$. The pair $U,W$ is called a (vertex) bipartition of $G$, and $U$ and $W$ are called the bipartition subsets \cite{graph1,graph2}.

Using the EP found in the first step one can associate a bipartite graph $G_F^\prime$ to the original graph $G_F$. The idea is simple, since on each vertex there are two incoming and two outgoing edges then one can split each vertex in two, one of them contains the incoming edges (we call them the blue vertices) and the other one contains the outgoing edges (we call them the yellow vertices). This procedure generates a bipartite graph $G_F^\prime$, where the bipartition is given by the blue and yellow vertices, i.e.
$$
U^\prime_F=\{{\rm Blue\,\, Vertices}\},\qquad W^\prime_F=\{{\rm Yellow\,\, Vertices}\},
$$
and therefore each edge of $G^\prime_F$ has one endpoint in $U^\prime_F$ and  one endpoint in $W^\prime_F$.

Using this procedure on the Eulerian paths given in the Figure (Fig.A.1) one obtains the following
\begin{center}
\includegraphics[scale=0.6]{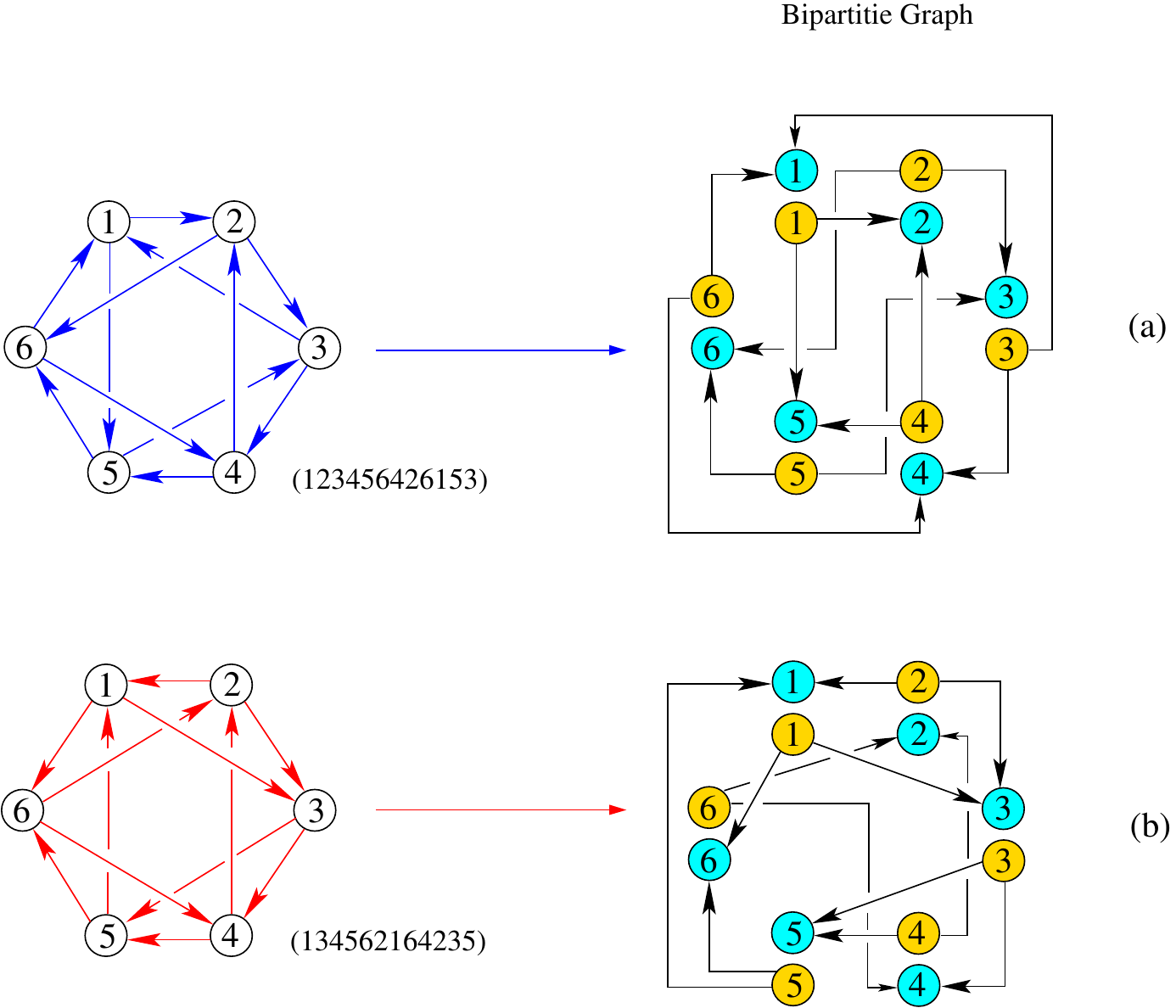},
\begin{center}
({\bf Fig.A.2}) {\small {\rm Bipartite graphs}}
\end{center}
\end{center}
where we have kept the orientation of the edges on the bipartite graph, but this is not necessary. Note that the two bipartite graphs in  (Fig.A.2) are 2-regular.

\item {\bf 1-Factor (Perfect Matching)}

The third step is to identify two edge-disjoint (disjoint sets of edges) 1-Factors  in the bipartite graph $G_F^\prime$. By definition, a 1-Factor in a graph $G$ is a 1-regular spanning subgraph, i.e. there is a set of edges without common vertices and connects all vertices of the graph  $G$ (perfect matching). Moreover, since we have found a bipartite 2-regular graph in the step 2, then there is a theorem which implies the existence of a perfect matching in it \cite{graph1,graph2}:

{\bf Theorem A.1}\\
Every $r$-regular bipartite graph $G$ with $r>0$ is 1-factorable,  i.e. there are $r$ edges-independent 1-Factors\footnote{This theorem is not proved in this paper.}.

This powerful theorem means that the bipartite 2-regular graphs $G_F^\prime$ obtained in the step 2 have two edge-disjoint 1-Factors.  These 1-Factors can be obtained easily. First of all, one should detect  all cycles in the bipartite 2-regular graph $G_F^\prime$ and after that, one should remove alternate edges in each cycles. This procedure generates one 1-factor and the other  one is given by the edges removed.

In the bipartite graph given by the (Fig.A.2)(a) we have found two edge-disjoint 1-Factors
\begin{center}
\includegraphics[scale=0.6]{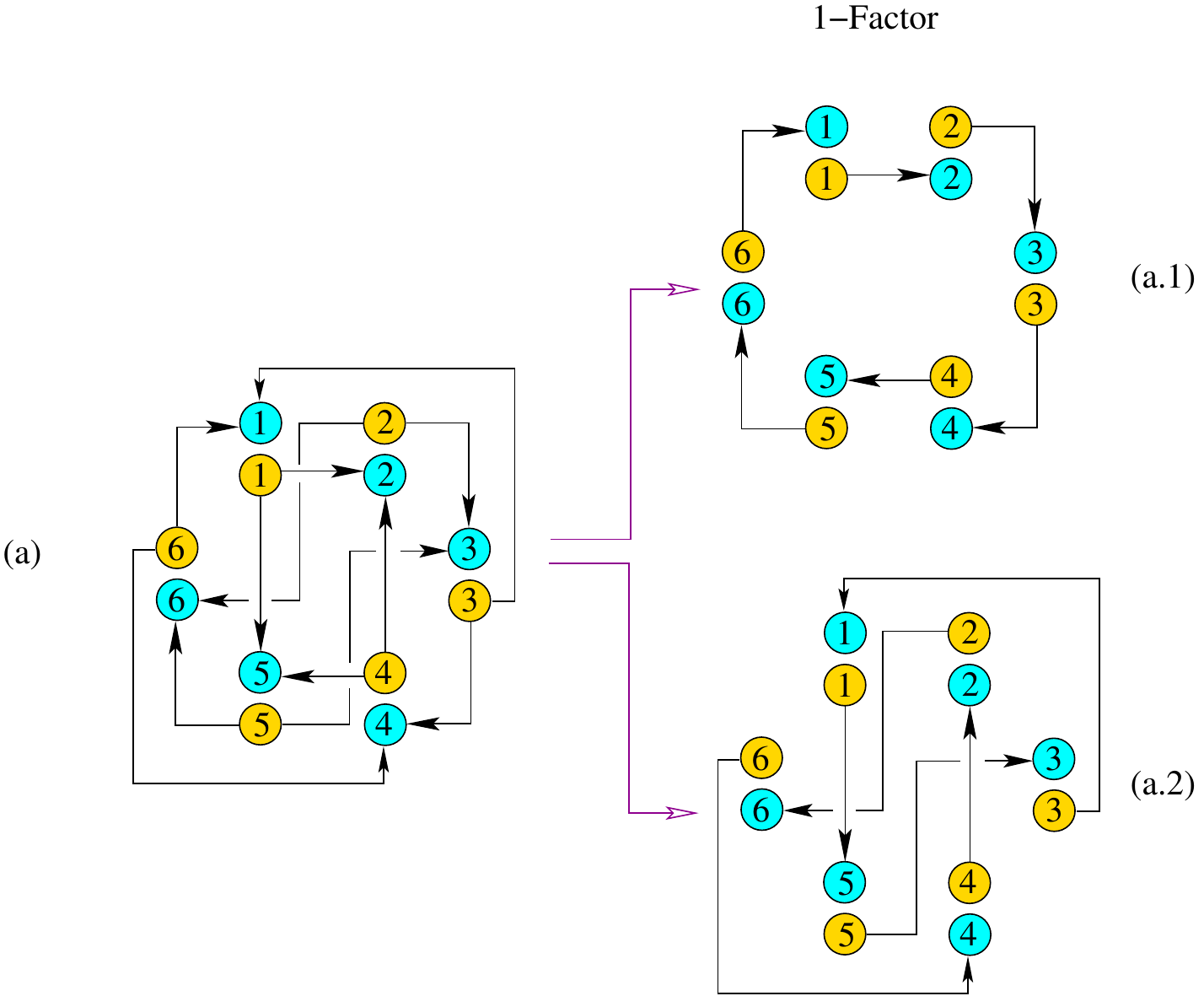},
\begin{center}
({\bf Fig.A.3a}) {\small {\rm Two edge-disjoint 1-Factor graphs}}
\end{center}
\end{center}
and in the bipartite graph (Fig.A.2)(b) we choose the following two edge-disjoint 1-Factors
\begin{center}
\includegraphics[scale=0.65]{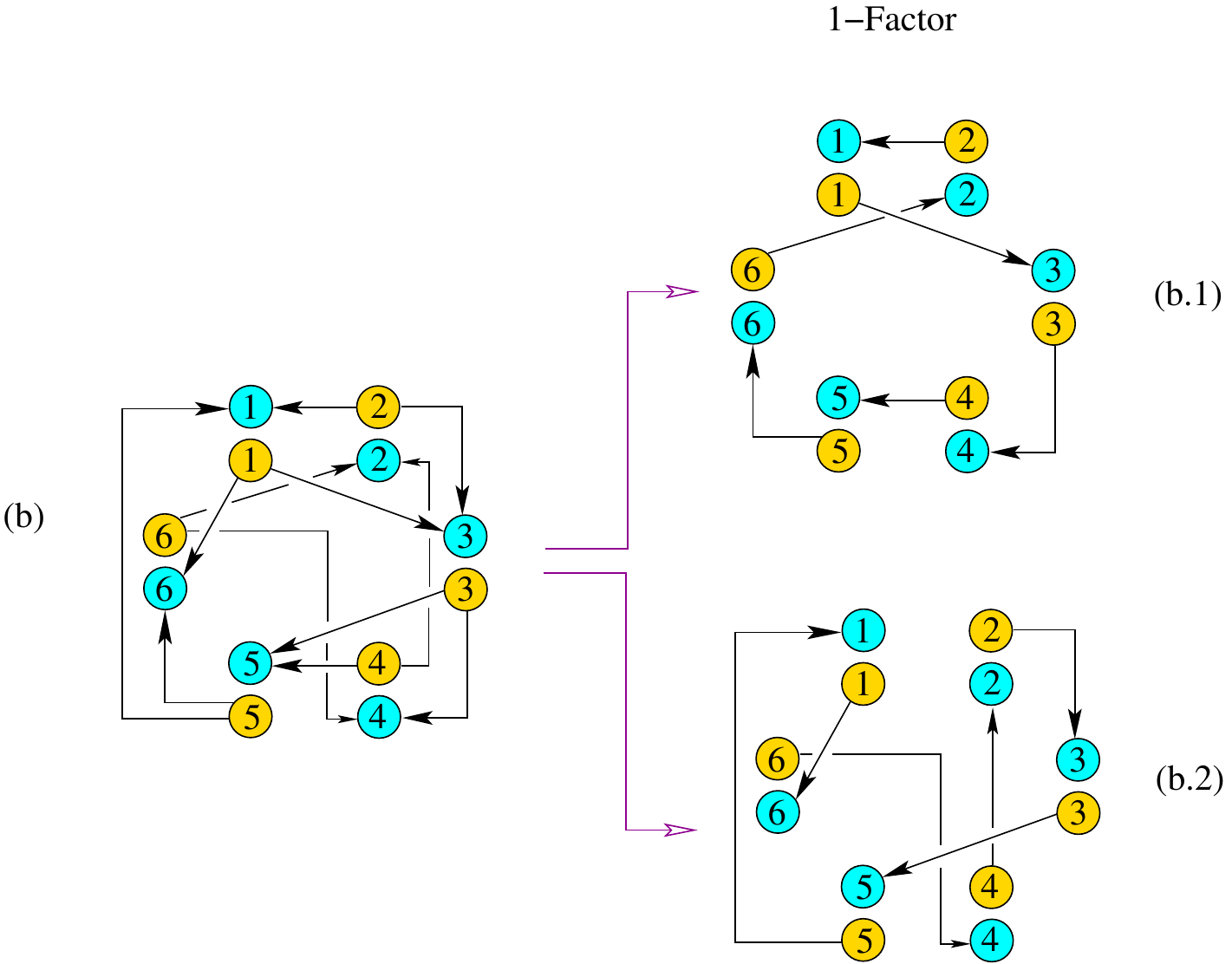}.
\begin{center}
({\bf Fig.A.3b}) {\small {\rm Two edge-disjoint 1-Factor graphs.}}
\end{center}
\end{center}

\item {\bf Identifying Vertices}

The last step is very simple. The idea is just to identify the blue and yellow vertex with the same label in each 1-Factor graph, i.e.
\begin{center}
\includegraphics[scale=1]{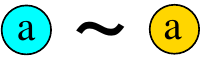}
\begin{center}
({\bf Fig.A.4}) {\small {\rm Identifying Vertices.}}
\end{center}
\end{center}
and so one obtains the decomposition of the 4-regular graph $G_F$ in two edge-disjoint 2-regular graphs ($G^L_F,G^R_F$).\\
In the figures (Fig.A.5a) and (Fig.A.5b) we show graphically the whole algorithm using the two Eulerian paths found in the example of the step 1.
\begin{center}
\includegraphics[scale=0.7]{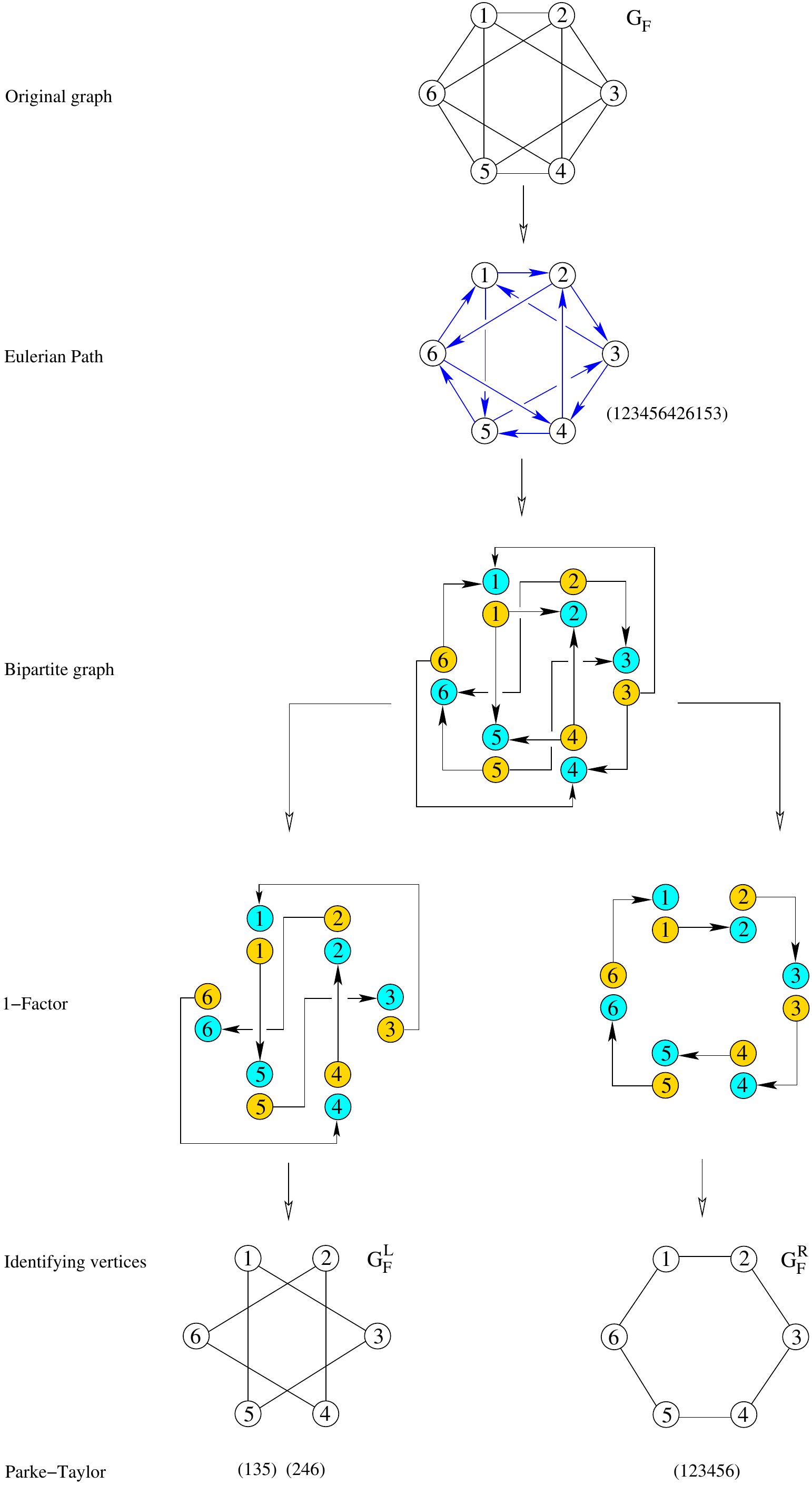}
\begin{center}
({\bf Fig.A.5a}) {\small {\rm }}
\end{center}
\end{center}
\begin{center}
\includegraphics[scale=0.7]{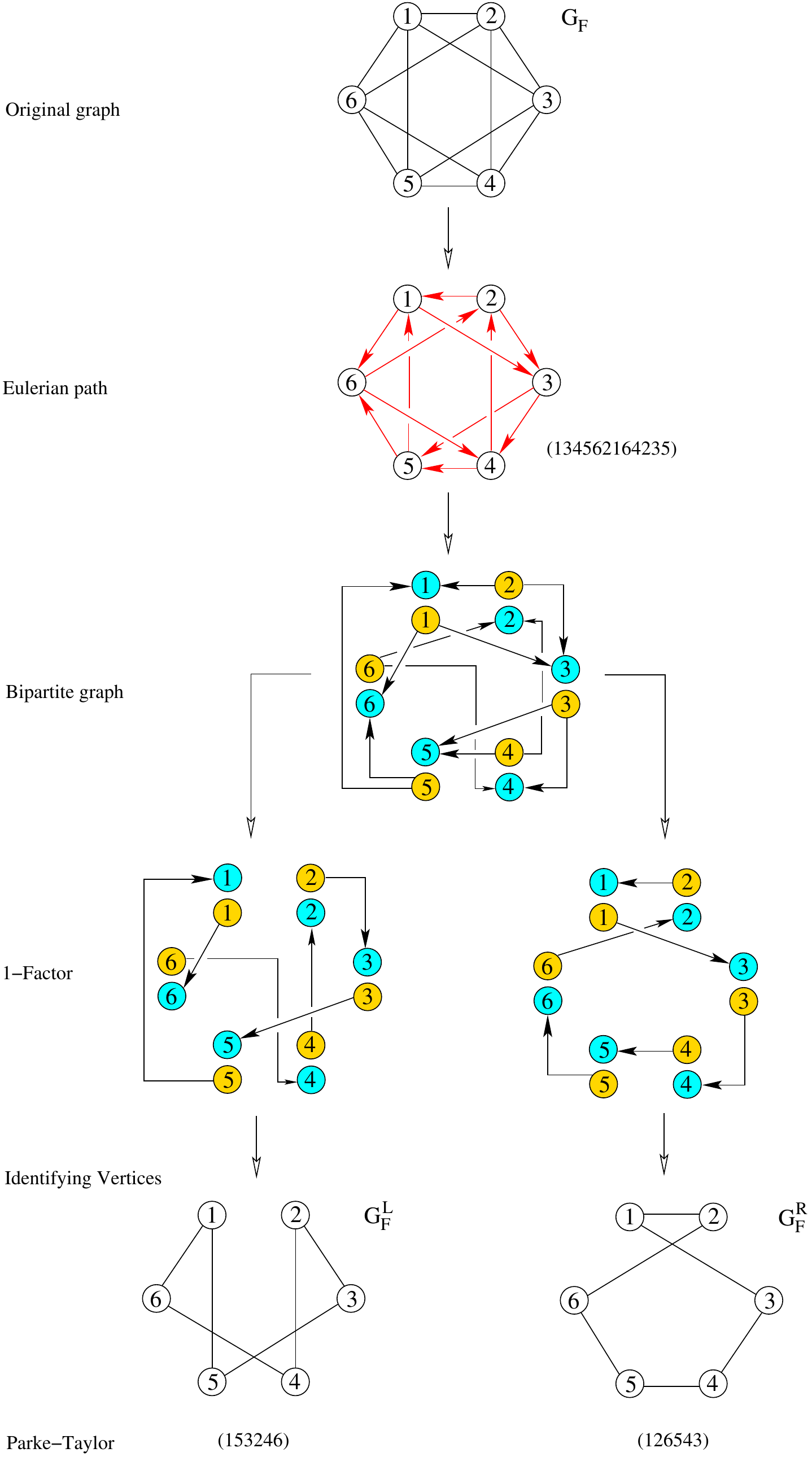}
\begin{center}
({\bf Fig.A.5b}) {\small {\rm }}
\end{center}
\end{center}
Note that we have obtained two different decompositions. The first one (Fig.A.5a) is one Parke-Taylor factor and two triangles
$$
F_d =\frac{1}{(123)(345)(561)(246)}= F_d^L\,F_d^R,\qquad F_d^L=\frac{1}{(135)(246)},\,\,\, F_d^R=\frac{1}{(123456)},
$$
and the second one (FigA.5b) is two Parke-Taylor factors
$$
F_d =\frac{1}{(123)(345)(561)(246)}= F_d^L\,F_d^R,\qquad F_d^L=\frac{1}{(153246)},\,\,\, F_d^R=\frac{1}{(126543)}.
$$
\end{itemize}
\subsection{Petersen's Theorem}

Now, we are able to state the Petersen's Theorem \cite{graph1,graph2}, which was sketching previously with the algorithm.

{\bf Petersen's Theorem}

Every $2k$-regular graph $G$, with $k\in \mathbb{N}_+$, is 2-factorable.

In the particular case when $k=2$ one has that every 4-regular graph is two factorable, i.e., there are two edge-disjoint 2-regular graphs.

\bibliography{Integrals_M_0n.bib}
\bibliographystyle{JHEP}

\end{document}